\documentclass[aop,seceqn,dvips]{arximspdf}
\usepackage{subenv}
\usepackage{graphicx}

\doi{10.1214/009117907000000231}
\volume{36}
\issue{2}
\pubyear{2008}
\firstpage{530}
\lastpage{593}

\makeatletter

\newtheorem{theorem}{Theorem}[section]
\newproclaim{remark}[theorem]{Remark}
\newproclaim{notation}{Notation}
\newtheorem{cor}[theorem]{Corollary}
\newtheorem{lemma}[theorem]{Lemma}
\newtheorem{prop}[theorem]{Proposition}

\makeatother

\begin{document}
\begin{frontmatter}

\title{Decay of correlations in nearest-neighbor
self-avoiding walk, percolation, lattice\\ trees
and animals}
\runtitle{Decay of correlations}
\pdftitle{Decay of correlations in nearest-neighbor
self-avoiding walk, percolation, lattice trees
and animals}

\begin{aug}
\author[A]{\fnms{Takashi} \snm{Hara}\corref{}\thanksref{t1}\ead[label=e1]{hara@math.kyushu-u.ac.jp}
\ead[label=u1,url]{http://www.math.kyushu-u.ac.jp/\texttildelow hara/}}
\thankstext{t1}{Supported in part by Grant-in-Aid for
Scientific Research (C) of the Ministry of Education,
Culture, Sports, Science and Technology of Japan.}
\runauthor{T. Hara}
\affiliation{Kyushu University}
\address[A]{Faculty of Mathematics \\
Kyushu University \\
Ropponmatsu, Chuo-ku, Fukuoka 810-8560\\
Japan \\
\printead{e1}\\
\printead{u1}}
\end{aug}

\received{\smonth{4} \syear{2005}}
\revised{\smonth{3} \syear{2007}}

%
\begin{abstract}
We consider nearest-neighbor self-avoiding walk, bond
percolation, lattice trees, and bond lattice animals on
${\mathbb Z}^d$.
The two-point functions of these models are respectively
the generating function for self-avoiding walks from the origin to
$x \in{\mathbb Z}^d$, the probability of a connection from the
origin to $x$, and the generating functions for lattice trees
or lattice animals containing the origin and $x$.
Using the lace expansion, we prove that the two-point
function at the critical point is asymptotic to
$\mathit{const.} |x|^{2-d}$ as $|x| \rightarrow\infty$, for $d \geq
5$ for
self-avoiding walk, for $d \geq19$ for percolation, and for
sufficiently large $d$ for lattice trees and animals.
These results are complementary to those of [\textit{Ann. Probab.}
\textbf{31} (2003) 349--408], where
spread-out models were considered.
In the course of the proof, we also provide
a sufficient (and rather sharp if $d > 4$) condition
under which the two-point function of a random walk
on ${ {\mathbb Z}^d }$ is asymptotic to $\mathit{const.} |x|^{2-d}$
as $|x| \rightarrow\infty$.
\end{abstract}

%
\begin{keyword}[class=AMS]
\kwd[Primary ]{82B27}
\kwd{82B41}
\kwd{82B43}
\kwd{82C41}
\kwd[; secondary ]{60K35}.
\end{keyword}
\begin{keyword}
\kwd{Critical behavior}
\kwd{two-point function}
\kwd{self-avoiding walk}
\kwd{percolation}
\kwd{lattice trees and animals}
\kwd{lace expansion}.
\end{keyword}

\end{frontmatter}

\section{Introduction}
\label{sc-int}

\subsection{The models and results}
\label{sb-results}

In this paper, we consider nearest-neighbor self-avoiding
walk, bond percolation, lattice trees, and bond lattice
animals on $d$-dimensional hypercubic lattice ${\mathbb Z}^d$,
and prove that their critical two-point
functions exhibit the Gaussian behavior, that is,
%
\begin{equation}
G_{p_c} (x) \sim\frac{\mathit{const.}}{|x|^{d-2}}\qquad
\mbox{as } |x| \rightarrow\infty,
\end{equation}
when $d$ is large.

We first define the models we consider.
A \textit{bond} is a pair of sites
$\{x,y\} \subset{\mathbb Z}^d$ with $|y-x| = 1$. For $n \geq0$,
an $n$-step \textit{walk} from $x$ to $y$ is a mapping
$\omega\dvtx \{0,1,\ldots, n\} \rightarrow{\mathbb Z}^d$ such that
$|\omega(i+1)-\omega(i)| = 1$ for $i=0,\ldots, n-1$,
with $\omega(0)=x$ and $\omega(n)=y$.
Let $\mathcal{W}(x,y)$ denote the set of walks
from $x$ to $y$, taking any number of steps.
An $n$-step \textit{self-avoiding walk} (SAW)
is an $n$-step walk $\omega$ such that
$\omega(i) \neq\omega(j)$ for each pair $i \neq j$.
Let $\mathcal{S}(x,y)$ denote the set of self-avoiding
walks from $x$ to $y$, taking any number of steps.
A \textit{lattice tree} (LT) is a finite connected set of bonds
which has no cycles. A \textit{lattice animal} (LA) is a finite
connected set of bonds which may contain cycles.
Although a tree $T$ is defined as a set of bonds,
we write $x \in T$ if $x$ is an endpoint of some bond of $T$,
and similarly for lattice animals.
Let $\mathcal{T}(x,y)$ denote the set of lattice trees containing
$x$ and $y$, and let $\mathcal{A}(x,y)$ denote
the set of lattice animals containing $x$ and $y$.
We often abbreviate lattice trees and animals as LTLA.

The random walk and self-avoiding walk \textit{two-point functions}
are defined respectively by
%
\begin{equation}
S_p(x) = \sum_{\omega\in\mathcal{W}(0,x)} p^{|\omega|},
\qquad
G_p(x) = \sum_{\omega\in\mathcal{S}(0,x)} p^{|\omega|},
\end{equation}
where $|\omega|$ denotes the number of steps of the
walk $\omega$.
For any $d>0$, $\sum_x S_p(x)$ converges for $p < (2d)^{-1}$
and diverges for $p>(2d)^{-1}$, and $p=(2d)^{-1}$
plays the role of a critical point.
It is well known \cite{Uchi98} that, for $d>2$,
%
\begin{equation}
\label{e:Cxdecay}
S_{1/2d}(x) \sim{\mathit{const.}}\frac{1}{|x|^{d-2}}\qquad
\mbox{as $|x| \rightarrow\infty$} .
\end{equation}
A standard subadditivity argument
\cite{HM54,Hugh95,MS93} implies that $\sum_x G_p(x)$
converges for $p<p_c$ and diverges for $p>p_c$, for some
finite positive critical value $p_c$.

The lattice tree and lattice animal two-point functions are
defined by
%
\begin{eqnarray}
\label{e:1.1}
G_p (x) & = & \sum_{T \in\mathcal{T}(0,x)} p^{|T|}\qquad
\mbox{(lattice trees)}, \nonumber\\[-8pt]\\[-8pt]
G_p (x) & = & \sum_{A \in\mathcal{A}(0,x)} p^{|A|}\qquad
\mbox{(lattice animals)},\nonumber
\end{eqnarray}
where $|T|$ and $|A|$ denote the number of bonds in $T$ and $A$,
respectively.
A standard subadditivity argument implies that there
are positive finite $p_c$ (depending on the model) such that
$\sum_x G_p(x)$ converges for $p<p_c$ and diverges for $p>p_c$
\cite{Klar67,Klei81}.

Turning now to bond percolation,
we associate independent Bernoulli random variables
$n_{\{x,y\}}$ to each bond $\{x,y\}$ (here $|x-y|=1$), with
%
\begin{equation}
{\mathbb P}\bigl(n_{\{x,y\}}=1\bigr) = p,\qquad
{\mathbb P}\bigl(n_{\{x,y\}}=0\bigr)=1- p ,
\end{equation}
where $p \in[0, 1]$.
A configuration is a realization of the bond
variables. Given a configuration,
a bond $\{x,y\}$ is called \textit{occupied} if
$n_{\{x,y\}}=1$ and otherwise is called \textit{vacant}.
The percolation \textit{two-point function} is defined by
%
\begin{equation}
G_p(x) = {\mathbb P}_p(0 \mbox{ and } x
\mbox{ are connected by occupied bonds}),
\end{equation}
where ${\mathbb P}_p$ is the probability measure on
configurations induced by the bond variables.
There is a critical value $p_c \in(0,1)$ such that
$\sum_{x} G_p(x) < \infty$ for $p \in[0,p_c)$
and $\sum_{x} G_p(x) = \infty$ for $p \geq p_c$ \cite{Grim99}.
This critical point can also be characterized by the fact that
the probability of existence of an infinite cluster of
occupied bonds is
$1$ for $p>p_c$ and $0$ for $p<p_c$ \cite{AB87,Mens86}.

We use $G_p$ and $p_c$ to denote the two-point function and
the critical point of these models,
although they are, of course, model-dependent. In what follows,
it will be clear from the context which model is intended.

Our main result is the following theorem.
\begin{theorem}
\label{thm-saw}
For nearest-neighbor self-avoiding walk in $d \geq5$ and for
percolation and lattice trees and animals in sufficiently high
dimensions, their critical two-point function $G_{p_c}(x)$
satisfies, as $|x| \rightarrow\infty$,
%
\begin{equation}
\label{e:Gxasmp.1}
G_{p_c}(x) = \frac{a_d A}{|x|^{d-2}}
+ O\biggl( \frac{1}{|x|^{d-2 + 2/d}} \biggr)\qquad
\mbox{with }
a_d = \frac{d\Gamma({d}/{2}-1)}{2\pi^{d/2}}
.
\end{equation}
Here $A$ is a model-dependent constant whose explicit form
is given in (\ref{e:A-value}) below,
in terms of quantities appearing in the lace expansion.
\end{theorem}

\begin{remark}
(i)
The error term of (\ref{e:Gxasmp.1}) is not optimal;
the error bound in the Gaussian lemma (Theorem \ref{th-gau})
is responsible for the current estimate. However, the author
has recently
succeeded in improving the error bound of Theorem \ref{th-gau}.
As a result, (\ref{e:Gxasmp.1}) has now been improved to
%
\begin{equation}
\label{e:Gxasmp.1_improved}
G_{p_c}(x) = \frac{a_d A}{|x|^{d-2}}
+ O\biggl( \frac{1}{|x|^{d}} \biggr)
.
\end{equation}
The proof of this improvement will be presented
elsewhere \cite{Hara07b}.

(ii)
For percolation, $d \geq19$ is sufficient for
the above theorem to hold.
The restriction $d \geq19$ comes from the fact that
convergence of the lace expansion has been proved
only in these dimensions.
This is far from the expected limit of $d$
($d > 6$ should be sufficient).
See, for example, \cite{HHS03} for the role played by the
critical dimension.

(iii)
The method of the present paper can also be applied to spread-out
models, and reproduces the asymptotic form proved in \cite{HHS03},
for self-avoiding walk in $d \geq5$, for percolation in $d \geq11$,
and for lattice trees/animals in $d \geq27$.
See the explanations around (\ref{e:perc-rec.1}) and
(\ref{e:LTLA-rec.1}) about how these restrictions
on the dimension arises.

(iv)
The method of the present paper can be applied to other models,
as long as we have a suitable lace expansion. An important
example is the Ising model in sufficiently
high dimensions \cite{Sakai06}.

(v)
The theorem provides a necessary input for a result of
Aizenman \cite{Aize97}, who proved, under certain assumptions on
the decay of critical two-point function, that
the largest percolation cluster present in
a box of side length $N$ are of size approximately
$N^4$ and are approximately $N^{d-6}$ in number.
Our theorem does prove the assumptions of Aizenman, and
thus establishes his result mentioned above for the
nearest-neighbor percolation in $d \geq19$.
(Similar input for spread-out models has been provided by \cite{HHS03}.)
\end{remark}

Results similar to the above have been proven
in \cite{HHS03}, where \textit{spread-out} models of
self-avoiding walk, bond percolation, and lattice trees/animals
were treated in a unified manner. (Spread-out models are
defined by considering all the pairs $\{x, y\}$ with
$0 < |x-y| \leq L$ as bonds, for some large $L$. This $L$
represents the range of the interaction, not
the system size.) However,
the method of \cite{HHS03} is not directly applicable to
nearest-neighbor models, for the following reason.
Critical two-point functions of spread-out models obey
as $|x| \rightarrow\infty$ \cite{HHS03}, Theorem 1.2:
%
\begin{equation}
\label{e:Gxasmp.L1}
G_{p_c}(x) \sim\frac{a_d A}{\sigma^2 |x|^{d-2}},
\end{equation}
where $A$ is a model-dependent constant close to $1$, and
$\sigma^2$ is a constant which is of the order of $L^2$.
For the spread-out model, by taking $L$ sufficiently large (for
fixed $d$), we can always make the coefficient $a_d A/\sigma^2$
as small as we want. Therefore, the lace expansion diagrams
converge if we assume $G(x)$ is bounded by, say,
twice of the right hand of (\ref{e:Gxasmp.L1}).
This makes it possible to prove convergence of the lace expansion
in a self-consistent way based on the asymptotic form;
the result of \cite{HHS03} was in fact proved in this manner.

In contrast, for the nearest-neighbor model, there is no
$\sigma^2$ to cancel $a_d$, which is quite large for large $d$
[$a_d \approx(d/2)!$].
This means the asymptotic form of
(\ref{e:Gxasmp.1}) is much bigger than the true behavior of $G(x)$
for small $x$, and it would be difficult to prove the convergence of
the lace expansion using this asymptotic form. In this paper, we
bypass this difficulty by borrowing convergence results
from previous works, and take a different approach
from that of \cite{HHS03}.

\begin{notation*}
For $a, b \in{\mathbb R}$, we write
$a \vee b = \max\{a, b\}$, and $a \wedge b = \min\{a, b\}$.
The greatest integer $n$ which satisfies $n \leq x$ is denoted by
$\lfloor x \rfloor$. The smallest integer $n$ which satisfies
$n \geq x$ is denoted by $\lceil x \rceil$.

The Euclidean norm of $x \in{ {\mathbb R}^d}$ is
denoted by $|x|$, and we write $|\!|\!|x |\!|\!|:= |x| \vee1$.

The indicator of an event $A$ is denoted by $I[A]$.

A convolution on ${ {\mathbb Z}^d }$ is denoted by $*\dvtx
(f*g)(x):= \sum_{y \in{ {\mathbb Z}^d }} f(x-y) g(y)$.

Given a function $f(x)$ on ${ {\mathbb Z}^d }$, we define its Fourier
transform as
%
\begin{eqnarray}
\label{e:Fourier-def.1}
{\hat{f}}(k) :\!&=& \sum_{x \in{ {\mathbb Z}^d }} f(x) e^{- i k\cdot x}
\qquad\mbox{so that}\nonumber\\[-8pt]\\[-8pt]
f(x) &=& \int_{[-\pi,\pi]^{d}}\frac{d^d k}{(2\pi)^d}e^{ik\cdot x}
{\hat{f}}(k) ,\nonumber
\end{eqnarray}
when both equations make sense.
When the sum defining ${\hat{f}}(k)$ is not well defined [i.e.,
when $f(x) = G_{p_c}(x)$], we interpret ${\hat{f}}(k)$
by the second identity above more details are given
in\vadjust{\goodbreak} Appendix \ref{sb-ap-Jcont}.

A function $f(x)$ on ${ {\mathbb Z}^d }$ is called
\textit{${ {\mathbb Z}^d }$-symmetric}, if
it is invariant under the ${ {\mathbb Z}^d }$-symmetries of
reflection in coordinate hyperplanes and rotation by $90^\circ$.

We denote a positive constant by $c$.
On each appearance $c$ may change
its value, even in a single equation.
We write $f(x) \approx g(x)$ when there are finite
and positive constants $c_1, c_2$ such
that $c_1 g(x) \leq f(x) \leq c_2 g(x)$ for all $x$.
We also use large-$O$ and small-$o$ notation:
$f(x) = O(g(x))$ means $f(x)/g(x)$ remains bounded,
while $f(x) = o(g(x))$ means $f(x)/g(x) \rightarrow0$,
as $x \rightarrow\infty$ (or $x \rightarrow0$, depending on the context).
Constants $c$ and large-$O/$small-$o$'s could depend
on other parameters. We explain these dependencies
on each occurrence if necessary.

We make use of the following quantities
($a, b \in{ {\mathbb Z}^d }$ and $\alpha, \beta, \gamma\geq0$,
and the summations run over ${ {\mathbb Z}^d }$):
\begin{eqnarray}\qquad\qquad\quad
G^{(\alpha)}(a)
 :\!&=& |a|^{\alpha} G(a),
\label{e:Gbardef}\\
B(a)
:\!&=&\sum_{y\neq0} G(y) G(a-y) , \label{e:Ba-def}\\
W^{(\beta,\gamma)}(a)
:\!&=&\sum_{y} G^{(\beta)}(y) G^{(\gamma)}(a-y)
= \bigl(G^{(\beta)} * G^{(\gamma)}\bigr)(a),
\label{e:Wa-def}\\
T^{(\beta, \gamma)}(a)
:\!&=&\sum_{x,y} G^{(\beta)}(x) G^{(\gamma)}(y-x) G(a-y)
\{ 1 - I[x=y=a=0] \}
\nonumber\label{e:Ta-def}\\[-8pt]\\[-8pt]
& =&
\bigl(G^{(\beta)}* G^{(\gamma)}* G\bigr)(a)
- I[ a = 0 \mbox{ and } \beta=\gamma=0] G(0)^3 ,
\nonumber\\
S^{(\gamma)}(a)
:\!&=&\bigl(G^{(\gamma)}* G* G *G\bigr)(a)
- I[ a = 0 \mbox{ and } \gamma=0] G(0)^4,
\label{e:Sa-def}\\
P(a)
:\!&=&(G*G* G *G * G)(a) ,\label{e:Pa-def}
\\
H^{(\beta)}(a, b)
:\!&=&\sum_{x, y, z, u, v} G(z) G(u) G(x-u)
G^{(\beta)}(x) G(y-x) G(v-u)\nonumber \label{e:Ha-def}\\[-8pt]\\[-8pt]
&&\hspace*{32.1pt}{}\times G(z+a-v) G(y+b-v)
.\nonumber
\end{eqnarray}
%
Diagrammatic representations for these quantities are given in
Figure \ref{fig-diag1}(a) of Section \ref{ss-diagram-notes}.
We denote suprema (over $a, b \in{ {\mathbb Z}^d }$) of these
quantities by bars, that is,
$\bar{G}^{(\alpha)} := \sup_{a} |a|^\alpha G(a)$,
$\bar{B} := \sup_{a} B(a)$,
$\bar{W}^{(\beta, \gamma)} := \sup_{a} W^{(\beta, \gamma)}(a)$,
and so on.
These of course depend on $p$, but we usually omit the subscript
$p$, because we almost always consider these quantities at
criticality, $p=p_c$.
\end{notation*}

\subsection{Framework of the proof}
\label{sb-methods}

In this section, we explain the framework of the proof
of our main result, Theorem \ref{thm-saw}, and reduce its
proof to several propositions.
We give a complete proof of Theorem \ref{thm-saw}
for self-avoiding walk in $d \geq5$, but only give a proof
for large $d$
(say $d \geq30$) for percolation. Results for percolation
in $d \geq19$ can be obtained by more detailed diagrammatic
estimates which slightly improve conditions in
Lemmas \ref{lem-Pix} and  \ref{lem-Pihat}.
The extra work required
for percolation near $d = 19$ is essentially the same as
the analysis\vadjust{\goodbreak} used to prove the convergence of the lace
expansion in $d = 19$ (announced in \cite{HS94}),
and is not reproduced here.

\subsubsection{The lace expansion}
\label{subsub-lace-expansion}
For self-avoiding walk in $d \geq5$, for percolation
in $d \geq19$, and for lattice trees$/$animals in $d \gg1$,
we have a convergent expansion, called the
\textit{lace expansion}, which provides a useful expression for
two-point functions.
The literature on lace expansion has increased rapidly,
and we here list only a few which will be directly relevant for the
present paper \cite{BS85,HS90a,HS90b,HS92b,HS92a,Slad87}.
Good reviews will be found in \cite{HS94,MS93,Slad04}.

\begin{prop}
\label{thm-TW-90a}
For self-avoiding walk in $ d\geq5$, for percolation in $d \geq19$,
and for lattice trees/animals in sufficiently high dimensions,
the two-point function for $p \leq p_c$ is represented as
%
\begin{equation}
\label{e:Gsawx.1}
G_p(x) = \int_{[-\pi,\pi]^{d}}\frac{d^d k}{(2\pi)^d}e^{ikx} {\hat
{G}}_p(k),\qquad
{\hat{G}}_p(k) :=
\frac{{\hat{g}}_p(k)}{1 - {\hat{J}}_p(k)}
.
\end{equation}
Here
%
\begin{eqnarray}
&D(x)
:= \dfrac{1}{2d} \delta_{|x|,1} ,\qquad
{\hat{D}}(k) = \displaystyle\sum_{x} e^{-ik\cdot x} D(x)
= \dfrac{1}{d} \displaystyle\sum_{j=1}^d \cos k_j ,&
\label{e:D-def.1}\\
&{\hat{J}}_p(k)
 :=
\cases{
2 d p {\hat{D}}(k) + {\hat{\Pi}}_p(k), &\quad (SAW), \cr
2 d p {\hat{D}}(k) \{1 + {\hat{\Pi}}_p(k)\}, &\quad (percolation),
\cr
2 d p {\hat{D}}(k) \{ G_p(0) + {\hat{\Pi}}_p(k) \}, &\quad(LTLA),
}&
\label{e:Jhat-def.1}\\
&{\hat{g}}_p(k)
 :=
\cases{
1, & \quad(SAW), \cr
1 + {\hat{\Pi}}_p(k), &\quad (percolation), \cr
G_p(0) + {\hat{\Pi}}_p(k), &\quad (LTLA),}&
\\
&\hat{\Pi}_p(k)
 := \displaystyle\sum_x \Pi_p(x) e^{-ik\cdot x} ,\qquad
\Pi_p(x) = \displaystyle\sum_{n=0}^\infty(-1)^n \Pi_p^{(n)} (x) ,&
\label{e:Pihat-def.1}\end{eqnarray}
and $\Pi_p^{(n)}(x)$ is a nonnegative function of $x$.
Moreover,
there are positive constants $c$, $c_1$ through $c_4$
which are independent of $p$ and $d$, a constant $\lambda\in(0,1)$
which is independent of $p$, and a positive function $h^{(n)}(x)$,
such that for $p \leq p_c$,
%
\begin{eqnarray}
&0 \leq\Pi_p^{(n)}(x) \leq h^{(n)}(x),\qquad
\displaystyle\sum_{x} \displaystyle\sum_{n=0}^\infty h^{(n)}(x) \leq\frac{c}{d} ,&\label{e:tr-Pi-90aa}
\\
&\displaystyle\sum_{x} |x|^2 | \Pi_p(x) | \leq\frac{c}{d} ,
\qquad
c_1 \frac{|k|^2}{d} \leq{\hat{J}}_p(0) - {\hat{J}}_p(k)
\leq c_2 \frac{|k|^2}{d} ,&\label{e:tr-Pi-90}
\\
&0 \leq{\hat{G}}_p(k) \leq\dfrac{c d}{|k|^2}\qquad
\mbox{(infrared bound)}&\label{e:tr-Ghat-90}
\end{eqnarray}
and
\begin{eqnarray}\qquad\qquad
 \bar{G}_p^{(2)} &<& \lambda,\
\bar{B}_p < \lambda,
\qquad\hspace*{95.9pt}
(\mbox{SAW in }d \geq5),\label{e:tr-GB-90}
\\
\bar{W}_p^{(2,0)} &<& \lambda,\qquad
\bar{T}_p^{(0,0)} < \lambda,\qquad
\bar{H}_p^{(2)} < c\qquad
(\mbox{percolation in } d \geq19),
\label{e:tr-TW-90aNN}\\
\bar{T}_p^{(2,0)} &<& \lambda,\
\bar{S}_p^{(0)} < \lambda,\
1 \leq G_{p}(0) \leq4
\qquad\hspace*{26.3pt}(\mbox{LTLA in } d \gg1).
\label{e:tr-LTLA-90aNN}
\end{eqnarray}
$\lambda$ satisfies
%
\begin{equation}
\lambda\leq
\cases{ 0.493,
& \quad(SAW in  $d \geq5$),
\cr
\dfrac{c_3}{d}, &\quad (percolation in  $d \geq19$
and LTLA in  $d \gg1$).
}
\end{equation}
The critical point $p=p_c$ is characterized by
%
\begin{equation}
\label{e:crit-char.1}
{\hat{J}}_{p_c} (0) = 1
,
\end{equation}
and satisfies
%
\begin{eqnarray}
\label{e:pc-bd.1}
1 & \leq&2 d p_c \leq1 + c_4 \lambda\qquad\hspace*{31.5pt}
(\mbox{SAW$/$percolation}), \nonumber\\[-8pt]\\[-8pt]
1 & \leq&2 d p_c G_{p_c}(0) \leq1 + c_4 \lambda\qquad
(\mbox{LTLA}).\nonumber
\end{eqnarray}
\end{prop}
For self-avoiding walk, $\Pi_p^{(0)}(x) \equiv0$ for all $x$,
and the sum over $n$ in (\ref{e:Pihat-def.1}) starts from $n = 1$.
For percolation, our ${\hat{\Pi}}_p(k)$ is the same as that of
\cite{HHS03}, but differs from that of
\cite{HS90a} by the factor $2 dp {\hat{D}}(k)$ and is equal
to ${\hat{g}}_p(k)$ of that paper.

The above proposition is a slightly improved version of
the results obtained previously.
We briefly explain how to prove the above proposition in
Appendix \ref{sb-ap-Jcont}.

\textit{In the following, we concentrate on quantities at $p=p_c$
\textup{(}except stated otherwise\textup{),} and omit the subscript $p$
or $p_c$ altogether.}

\subsubsection{Gaussian lemma}
\label{subsub-gaussian}
Our main results are proved by making use of the following theorem
and its corollary, which give sufficient conditions for
the Gaussian behavior, $G(x) \sim\mathit{const.} |x|^{2-d}$,
for two-point functions of random walks and related models.
For ${\mathbb Z}^d$-symmetric (not necessarily positive)
functions $J(x)$ and $g(x)$, we define
%
\begin{eqnarray}
\label{e:GxHxgau.def}
C(x) &:= &\int_{[-\pi, \pi]^{d}} \frac{d^d k}{(2\pi)^d}
\frac{e^{i k\cdot x}}{1 - {\hat{J}}(k)} ,\nonumber\\[-8pt]\\[-8pt]
H(x) &:=& \int_{[-\pi, \pi]^{d}} \frac{d^d k}{(2\pi)^d}e^{i
k\cdot x}
\frac{{\hat{g}}(k)}{1 - {\hat{J}}(k)} .\nonumber
\end{eqnarray}

\begin{theorem}
\label{th-gau}
Let $d \geq3$.
Suppose ${ {\mathbb Z}^d }$-symmetric $J(x)$
satisfies\vadjust{\goodbreak} with finite positive $K_{0}$ through $K_{3}$
\begin{eqnarray}\qquad\qquad
&{\hat{J}}(0) := \displaystyle\sum_{x} J(x) = 1 ,\qquad
{\hat{J}}(0) - {\hat{J}}(k) \geq K_{0} \dfrac{|k|^{2}}{2d}
\qquad(k \in[-\pi, \pi]^d),&\label{e:J-cnd.1}
\\
&\displaystyle\sum_{x} |x|^{2} J(x) := K_{1},\qquad
\displaystyle\sum_{x} | x |^{2} | J(x) | \leq K_{2} ,&
\label{e:J-cnd.2}\\
&| J(x) | \leq\dfrac{K_{3}}{|\!|\!|x |\!|\!|^{d+2}}.&\label{e:J-cnd.3}
\end{eqnarray}
Then $C(x)$ of (\ref{e:GxHxgau.def})
is well defined and satisfies as $|x| \rightarrow\infty$:
%
\begin{equation}
\label{e:gauss-asmp.1}
C(x) \sim\frac{a_{d}}{K_{1}} \frac{1}{|x|^{d-2}} .
\end{equation}

Suppose further that $J(x)$ satisfies
%
\begin{equation}
\label{e:J-cnd.3'}
\sum_{x} |x|^{2+\rho} |J(x) | < K_2',\qquad
| J(x) | \leq\frac{K_3'}{|\!|\!|x |\!|\!|^{d+2+\rho}}
\end{equation}
with finite positive $\rho, K_2', K_3'$.
Then $C(x)$ satisfies
%
\begin{equation}
\label{e:gauss-asmp.1a}
C(x) = \frac{a_{d}}{K_{1}} \frac{1}{|\!|\!|x |\!|\!|^{d-2}}
+ O \biggl( \frac{1}{|\!|\!|x |\!|\!|^{d-2+(\rho\wedge2)/d}} \biggr)
.
\end{equation}
\end{theorem}
Section \ref{sc-gau} gives a complete proof of the theorem.

\begin{remark}
(i)
The above $C(x)$ is the two-point function of the
Gaussian spin system
whose spins at $x$ and $y$ interact with $J(x-y)$.
When $J(x) \geq0$, $C(x)$ can also be interpreted as the Green's
function of the random walk whose transition probability from
$x$ to $y$ is given by $J(x-y)$.
We are allowing $J(x) <0$, because $\Pi(x)$ is not
necessarily positive in our lace expansion (\ref{e:Gsawx.1}).

(ii)
The pointwise bound (\ref{e:J-cnd.3}) is sharp in $d > 4$,
in the sense that there are models which mildly violate
this condition and which do not exhibit the Gaussian behavior
of (\ref{e:gauss-asmp.1}).
Details will be given in Section \ref{sub-counter-gauss}.
For $d = 4$ and $5$, the fact that (\ref{e:J-cnd.3}) is sufficient
for nonnegative $J$'s
has been pointed out by Uchiyama \cite{Uchi98}.
The author has recently learned that Lawler \cite{Lawl04a} has also
shown that (\ref{e:J-cnd.3}) is sufficient for $d > 4$ for
nonnegative $J$.

(iii)
For $d \leq4$, the uniform bound (\ref{e:J-cnd.3}) will not be
sharp. Sharp conditions when $J \geq0$ are
$\sum_x |x|^2 J(x) < \infty$ for $d < 4$ \cite{Uchi98},
and $\sum_{x\dvtx |x|\geq r} J(x) = o(\frac{1}{r^2 \log r})$
for $d =4$ \cite{Lawl94a}.

(iv)
The error bound in (\ref{e:gauss-asmp.1a}) is not optimal.
However, the author has recently succeeded in proving a
better (and hopefully optimal) error bound, according to
which (\ref{e:gauss-asmp.1a}) is improved to
%
\begin{equation}
\label{e:gauss-asmp.1a_improved}
C(x) = \frac{a_{d}}{K_{1}} \frac{1}{|\!|\!|x |\!|\!|^{d-2}}
+ O \biggl( \frac{1}{|\!|\!|x |\!|\!|^{d-2+(\rho\wedge2)}} \biggr)
.
\end{equation}
The proof of this improvement is somewhat lengthy, and
will be presented elsewhere \cite{Hara07b}.
For nonnegative $J$'s, the error bound like
(\ref{e:gauss-asmp.1a_improved}) has been obtained
by Lawler \cite{Lawl04a}.
\end{remark}

\begin{cor}
\label{prop-gau}
Let $d \geq3$.
Let $J(x)$ and $g(x)$ be ${ {\mathbb Z}^d }$-symmetric
functions. Suppose $J(x)$ satisfies
(\ref{e:J-cnd.1})--(\ref{e:J-cnd.3}), and $g(x)$ satisfies
\begin{equation}
\label{e:g-cnd.1aa}
\sum_{x} | g(x) | < \infty,\qquad
| g(x) | \leq\frac{K_4}{|\!|\!|x |\!|\!|^{d}}
\end{equation}
with finite positive $K_4$.
Then, $H(x)$ of (\ref{e:GxHxgau.def}) is well defined and
satisfies as \mbox{$|x| \rightarrow\infty$}
%
\begin{equation}
\label{e:Hx-asmp}
H(x) \sim\frac{\sum_{y} g(y)}{\sum_{y} |y|^2 J(y)}
\frac{a_{d}}{|x|^{d-2}} .
\end{equation}
Suppose further $J(x)$ satisfies (\ref{e:J-cnd.3'})
and $g(x)$ satisfies
%
\begin{equation}
\label{e:J-cnd.3'aaa}
| g(x) | \leq\frac{K_4'}{|\!|\!|x |\!|\!|^{d+\rho}}
\end{equation}
with finite positive $\rho, K_4'$.
Then $H(x)$ satisfies as $|x| \rightarrow\infty$
%
\begin{equation}
\label{e:gauss-asmp.1aaa}
H(x) = \frac{\sum_{y} g(y)}{\sum_{y} |y|^2 J(y)}
\frac{a_{d}}{|x|^{d-2}}
+ O \biggl( \frac{1}{|x|^{d-2+(\rho\wedge2)/d}} \biggr)
.
\end{equation}
\end{cor}

Corollary \ref{prop-gau} follows immediately
from Theorem \ref{th-gau} and a basic property of
convolutions, Lemma \ref{lem-conv}(iv).
This is because $H(x) = (C*g)(x)$, where $*$ denotes convolution.

We intend to apply the above corollary to the representation
of two-point functions by the lace expansion, (\ref{e:Gsawx.1}).
If the corollary can in fact be applied (with $\rho= 2$),
then it proves Theorem \ref{thm-saw} with
%
\begin{equation}
\label{e:A-value}
A :=
\frac{\sum_{y} g(y)}{\sum_{y} |y|^2 J (y)} ,
\end{equation}
with $J$ and $g$ given by (\ref{e:Jhat-def.1}).
The question is whether we can really apply the proposition.
For this, note that (\ref{e:J-cnd.1}) and
(\ref{e:J-cnd.2}) follow
directly from Proposition \ref{thm-TW-90a} at $p = p_c$.
Therefore, it suffices to prove
pointwise $x$-space bound (\ref{e:J-cnd.3}) and (\ref{e:J-cnd.3'}).
[Because $J(x)$ and $g(x)$ are essentially the same,
(\ref{e:g-cnd.1aa}) and
(\ref{e:J-cnd.3'aaa}) for $g(x)$ are automatically satisfied if
$J(x)$ satisfies (\ref{e:J-cnd.3}) and (\ref{e:J-cnd.3'}).]

\subsubsection{Reduction of the proof to an estimate on the
two-point function}
\label{subsub-reduce-to-2pt}

The condition (\ref{e:J-cnd.3}) is about the decay of $J(x)$,
but its sufficient condition can be
given in terms of $G(x)$ with the help of the following lemma,
which turns an $x$-space bound on $G(x)$ into that on $\Pi(x)$.

\begin{lemma}
\label{lem-Pix}
Consider SAW, percolation, or LTLA for which
Proposition \textup{\ref{thm-TW-90a}} holds. Suppose we have a bound
%
\begin{equation}
\label{e:PixS-ass}
G(x) \leq\frac{\beta}{|\!|\!|x |\!|\!|^{\alpha}}
\end{equation}
with $\beta> 0$ and $0 < \alpha< d$.
Then for $x \neq0$,
%
\begin{equation}
\label{e:Pix-saw-bd}
| \Pi(x) | \leq
\cases{
c \beta^3 |\!|\!|x |\!|\!|^{-3 \alpha},
 \cr
\qquad\mbox{(SAW with  $\lambda< 1$),} \cr
c \beta^{2} |\!|\!|x |\!|\!|^{- 2 \alpha},\cr
\qquad\mbox{(percolation in  $d > 8$ with  $\lambda\ll1$),} \cr
c (\beta^{2} \vee\beta^{4}) |\!|\!|x |\!|\!|^{-(4 \alpha-2d)},\cr
\qquad\mbox{(LTLA in  $d > 10$ with  $\lambda\ll1,
\alpha> d/2$)}
}
\end{equation}
with a $\lambda$-dependent constant $c$.
\end{lemma}

This lemma is proved in
Sections \ref{ss-prf-PixS}, \ref{ss-prf-lem-Pix}
and \ref{ss-prf-lem-PixLTLA}.
The restriction $d > 8$ (for percolation) and $d >10$ (for LTLA)
is unnatural, but is present for technical reasons.
Also the exponent $4\alpha-2d$ for LTLA will
not be optimal; the optimal result would give $|\Pi(x)| =
O(|\!|\!|x |\!|\!|^{3\alpha-d})$, as proved for spread-out models
in Proposition 1.8 of \cite{HHS03}.
These facts will reflect some limitations of our current method,
but the lemma still suffices for our purpose.

Employing Lemma \ref{lem-Pix}, one can immediately conclude
that a sufficient condition for (\ref{e:J-cnd.3}) is
%
\begin{equation}
\label{etau3saw}
G(x) \leq\frac{c}{|\!|\!|x|\!|\!|^{\alpha}}\qquad
\mbox{with }
\alpha=
\cases{
\dfrac{d+2}{3}, & \quad(SAW), \cr
\dfrac{d+2}{2}, &\quad (percolation), \cr
\dfrac{3d+2}{4}, &\quad (LTLA)
}
\end{equation}
with some constant $c$, together with $d > 4$ (SAW),
$d > 8$ (percolation), and $d > 10$ (LTLA)
[and $\lambda<1$ for SAW, $\lambda\ll1$ for percolation$/$LTLA].
We now show that this is sufficient for (\ref{e:J-cnd.3'})
as well.
Once we have (\ref{e:J-cnd.3}), Theorem \ref{th-gau}
establishes (\ref{e:gauss-asmp.1}).
We can then use (\ref{e:gauss-asmp.1}) as an input to
Lemma \ref{lem-Pix}, and get
%
\begin{equation}
|\Pi(x)| \leq
\cases{
c |\!|\!|x|\!|\!|^{-3(d-2)}, &\quad (SAW), \cr
c |\!|\!|x|\!|\!|^{-2(d-2)}, &\quad (percolation), \cr
c |\!|\!|x|\!|\!|^{-(2d-8)}, &\quad (LTLA).
}
\end{equation}
This in turn implies $| J(x) | , |g(x)|
\leq c |\!|\!|x|\!|\!|^{-(d+2 + \rho)}$, with
%
\begin{equation}
\rho=
\cases{
2(d-4), &\quad (SAW), \cr
d-6, &\quad (percolation), \cr
d-10, &\quad (LTLA).
}
\end{equation}
This establishes (\ref{e:J-cnd.3'}) with $\rho=2$
(for $d \geq5$ for SAW and for sufficiently high $d$ for
percolation$/$LTLA).

Our task has thus been reduced to proving (\ref{etau3saw}).

\subsubsection[Proving the estimate (1.47) on two-point functions from two lemmas]{Proving the estimate \textup{(\protect\ref{etau3saw})}
on two-point functions from two lemmas}
\label{subsub-prove-2pt}

To prove (\ref{etau3saw}), we use two lemmas. The first one
is our second diagrammatic lemma, which turns bounds on
weighted quantities of (\ref{e:Gbardef})--(\ref{e:Ha-def})
into those on $\sum_x |x|^{\alpha} | \Pi(x) |$
with some power $\alpha$.
\begin{lemma}
\label{lem-Pihat}
Consider SAW, percolation or LTLA for which
Proposition \textup{\ref{thm-TW-90a}} holds\textup{:}

\textup{(i)} For SAW with $\lambda< 1$, suppose $\bar{G}^{(\alpha)}$ and
$\bar{W}^{(\beta, \gamma)}$ are finite for some
$\alpha, \beta, \gamma\geq0$.
Then,
%
\begin{equation}
\label{e:PixS-wght.1}
\sum_{x} |x|^{\alpha+\beta+\gamma}
| \Pi(x) | < \infty
.
\end{equation}

\textup{(ii)} For percolation with $\lambda$ sufficiently small,
suppose $\bar{W}^{(\beta, \gamma)}$, $\bar{T}^{(0,\gamma)}$
and $\bar{H}^{(\beta)}$ are finite for some
$\beta, \gamma\geq0$.
Then,
%
\begin{equation}
\label{e:Pix-wght.1}
\sum_{x} |x|^{\beta+\gamma} | \Pi(x) | < \infty.
\end{equation}

\textup{(iii)} For LTLA with $\lambda$ sufficiently small,
suppose $\bar{T}^{(\beta, \gamma)}$, $\bar{S}^{(\gamma)}$
are finite for some $\beta, \gamma\geq0$.
Then,
%
\begin{equation}
\label{e:PixLTLA-wght.1}
\sum_{x} |x|^{\beta+\gamma} | \Pi(x) | < \infty.
\end{equation}
\end{lemma}

This lemma is proved in Sections \ref{ss-prf-PihatS},
\ref{ss-prf-lem-Pihat} and \ref{ss-prf-lem-PihatLTLA}.

Our second lemma is complementary to Lemma \ref{lem-Pihat},
and turns a bound on $\sum_x |x|^* | \Pi(x) |$ into
those on $\bar{G}^{(\alpha)}$, $\bar{W}^{(\beta, \gamma)}$,
$\bar{T}^{(\beta, \gamma)}$ and $\bar{S}^{(\gamma)}$.

\begin{lemma}
\label{lem-PiToGbarWT}
Suppose we have the expression
(\ref{e:Gsawx.1})--(\ref{e:Pihat-def.1}) of $G$ in terms
of the lace expansion.
Suppose further
%
\begin{equation}
\label{e:PiToGbarWT.1}
\sum_{x} |x|^{\phi} | \Pi(x) | < \infty
\end{equation}
for some $\phi> 1$.
Then, we have for nonnegative $\alpha, \beta, \gamma$
which are not odd integers\textup{:}
\begin{eqnarray}
\bar{G}^{(\alpha)} &<& \infty
\qquad \mbox{if }
\alpha\leq\phi\mbox{ and }
\alpha< d -2,
\label{e:PiToGbarWT.2}\\
\bar{W}^{(\beta, \gamma)} &<& \infty
\qquad \mbox{if }
\beta, \gamma\leq\lfloor\phi\rfloor,
\beta+ \gamma< d-4
\nonumber\label{e:PiToGbarWT.3p}\\[-8pt]\\[-8pt]
&&\hspace*{34.1pt} \mbox{and }
\beta+\gamma- (\lfloor\beta\rfloor+ \lfloor\gamma\rfloor)
< 1, \nonumber
\\
\bar{T}^{(\beta, \gamma)} &<& \infty
\qquad \mbox{if }
\beta, \gamma\leq\lfloor\phi\rfloor,
\beta+ \gamma< d-6
\nonumber\label{e:PiToGbarWT.4p}\\[-8pt]\\[-8pt]
&&\hspace*{34.1pt} \mbox{and }
\beta+\gamma- (\lfloor\beta\rfloor+ \lfloor\gamma\rfloor)
< 1,
\nonumber
\\
\bar{S}^{(\gamma)} &<& \infty
\qquad \mbox{if }
\gamma\leq\lfloor\phi\rfloor\mbox{ and }
\gamma< d-8,
\label{e:PiToGbarS.4p}\\
\bar{H}^{(\beta)} &<& \infty
\qquad \mbox{if }
\beta\leq\lfloor\phi\rfloor,
\beta< d-4, \mbox{ and }
d > 6.\label{e:PiToGbarH.4p}
\end{eqnarray}
\end{lemma}

Odd integers are excluded to make the proof simpler.
This restriction could be removed with some extra work,
but the lemma is sufficient for our purpose in its current form.

We now explain how to prove (\ref{etau3saw}) based on these
lemmas.
The basic idea is to use these lemmas repeatedly, and prove
$\bar{G}^{(\alpha)}$ is finite for $\alpha$ required
in (\ref{etau3saw}).
Consider SAW in $d \gg1$.
From (\ref{e:tr-Pi-90}) of Proposition \ref{thm-TW-90a},
$\sum_{x} |x|^2 |\Pi(x)|$ is finite. We start from this and
use Lemmas \ref{lem-PiToGbarWT} and \ref{lem-Pihat}
repeatedly, and see the quantities in the following sequence
are all finite (we choose $\beta=0$):
%
\begin{eqnarray}\hspace*{37pt}
\sum_{x} |x|^2 |\Pi(x)|
\stackrel{\mathrm{Lem\ \fontsize{8.36}{8.36}\selectfont{\ref{lem-PiToGbarWT}}}}{\longrightarrow}
\bar{G}^{(2)} ,
&&\hspace*{-1pt}\bar{W}^{(0,2)}
\stackrel{\mathrm{Lem\ \fontsize{8.36}{8.36}\selectfont{\ref{lem-Pihat}}}}{\longrightarrow}
\sum_{x} |x|^4 |\Pi(x)|\nonumber\\[-8pt]\\[-8pt]
&&
\stackrel{\mathrm{Lem\ \fontsize{8.36}{8.36}\selectfont{\ref{lem-PiToGbarWT}}}}{\longrightarrow}
\bar{G}^{(4)} , \ \bar{W}^{(0,4)}
\stackrel{\mathrm{Lem\ \fontsize{8.36}{8.36}\selectfont{\ref{lem-Pihat}}}}{\longrightarrow}
\cdots\,.\nonumber
\end{eqnarray}
The exponents $\phi, \alpha, \gamma$ are doubled in each
iteration, and we can continue as far as the exponents satisfy
the conditions of Lemma \ref{lem-PiToGbarWT}, that is,
$\alpha< d-2$ and $\gamma< d-4$.
For large $d$, $\alpha$ eventually exceeds $\frac{d+2}{3}$
required in (\ref{etau3saw}), and we are done.
For small $d$, it may not be so clear that $\alpha$ can
exceed $\frac{d+2}{3}$, still satisfying $\gamma< d-4$.
In the following we give a rigorous proof, focusing on this point.

\begin{pf*}{Proof of (\protect\ref{etau3saw}), assuming
Lemmas \protect\ref{lem-Pihat} and \protect\ref{lem-PiToGbarWT}}
We begin with SAW in $d > 4$.
Suppose $\sum_x |x|^{\phi_{i}} |\Pi(x)|$ is finite
for some $\phi_{i} \geq2$, and define
%
\begin{eqnarray}
\label{e:saw-rec.1}
 \alpha_{i+1} &=& 2,\qquad
\gamma_{i+1} = \{(d-4) \wedge\lfloor\phi_{i} \rfloor\}
-\varepsilon, \nonumber\\[-8pt]\\[-8pt]
 \phi_{i+1} &=& \alpha_{i+1} + \gamma_{i+1}
= \{(d-2) \wedge( \lfloor\phi_{i} \rfloor+ 2 ) \}
-\varepsilon
,\nonumber
\end{eqnarray}
with $0 < \varepsilon\ll1$. Then
Lemma \ref{lem-PiToGbarWT} shows that $\bar{G}^{(\alpha_{i+1})}$
and $W^{(0, \gamma_{i+1})}$ are finite.
Using this as an input to Lemma \ref{lem-Pihat}, we see
that $\sum_x |x|^{\phi_{i+1}} |\Pi(x)|$ is finite as well,
as long as $\phi_{i+1}$ is given by (\ref{e:saw-rec.1}).

We start from $\phi_{0} = 2$, and repeat the above procedure.
First three iterations for $\phi_i$ read:
%
\begin{eqnarray}
\label{e:saw-rec.5}
\phi_0 &=& 2,\qquad\hspace*{71.3pt}
\phi_1 = \{(d-2) \wedge4\} -\varepsilon,\nonumber\\[-8pt]\\[-8pt]
\phi_2 &=& \{(d-2) \wedge5 \} - \varepsilon,\qquad
\phi_3 = \{(d-2) \wedge6 \} - \varepsilon
.\nonumber
\end{eqnarray}
As the above shows, $\phi_{i}$ is increased by one in each
iteration, until it finally reaches $d-2-\varepsilon$.
This in particular means $\sum_{x} |x|^{\phi} |\Pi(x)|$ is
finite with $\phi= d-2 -\varepsilon$.

Using Lemma \ref{lem-PiToGbarWT} with $\phi= d-2-\varepsilon$
then implies that $\bar{G}^{(\alpha)}$ is finite with
$\alpha= d-2-\varepsilon$, or
$G(x) = O(|x|^{-(d-2-\varepsilon)})$.
This is sufficient for (\ref{etau3saw}), as long as
$\frac{d+2}{3} < d-2-\varepsilon$, or $d > 4 + 3\varepsilon/2$.
Because $\varepsilon> 0$ is arbitrary,
this proves (\ref{etau3saw}) for SAW in $d > 4$.\vadjust{\goodbreak}

The proof proceeds in a similar fashion for percolation,
using $\bar{W}^{(\beta, \gamma)}$, $\bar{T}^{(0, \gamma)}$,
and $\bar{H}^{(\beta)}$.
We start from $\phi_0=2$ and choose, instead of (\ref{e:saw-rec.1}),
%
\begin{eqnarray}
\label{e:perc-rec.1}
 \beta_{i+1} &=& 2,\qquad
\gamma_{i+1} = \{(d-6) \wedge\lfloor\phi_{i} \rfloor\}
-\varepsilon, \nonumber\\[-8pt]\\[-8pt]
\phi_{i+1} &=& \beta_{i+1} + \gamma_{i+1}
= \{(d-4) \wedge( \lfloor\phi_{i} \rfloor+ 2 ) \}
-\varepsilon
.\nonumber
\end{eqnarray}
For $d > 6$, repeating this recursion increases $\phi_{i}$ until
it reaches $d-4-\varepsilon$.
Using Lemma \ref{lem-PiToGbarWT} with $\phi=d-4-\varepsilon$
implies $\bar{G}^{(\alpha)}$ is finite with
$\alpha= d-4-\varepsilon$.
This is sufficient for (\ref{etau3saw}), as long as
$\frac{d+2}{2} < d-4-\varepsilon$, or $d > 10$.

Finally we deal with LTLA, this
time using $\bar{T}^{(\beta, \gamma)}$ and $\bar{S}^{(\gamma)}$.
We start from $\phi_0=2$ and choose
%
\begin{eqnarray}
\label{e:LTLA-rec.1}
 \beta_{i+1} &=& 2,\qquad
\gamma_{i+1} = \{(d-8) \wedge\lfloor\phi_{i} \rfloor\}
-\varepsilon, \nonumber\\[-8pt]\\[-8pt]
\phi_{i+1} &=& \beta_{i+1} + \gamma_{i+1}
= \{(d-6) \wedge( \lfloor\phi_{i} \rfloor+ 2 ) \}
-\varepsilon
.\nonumber
\end{eqnarray}
For $d > 8$, repeating this recursion increases $\phi_{i}$ until
it reaches $d-6-\varepsilon$.
Lemma \ref{lem-PiToGbarWT} now implies
$\bar{G}^{(\alpha)}$ is finite with $\alpha= d-6-\varepsilon$.
This is sufficient for (\ref{etau3saw}), as long as
$\frac{3d+2}{4} < d-6-\varepsilon$, or $d > 26$.
\end{pf*}

\begin{remark}
The condition (\ref{etau3saw}) follows
immediately (for SAW in \mbox{$d > 4$,}
for percolation in $d > 6$, and for LTLA in $d > 10$),
\textit{if} we can prove the $x$-space infrared bound,
$G(x) \leq c |\!|\!|x |\!|\!|^{2-d}$.
Although there are models (e.g., nearest-neighbor Ising model)
for which the $k$-space infrared bound (\ref{e:tr-Ghat-90}) does
imply its $x$-space counterpart
(\cite{Soka82}, Appendix A), it is not clear whether the
same is true for models considered in this paper.
The argument in this subsection has been employed to circumvent
this difficulty.
\end{remark}

\section[Proof of a Gaussian lemma, Theorem 1.4]{Proof of a Gaussian lemma, Theorem \protect\ref{th-gau}}
\label{sc-gau}

In this section, we prove Theorem \ref{th-gau}.
The proof is rather long, so we first present in
Section \ref{sb-gauss-pf} the framework of the proof,
in particular that of (\ref{e:gauss-asmp.1}),
assuming some lemmas which are proven later in
Section \ref{sb-Jhat} through Section \ref{sb-small-t}.
In Section \ref{sub-counter-gauss}, we
give an example which shows that the pointwise bound
(\ref{e:J-cnd.3}) is sharp in $d > 4$.
Finally in Section \ref{sub-gau-err},
we comment on how to prove (\ref{e:gauss-asmp.1a}).

\subsection[Overview of the Proof of Theorem 1.4, (1.36)]{Overview of the Proof of Theorem \textup{\protect\ref{th-gau},}
\textup{(\protect\ref{e:gauss-asmp.1})}}
\label{sb-gauss-pf}

Here we explain the framework of the proof of Theorem \ref{th-gau},
in particular (\ref{e:gauss-asmp.1}).
The proof of (\ref{e:gauss-asmp.1a}) is similar, and is
briefly explained in Section \ref{sub-gau-err}.

We first introduce an integral representation for $C(x)$,
which was also used in \cite{HHS03}.
The integrability of $\{1 - \hat{J}(k)\}^{-1}$ by
(\ref{e:J-cnd.1}), and a trivial identity
$\frac{1}{A} = \int_{0}^{\infty} dt e^{-t A}$ ($A > 0$)
immediately imply
\begin{eqnarray}\qquad\quad
\label{e:int-rep.1}
C(x)
& = &\int_{[-\pi, \pi]^{d}} \frac{d^d k}{(2\pi)^d}
\frac{e^{i k\cdot x}}{1 - \hat{J}(k)}
= \int_{[-\pi, \pi]^{d}} \frac{d^d k}{(2\pi)^d}
e^{i k\cdot x} \int_{0}^{\infty} dt e^{ -t\{1 - \hat{J}(k)\} }
\nonumber\\[-8pt]\\[-8pt]
&
= &\int_{0}^{\infty} dt I_{t}(x),\nonumber
\end{eqnarray}
with
%
\begin{equation}
I_{t}(x) :=
\int_{[-\pi, \pi]^{d}} \frac{d^d k}{(2\pi)^d}e^{i k\cdot x}
e^{-t\{1 - \hat{J}(k)\}} .
\end{equation}
Our task is to estimate this integral in detail.

We divide (\ref{e:int-rep.1}) into two parts.
We define, depending on $x$,
%
\begin{equation}
\label{e:T-choice}
T := \varepsilon|x|^{2}
\end{equation}
and
%
\begin{equation}
\label{e:G><-def}
C_{<}(x) := \int_{0}^{T} dt I_{t}(x) ,\qquad
C_{>}(x) := \int_{T}^{\infty} dt I_{t}(x)
\end{equation}
so that
%
\begin{equation}
\label{e:int-rep.2}
C(x) = C_{<}(x) + C_{>}(x) .
\end{equation}
In the above, $\varepsilon$ is a small positive number, and
will be sent to zero at the last step.
The choice of $T$ is suggested by the fact that the
variable $t$ roughly corresponds to the number of
steps of random walks; compare with the method of Lawler
\cite{Lawl91}, Chapter 1.

Now, for our choice of $T = \varepsilon|x|^{2}$, and for $x$
satisfying $|x| \geq1/\varepsilon$,
we have the following estimates, which
are proven in Sections \ref{sb-large-t} and \ref{sb-small-t}, respectively:
\begin{eqnarray}
\hspace*{-30mm}C_{>}(x)
& = &\frac{a_{d}}{K_{1}} \frac{1}{|x|^{d-2}} + R_1 (x)
\nonumber\label{e:Gg-est}\\[-8pt]\\[-8pt]
&&\eqntext{\mbox{with }
| R_1 (x) | \leq
o \biggl(\dfrac{1}{|x|^{d-2}} \biggr)
+ \dfrac{c_1 \varepsilon^{-d/2+1}
e^{- c_2 / \varepsilon} }{|x|^{d-2}}
+ \dfrac{c_3 \varepsilon}{|x|^{d-2}},}
\\
\hspace*{-30mm}| C_{<}(x) |
& \leq&\frac{c_4 \varepsilon} {|x|^{d-2} }.\label{e:Gl-est}
\end{eqnarray}
Here $c_1$ through $c_4$ (given explicitly in the proof)
are finite positive constants
which can be expressed in terms of $d$ and $K_{i}$ but are
independent of $\varepsilon$ and $x$.
The error term $o(|x|^{2-d})$ \textit{does} depend on $\varepsilon$.

These two estimates, together with (\ref{e:int-rep.2}), immediately
prove (\ref{e:gauss-asmp.1}). That is, for fixed
$\varepsilon> 0$,
%
\begin{eqnarray}
\limsup_{|x| \rightarrow\infty} |x|^{d-2} C(x)
& \leq&
\frac{a_{d}}{K_{1}}
+ [
c_1 \varepsilon^{-d/2+1} e^{- c_2 / \varepsilon}
+ c_3 \varepsilon+ c_4 \varepsilon
] ,
\nonumber\\[-8pt]\\[-8pt]
\liminf_{|x| \rightarrow\infty} |x|^{d-2} C(x)
& \geq&
\frac{a_{d}}{K_{1}}
- [
c_1 \varepsilon^{-d/2+1} e^{- c_2 / \varepsilon}
+ c_3 \varepsilon+ c_4 \varepsilon
] .\nonumber
\end{eqnarray}
Now letting $\varepsilon\downarrow0$ establishes
%
\begin{equation}\quad
\limsup_{|x| \rightarrow\infty} |x|^{d-2} C(x)
= \liminf_{|x| \rightarrow\infty} |x|^{d-2} C(x)
= \lim_{|x| \rightarrow\infty} |x|^{d-2} C(x)
= \frac{a_{d}}{K_{1}} .
\end{equation}

We in the following prove (\ref{e:Gg-est})
and (\ref{e:Gl-est}) step by step.

\subsection{Estimates on ${\hat{J}}(k)$}
\label{sb-Jhat}

We start from some estimates on $1 - {\hat{J}}(k)$.
%
\begin{lemma}
\label{lem-Jhatbd.1}
Assume \textup{(\ref{e:J-cnd.1})--(\ref{e:J-cnd.2})} of Theorem
\textup{\ref{th-gau}}
are satisfied. Then ${\hat{J}}(k)$ satisfies for $k \in[-\pi, \pi]^d$
%
\begin{equation}
\label{e:Jbd.2a}
0 \leq1 - {\hat{J}}(k) \leq K_{2} \frac{|k|^{2}}{2d}
\end{equation}
and
%
\begin{equation}
\label{e:Jbd.2}
1 - {\hat{J}}(k) = K_{1} \frac{|k|^{2}}{2d}
+ \hat{R}_2(k)\qquad
\mbox{with }
| \hat{R}_2(k) |
= o(|k|^{2})
.
\end{equation}
In the above, $o(|k|^{2})$ depends only on $d$ and $K_{i}$\textup{'}s.
Also, we have
%
\begin{equation}
\label{e:Jderbd.1}
\biggl| \frac{\partial}{\partial k_{1}} {\hat{J}}(k) \biggr|
\leq\frac{K_{2}}{d} |k_{1}| ,\qquad
\biggl| \frac{\partial^{2}}{\partial k_{1}^{2}}
{\hat{J}}(k) \biggr|
\leq\frac{K_{2}}{d}.
\end{equation}
\end{lemma}

\begin{pf}
We first note by (\ref{e:J-cnd.1})
\begin{eqnarray}\quad
\label{e:Jhat-rwrt.1}
1 - \hat{J}(k) &=& {\hat{J}}(0) - {\hat{J}}(k)
= \sum_{x} \{ 1 - \cos(k\cdot x) \} J(x)
\nonumber\\[-8pt]\\[-8pt]
&=& \sum_{x} \frac{(k\cdot x)^{2}}{2} J(x)
+ \sum_{x} \biggl\{ 1 - \cos(k\cdot x) - \frac{(k\cdot x)^{2}}{2}
\biggr\} J(x) .\nonumber
\end{eqnarray}
Using $0 \leq1 -\cos t \leq t^{2}/2$ and ${ {\mathbb Z}^d }$-symmetry,
we have from the first line of (\ref{e:Jhat-rwrt.1})
%
\begin{equation}\quad
\label{e:Jhat-rwrt.1a}
0 \leq1 - \hat{J}(k) \leq\sum_{x} \frac{(k\cdot x)^{2}}{2} | J(x) |
= \sum_{x} \frac{|k|^{2}}{2d} |x|^{2} | J(x) |
\leq\frac{|k|^{2}}{2d} K_{2}.
\end{equation}
Also, by ${ {\mathbb Z}^d }$-symmetry of $J(x)$, the first term on the
second line
of (\ref{e:Jhat-rwrt.1}) is equal to
%
\begin{equation}
\label{e:Jhat-main1}
\frac{|k|^{2}}{2d} \sum_{x} |x|^{2} J(x)
= K_{1} \frac{|k|^{2}}{2d}
.
\end{equation}

Next we proceed to deal with the second term of
(\ref{e:Jhat-rwrt.1}),
that is, $\hat{R}_2(k)$. We want to show that it
is of smaller order than $|k|^{2}$, so we consider
$|k|^{-2} \hat{R}_2(k)$:
%
\begin{equation}
\label{e:R4def}
\frac{\hat{R}_2(k)}{|k|^{2}} =
\sum_{x} \frac{1 - \cos(k\cdot x) - (k\cdot x)^{2}/2 }{|k|^{2}}
J(x) .
\end{equation}
Now we note for all $t \in{\mathbb R}$
%
\begin{equation}
\label{e:con.bd1}
\biggl| 1 - \cos t - \frac{t^{2}}{2} \biggr|
\leq
\frac{t^{2}}{2} \wedge
\frac{t^{4}}{24} .
\end{equation}
The first bound of (\ref{e:con.bd1}) can be used to show that the
sum in (\ref{e:R4def}) is uniformly bounded in $k$:
\begin{eqnarray}\quad
\label{e:R4bd}
\frac{| \hat{R}_2(k)| }{|k|^{2}}
& \leq&\sum_{x}
\biggl| \frac{1 - \cos(k\cdot x) - (k\cdot x)^{2}/2 }{|k|^{2}}
J(x)\biggr |
\leq
\sum_{x} \frac{(k\cdot x)^{2}}{2 |k|^{2}} | J(x) |
\nonumber\\[-8pt]\\[-8pt]
&
=&
\frac{1}{2d} \sum_{x} |x|^{2} |J(x) |
\leq\frac{K_{2}}{2d},
\nonumber
\end{eqnarray}
where on the second line we used ${ {\mathbb Z}^d }$-symmetry
as we did in (\ref{e:Jhat-rwrt.1a}).
The second bound of (\ref{e:con.bd1}) shows that the summand of
(\ref{e:R4def}) goes to zero (as $|k| \rightarrow0$)
for each fixed $x \in{ {\mathbb Z}^d }$.
Therefore, by dominated convergence,
the sum of (\ref{e:R4def}) goes to zero as
$|k| \rightarrow0$, that is, we have the bound of (\ref{e:Jbd.2}).

To prove (\ref{e:Jderbd.1}), we observe by ${ {\mathbb Z}^d }$-symmetry
\begin{eqnarray}
\biggl| \frac{\partial}{\partial k_{1}} {\hat{J}}(k)\biggr |
&
= &\Biggl| \sum_{x} x_{1} \sin(k_{1}x_{1})
\cos(k_{2}x_{2}) \cdots\cos(k_{d} x_{d}) J(x) \Biggr|
\nonumber\\[-8pt]\\[-8pt]
&
\leq&
\sum_{x} |k_{1}| |x_{1} |^{2} |J(x)|
\leq\frac{K_{2}}{d} |k_{1}|
.
\nonumber
\end{eqnarray}
Similarly, we note
%
\begin{equation}
\biggl| \frac{\partial^{2}}{\partial k_{1}^{2}} {\hat{J}}(k) \biggr|
=\Biggl | \sum_{x} \cos(k\cdot x) |x_{1}|^{2} J(x) \Biggr|
\leq
\sum_{x} |x_{1}|^{2} |J(x)|
\leq\frac{K_{2}}{d} .
\end{equation}
\upqed\end{pf}

\subsection[Contribution from $t \geq T$\textup{:} Proof of (2.6)]{Contribution from $t \geq T$\textup{:} Proof of
\textup{(\protect\ref{e:Gg-est})}}
\label{sb-large-t}

In this section, we prove (\ref{e:Gg-est}),
which gives an estimate on $C_{>}(x)$. The estimate itself is an
immediate consequence of the following lemma.
Note that no pointwise bound (\ref{e:J-cnd.3}) is needed for this
lemma.

\begin{lemma}
\label{lem-large-t}
Fix $\varepsilon>0$ and assume (\ref{e:J-cnd.1})--(\ref{e:J-cnd.2})
of Theorem \textup{\ref{th-gau}}.
Then we have for $t \geq1/\varepsilon$
%
\begin{eqnarray}
\label{e:Ibd.1}
&&I_{t}(x) = \biggl( \frac{d}{2 \pi K_{1} t} \biggr)^{{d/2}}
\exp\biggl( - \frac{d |x|^{2}}{2 t K_{1}} \biggr)
+ R_3(t) \nonumber\\[-8pt]\\[-8pt]
&&
\eqntext{\mbox{with }
| R_3(t) | \leq
o(t^{-d/2}) +
c_5 e^{- c_2 / \varepsilon} t^{-d/2}.}
\end{eqnarray}

In the above, $c_5 := 2 (d /\pi K_{0})^{d/2}$
and $c_2 := K_0/(4d)$.
The term $o(t^{-d/2})$ may depend on $\varepsilon$.
\end{lemma}

\begin{pf*}{Proof of (\ref{e:Gg-est}), assuming
Lemma \ref{lem-large-t}}
We just integrate (\ref{e:Ibd.1}) from
$t = T := \varepsilon|x|^2$ to $t = \infty$.
(We can apply Lemma \ref{lem-large-t}
because of our choice of $|x|$ and~$T$.)
The integral of
$R_3(t)$ is bounded as
\begin{eqnarray}\qquad
\label{e:pr2.8-1}
\biggl| \int_{T}^{\infty} dt R_3(t) \biggr|
&\leq&
\int_{T}^{\infty} dt [
o(t^{-d/2}) + c_5 e^{- c_2 / \varepsilon} t^{-d/2}
]
\nonumber\\
&
= &o(T^{-d/2+1}) +
\frac{2 c_5 e^{- c_2 / \varepsilon}}{d -2}
T^{-d/2+1}
\nonumber\\[-8pt]\\[-8pt]
&
=&
\varepsilon^{-d/2+1} o\bigl(|x|^{-(d-2)}\bigr)
+ c_1 \varepsilon^{-d/2+1} e^{- c_2 / \varepsilon}
|x|^{-(d-2)},\nonumber\\
&&\eqntext{c_1 := \dfrac{2 c_5}{d-2},}
\end{eqnarray}
where on the second line, we used our choice of $T$, (\ref{e:T-choice}).
On the other hand, the first term of (\ref{e:Ibd.1}) gives
\begin{eqnarray}
\label{e:pr2.8-2}
&&
\int_{T}^{\infty} dt
\biggl( \frac{d}{2 \pi K_{1} t} \biggr)^{{d/2}}
\exp\biggl( - \frac{d |x|^{2}}{2 t K_{1}} \biggr)\nonumber\\
&&\qquad=
\int_{0}^{\infty} dt (\cdots)
-
\int_{0}^{T} dt (\cdots)
\nonumber\\[-8pt]\\[-8pt]
&&\qquad
=
\frac{\Gamma({d}/{2}-1)d}{2 \pi^{d/2} K_{1}} |x|^{2-d}
- \int_{0}^{T} dt
\biggl( \frac{1}{2 \pi K_{1} t} \biggr)^{d/2}
\exp\biggl( - \frac{d |x|^{2}}{2 t K_{1}} \biggr)
\nonumber\\
&&\qquad =:
\frac{\Gamma({d}/{2}-1)d}{2 \pi^{d/2} K_{1}} |x|^{2-d}
- R_4(x) .\nonumber
\end{eqnarray}
For the integrand of $R_4(x)$, we use an inequality
%
\begin{equation}
\label{e:el.ineq.2a}
y^{-\beta} e^{-\alpha/y} \leq
\biggl( \frac{\beta}{\alpha e} \biggr)^{\beta}
\qquad(\mbox{valid for } \alpha, \beta, y > 0)
\end{equation}
with $\alpha= d |x|^{2}/(2K_{1})$, $\beta= d/2$ and $y = t$.
The result is
%
\begin{equation}\qquad
\label{e:inR23.bd}
[\mbox{integrand of }R_4(x)]
\leq
\biggl( \frac{1}{2 \pi K_{1} } \biggr)^{d/2}
\times
\biggl( \frac{K_{1}}{e |x|^{2}} \biggr)^{d/2}
=
\biggl( \frac{1}{2 \pi e |x|^{2}} \biggr)^{d/2} ,
\end{equation}
and thus, $R_4(x)$ is bounded as
%
\begin{equation}
\label{e:pr2.8-3}
R_4(x)
\leq T\times\biggl( \frac{1}{2 \pi e |x|^{2}} \biggr)^{d/2}
=
\biggl( \frac{1}{2 \pi e} \biggr)^{d/2}
\frac{\varepsilon}{|x|^{d-2}}
:=
\frac{c_3 \varepsilon}{|x|^{d-2}} .
\end{equation}

Combining (\ref{e:pr2.8-1})--(\ref{e:pr2.8-3}),
we get (\ref{e:Gg-est}).
\end{pf*}

\begin{pf*}{Proof of Lemma \protect\ref{lem-large-t}}
We introduce $k_{t} > 0$ by
%
\begin{equation}
k_t := (\varepsilon t)^{-1/2},\qquad ( \leq1)
\end{equation}
and divide $I_{t}(x)$ into four parts:
%
\begin{equation}
\label{e:Itdiv.00}
I_{t}(x) = I_{t,1}(x) + I_{t,2}(x) + I_{t,3}(x) + I_{t,4} (x),
\end{equation}
with
\begin{subequation}
\begin{eqnarray}
I_{t,1}(x) & :=& \int_{{ {\mathbb R}^d}} \frac{d^d k}{(2\pi)^d}
e^{i k\cdot x- t K_{1} |k|^{2}/(2d)}
,
\label{e:Itdiv.1}\\
I_{t,2}(x) & :=&
- \int_{|k| > k_{t}} \frac{d^d k}{(2\pi)^d}e^{i k\cdot x- t K_{1}
|k|^{2}/(2d)} ,
\label{e:Itdiv.2}\\
I_{t,3}(x) & :=&
\int_{|k| \leq k_{t}} \frac{d^d k}{(2\pi)^d}e^{i k\cdot x}
\bigl\{
e^{- t\{1 - {\hat{J}}(k)\}} - e^{- t K_{1} |k|^{2}/(2d)}
\bigr\} ,
\label{e:Itdiv.3}\\
I_{t,4}(x) & :=&
\int_{|k| > k_{t}, k \in[-\pi, \pi]^d} \frac{d^d k}{(2\pi)^d}e^{i
k\cdot x}
e^{-t\{1 - {\hat{J}}(k)\}} .\label{e:Itdiv.4}
\end{eqnarray}
\end{subequation}
Integrals $I_{t,1}(x)$ through $I_{t,3}(x)$ sum up to contributions
to $I_{t}(x)$ from $|k| \leq k_{t}$, and $I_{t,4}(x)$
represents the contribution from $|k| > k_{t}, k \in[-\pi, \pi]^d$.
The choice of $k_t$ is motivated so that we can use
(\ref{e:J-cnd.1}) for $I_{t,3}(x)$ (because of our choice
$t \geq1/\varepsilon$, we have $|k| \leq k_{t} \leq1$).
We estimate the above integrals one by one.

The first integral $I_{t,1}(x)$ gives the main contribution, and is
calculated exactly by completing the square:
%
\begin{equation}
\label{e:I1r}
I_{t,1}(x) = \biggl( \frac{d}{2 \pi K_{1} t} \biggr)^{{d/2}}
\exp\biggl( - \frac{d |x|^{2}}{2 t K_{1}} \biggr) .
\end{equation}

Second, for $I_{t,3}(x)$, we first change the integration variable
from $k$ to $l := \sqrt{t} k$ to obtain
\begin{eqnarray}\qquad\quad
\label{e:I3.1}
I_{t,3}(x)
& =&
t^{-d/2}
\int_{|l|^{2} \leq1/ \varepsilon}
\frac{d^{d}l}{(2\pi)^{d}}
\exp\biggl( \frac{i l \cdot x}{\sqrt{t}} \biggr)\nonumber\\
&&\hspace*{60.1pt}{}\times\biggl\{
\exp\biggl( - t \biggl\{ 1 -
{\hat{J}}\biggl( \frac{l}{\sqrt{t}} \biggr) \biggr\}
\biggr)
\\
&&\hspace*{102.7pt}{}
-
\exp\biggl( - t \frac{K_{1} |l|^{2}}{2d t} \biggr)
\biggr\}
=: t^{-d/2} \tilde{I}_{t,3}(x).\nonumber
\end{eqnarray}
Now the integral $\tilde{I}_{t,3}(x)$ is seen to be $o(1)$ as
$t \rightarrow\infty$, as follows. (i) We can get a uniform bound as
\begin{eqnarray}\qquad
\label{e:I3.2}
| \tilde{I}_{t,3}(x) |  \leq
\int_{|l|^{2} \leq1/ \varepsilon}
\frac{d^{d}l}{(2\pi)^{d}}
\biggl[
\exp\biggl( - \frac{K_{0} |l|^{2}}{2d} \biggr)
+
\exp\biggl( - \frac{K_{1} |l|^{2}}{2d} \biggr)
\biggr]
< \infty,
\end{eqnarray}
where we used the lower bound
(\ref{e:J-cnd.1}) for the first term.
(ii) For each fixed $l \in{ {\mathbb R}^d}$, the integrand
of $\tilde{I}_{t,3}(x)$ goes to zero (as $t \rightarrow\infty$).
This is
because we can write
[recalling the definition (\ref{e:Jbd.2}) of $\hat{R}_2$]
\begin{eqnarray}
\label{e:I3.3}
&&\tilde{I}_{t,3}(x)  =
\int_{|l|^{2} \leq1/ \varepsilon}
\frac{d^{d}l}{(2\pi)^{d}}
\exp\biggl( \frac{i l \cdot x}{\sqrt{t}}
- \frac{K_{1} |l|^{2}}{2d} \biggr)\nonumber\\[-8pt]\\[-8pt]
&&\hspace*{79.7pt}{}\times\biggl\{
\exp\biggl(
- t \hat{R}_2 \biggl( \frac{l}{\sqrt{t}} \biggr)
\biggr)
-
1
\biggr\} .\nonumber
\end{eqnarray}
Then, in view of our bound of (\ref{e:Jbd.2}), the integrand goes
to zero as $t \rightarrow\infty$ for fixed $l$.
By (i) and (ii) above, we can use the dominated convergence theorem
to conclude that the integral $\tilde{I}_{t,3}(x)$
of (\ref{e:I3.1}) is $o(1)$ as $t \rightarrow\infty$
[this $o(1)$ can depend on $\varepsilon$], and therefore
%
\begin{equation}
\label{e:I3r}
I_{t,3}(x) = o(t^{-d/2})\qquad
\mbox{[$o(t^{-d/2})$ can depend on $\varepsilon$]}.
\end{equation}

Finally we estimate $I_{t,2}(x)$ and $I_{t,4}(x)$. By definition,
%
\begin{equation}
\label{e:I2r}
| I_{t,2}(x) |
\leq
\int_{|k| > k_{t} } \frac{d^d k}{(2\pi)^d}
e^{- t K_{1} |k|^{2}/(2d)} .
\end{equation}
For $I_{t,4}(x)$, we use (\ref{e:J-cnd.1})
to bound ${\hat{J}}(0) - {\hat{J}}(k)$ for $k \in[-\pi, \pi]^d$ as
%
\begin{eqnarray}
\label{e:I4r}
| I_{t,4} (x) | :\!&=&
\biggl|
\mathop{\mathop{\int}_{{|k| > k_{t}}}}_{k \in[-\pi, \pi]^d}
\frac{d^d k}{(2\pi)^d}e^{i k\cdot x} e^{-t\{{\hat{J}}(0) - {\hat
{J}}(k)\}}
\biggr|\nonumber\\[-8pt]\\[-8pt]
&\leq&
\mathop{\mathop{\int}_{{|k| > k_{t}}}}_{k \in{{\mathbb R}^d}}
\frac{d^d k}{(2\pi)^d}e^{-t K_{0} |k|^{2}/(2d)} .\nonumber
\end{eqnarray}
Therefore, using $K_{0} \leq K_{1}$ and
%
\begin{equation}\quad
\label{e:intgauss2.2}
\int_{|k|\geq b} \frac{d^d k}{(2\pi)^d}e^{-a |k|^{2}}
\leq\biggl( \frac{1}{2 \pi a} \biggr)^{d/2}
e^{-a b^{2}/2}\qquad
(\mbox{valid for } a, b > 0) ,
\end{equation}
we get
\begin{eqnarray}\qquad\quad
\label{e:I24r}
| I_{t,2} (x) + I_{t,4} (x) |
& \leq&
2 \biggl( \frac{d}{ \pi K_{0} t} \biggr)^{d/2}
e^{-t K_{0} k_{t}^{2}/(4d)}
\nonumber\\[-8pt]\\[-8pt]
&
=&
2\biggl ( \frac{d}{ \pi K_{0}} \biggr)^{d/2}
t^{-d/2}
e^{- K_{0} /(4d \varepsilon)}
=: c_5 t^{-d/2}
e^{- c_2 / \varepsilon} .\nonumber
\end{eqnarray}
The above (\ref{e:I1r}), (\ref{e:I3r}) and (\ref{e:I24r}) establish
(\ref{e:Ibd.1}) and prove the lemma.
\end{pf*}

\subsection[Contribution from $t < T$\textup{:} Proof of (2.7)]{Contribution from $t < T$\textup{:} Proof of
\textup{(\protect\ref{e:Gl-est})}}
\label{sb-small-t}
In this section, we prove (\ref{e:Gl-est}),
which gives an estimate on $C_{<}(x)$.
This is the place where we have to make use of our
assumption on \textit{pointwise} $x$-space bound on $J(x)$,
(\ref{e:J-cnd.3}). We do need something
like this, to exclude pathological examples which violate
(\ref{e:Gl-est}) for infinitely many $x$'s (see, e.g., page 32
of \cite{MS93} and Section \ref{sub-counter-gauss}
of the present paper).

The estimate (\ref{e:Gl-est}) itself is an
immediate consequence of the following lemma. To state the lemma,
we introduce some notation. For a function $f(x)$
on ${ {\mathbb Z}^d }$,
we write $f^{(*n)}$ for the $n$-fold convolution of $f$
and use ${\prod_{\ell=1}^{n}}{^{(*)}} f_{\ell}$
to denote the
convolution of functions $f_{\ell}$, $\ell= 1, 2, \ldots, n$:
%
\begin{eqnarray}
f^{(*n)}(x) &:=& (\underbrace{f*f* \cdots* f}_{n})(x) ,
\nonumber\\[-8pt]\\[-8pt]
{\prod_{\ell=1}^{n}}{^{(*)}} f_{\ell}
&:=& ( f_1 * f_2 * f_3 * \cdots* f_n ) (x) .\nonumber
\end{eqnarray}
Also, in this subsection and for the function $J(x)$ only,
we define for $j, \ell= 1, 2, \ldots, d$,
%
\begin{equation}
\label{e:Jndef}
J_{j}(x) := x_{j} J(x) ,\qquad
J_{j, \ell}(x) := x_j x_\ell J(x) .
\end{equation}
$J_j^{(*n)}$ denotes the $n$-fold convolution of $J_j$, \textit{not}
$x_j$ times $J^{(*n)}$.

\begin{lemma}
\label{lem-small-t}
Under the assumption of Theorem \textup{\ref{th-gau}}, we have
for integers $m \in[0, d]$ and $n_1, n_2, \ldots,
n_d \geq0$
%
\begin{equation}\qquad\quad
\label{e:Isbd.1}
\Biggl|\Biggl ( I_{t} * {\prod_{j=1}^d}{^{(*)}}
J_{j}^{(*n_j)} \Biggr)(x) \Biggr|
\leq c_6(m,\vec{n}) \frac{t^{-(d+n-m)/2}}{|\!|\!|x |\!|\!|^{m}}\qquad
\mbox{with } n := \sum_{j=1}^d n_j,
\end{equation}
where $c_6(m, \vec{n})$ is a calculable constant depending
on $K_{i}, d, m$ and $\vec{n} := (n_1, n_2,\break \ldots, n_d)$.
\end{lemma}

\begin{pf*}{Proof of (\protect\ref{e:Gl-est}), given Lemma \protect\ref{lem-small-t}}
This is easy. By (\ref{e:Isbd.1}) with $\vec{n}=\vec{0}$ and
$m=d$, we have $| I_{t}(x) | \leq c_6(d,\vec{0}) |\!|\!|x |\!|\!|^{-d}$.
Integrating this from $t=0$ to $t=T$ gives
%
\begin{equation}
\label{e:Gsbd.1}
| C_{<}(x) | \leq T \times
\frac{c_6(d,\vec{0})}{|\!|\!|x |\!|\!|^{d}}
= \frac{c_6(d,\vec{0}) \varepsilon}{|\!|\!|x |\!|\!|^{d-2}},
\end{equation}
where the last equality follows from our choice of
$T$, (\ref{e:T-choice}). This proves (\ref{e:Gl-est}),
with $c_4 = c_6(d,\vec{0})$.
\end{pf*}

\begin{pf*}{Proof of Lemma \protect\ref{lem-small-t}}
We first prove the lemma for $m = 0$ by estimating Fourier
integrals directly. We then proceed to prove
the lemma for $m\geq1$ by induction in $m$.
To simplify notation,
we abbreviate the left-hand side of (\ref{e:Isbd.1}) (without the
absolute value) as $F_{\vec{n}}(x; t)$.

\textit{The case $m=0$.}
In terms of Fourier transform, we have
%
\begin{equation}
\label{e:IJn-1}
F_{\vec{n}}(x; t)
= \int_{[-\pi, \pi]^d} \frac{d^d k}{(2\pi)^d}e^{i k x} e^{-t\{
1-{\hat{J}}(k)\}}
\prod_{j=1}^d
\{ i \partial_j {\hat{J}}(k) \}^{n_j},
\end{equation}
where (and in the following) $\partial_j$ denotes
$\partial/ \partial k_{j}$.
Using bounds (\ref{e:J-cnd.1}) and (\ref{e:Jderbd.1}), we can bound
(\ref{e:IJn-1}) as
%
\begin{eqnarray}
\label{e:eJbd.m0.2}
| F_{\vec{n}}(x; t) |
&\leq&
\biggl( \frac{K_{2}}{d} \biggr)^{n}
\int_{{\mathbb R}^d} \frac{d^d k}{(2\pi)^d}e^{-t K_{0} |k|^{2}/(2d)}
|k|^{n}\nonumber\\[-8pt]\\[-8pt]
&=& c t^{-(d+n)/2} ,\qquad
n := \sum_{j=1}^d n_j\nonumber
\end{eqnarray}
with some constant $c$.
This proves (\ref{e:Isbd.1}) for $m=0$,
if we take $c_6(0,\vec{n}) \geq c$.

\textit{The case $x=0$.}
The above (\ref{e:eJbd.m0.2}) also proves the lemma for
$x =0$, $t \geq1$ and
for all $m \geq0$, because in this case the right-hand side of
(\ref{e:Isbd.1}) increases as $m$ increases and thus the bound
for $m=0$ takes care of those for $m >0$ as well.

For $x=0$ and $t \leq1$, we first note a trivial bound
%
\begin{equation}
\label{e:eJbd.x0.2}
| F_{\vec{n}}(0) |
\leq
\biggl( \frac{K_{2}}{d} \biggr)^{n}
\int_{[-\pi,\pi]^{d}}\frac{d^d k}{(2\pi)^d}e^{-t K_{0}
|k|^{2}/(2d)} |k|^{n}
\leq\biggl( \frac{\pi K_{2}}{\sqrt{d}} \biggr)^{n},
\end{equation}
where we just bounded the exponential by $1$ and
$|k|^{n}$ by $(\sqrt{d} \pi)^{n}$.
Multiplying the right-hand side by $t^{-(d+n-m)/2} \geq1$ (valid
for $0 \leq m \leq d+n$) proves (\ref{e:Isbd.1}) for
$x=0, t < 1$ and $0 \leq m \leq d+n$.

We have thus proved (\ref{e:Isbd.1}) for $x=0$ and
$n \geq0$, $m \in(0, d+n]$.
Having treated $x=0$, we in the following focus on $x \neq0$.

\textit{The case $m \geq1$.}
Suppose we have proved (\ref{e:Isbd.1}) for $m -1$;
we now prove it for $m$ by induction.
With the help of Fourier transform, we see for
$l = 1, 2, \ldots, d$:
\begin{eqnarray}\hspace*{15pt}
\label{e:IJn-5}
 x_{l} F_{\vec{n}} (x)
&=& i^{n+1} \int_{[-\pi,\pi]^{d}}\frac{d^d k}{(2\pi)^d}e^{i k x}
\partial_l
\Biggl[ e^{-t\{1-{\hat{J}}(k)\}}
\prod_{j=1}^d
\{ \partial_j {\hat{J}}(k) \}^{n_j} \Biggr]
\nonumber\\
&
=& i^{n+1} \int_{[-\pi,\pi]^{d}}\frac{d^d k}{(2\pi)^d}e^{i k x}
e^{-t\{1-{\hat{J}}(k)\}}\nonumber\\
&&\hspace*{57pt}{}\times
\Biggl[ t \{ \partial_l {\hat{J}}(k) \}^{n_l+1}
\prod_{j \neq l}
\{ \partial_j {\hat{J}}(k) \}^{n_j}
\\
&&\hspace*{73pt}{}
+ \sum_{p=1}^d n_p
\{ \partial_p {\hat{J}}(k) \}^{n_p-1}
\{ \partial_l \partial_p {\hat{J}}(k) \}
\prod_{j\neq p}
\{ \partial_j {\hat{J}}(k) \}^{n_j}
\Biggr]
\nonumber\\
& = &t F_{\vec{n}'}(x) + \sum_{p=1}^d n_p
( F_{\vec{n}''}*J_{p, l} ) (x)
,
\nonumber
\end{eqnarray}
where $\vec{n}' = (n_1, n_2, \ldots, n_{l-1}, n_l +1,
n_{l+1}, \ldots)$ and
$\vec{n}'' = (n_1, n_2, \ldots, n_{p-1}, n_p-1, n_{p+1}, \ldots)$.
The first term on the right-hand side
is simply bounded by our inductive assumption as
(note: now $\sum_{j} n_j' = n+1$)
%
\begin{eqnarray}
\label{e:IJn-7}
| t F_{\vec{n}'}(x) |
&\leq& t \times
c_6(m-1,\vec{n}') \frac{t^{-(d+n+1-m+1)/2}}{|\!|\!|x
|\!|\!|^{m-1}}\nonumber\\[-8pt]\\[-8pt]
&=& c_6(m-1,\vec{n}') \frac{t^{-(d+n-m)/2}}{|\!|\!|x |\!|\!|^{m-1}}
.\nonumber
\end{eqnarray}
For the second term, we again use our inductive assumption on
$F_{\vec{n}''}$ (now $\sum_{j} n_j'' = n-1$):
%
\begin{eqnarray}
\label{e:IJn-11}
| F_{\vec{n}''}(x) |
&\leq&
c_6(m-1,\vec{n}'') \frac{t^{-(d+n-1-m+1)/2}}{|\!|\!|x |\!|\!|^{m-1}}\nonumber\\[-8pt]\\[-8pt]
&=& c_6(m-1,\vec{n}'') \frac{t^{-(d+n-m)/2}}{|\!|\!|x |\!|\!|^{m-1}}
.\nonumber
\end{eqnarray}
We have to take the convolution with $J_{p, l}$, and we
argue separately for $m=1$ and $m > 1$. For $m=1$, we estimate as
\begin{eqnarray}
\label{e:IJn-11b}
| ( F_{\vec{n}''}*J_{p, l} ) (x) |
& \leq&
c_6(0,\vec{n}'') t^{-(d+n-1)/2} \sum_y
| J_{p, l}(y) |
\nonumber\\
&
\leq& c_6(0,\vec{n}'') t^{-(d+n-1)/2}
\sum_y |y|^2 |J(y)|
\\
&
\leq& c_6(0,\vec{n}'') t^{-(d+n-1)/2} \times K_2
,
\nonumber
\end{eqnarray}
where we used our assumption (\ref{e:J-cnd.2}) in the last step.
For $m >1$, we take the convolution of (\ref{e:IJn-11})
with
%
\begin{equation}
\label{e:IJn-13}
| J_{p, l} (x) |
\leq\frac{K_3 | x_p x_l |} { |\!|\!|x |\!|\!|^{d+2}}
\leq\frac{K_3} { |\!|\!|x |\!|\!|^{d}} ,
\end{equation}
which satisfies
$\sum_x |J_{p,l}(x)| \leq\sum_x |x|^2 |J(x)| < \infty$.
The power $(m-1)$ of (\ref{e:IJn-11}) is not changed by the
convolution as long as $0 < m-1 < d$
[see Lemma \ref{lem-conv}(iii)], and we get
%
\begin{equation}
\label{e:IJn-15}
| ( F_{\vec{n}''}*J_{p, l} ) (x)
|
\leq c
c_6(m-1,\vec{n}'') \frac{t^{-(d+n-m)/2}}{|\!|\!|x |\!|\!|^{m-1}}
\end{equation}
with some constant $c$ arising from convolution.
Thus for both $m=1$ and $1 < m \leq d$, we get a bound of
the form of (\ref{e:IJn-15}) for the second term of (\ref{e:IJn-5}).

Combining (\ref{e:IJn-7}) and (\ref{e:IJn-15}),
we get
%
\begin{eqnarray}
\label{e:IJn-17}
| x_l F_{\vec{n}} (x) |
&\leq&
c_6(m,\vec{n})' \frac{t^{-(d+n-m)/2}}{|\!|\!|x |\!|\!|^{m-1}}
\quad\mbox{or}\nonumber\\[-8pt]\\[-8pt]
| F_{\vec{n}} (x) |
&\leq&
c_6(m,\vec{n})'
\frac{t^{-(d+n-m)/2}}{|x_l| |\!|\!|x |\!|\!|^{m-1}}\nonumber
\end{eqnarray}
with $c_6(m,\vec{n})' = c_6(m-1,\vec{n}')
+ c c_6(m-1,\vec{n}'')$. Because the above holds for all
$l = 1, 2, \ldots, d$, we can replace $|x_l|$
by $\|x\|_\infty$ in the above, and we get (\ref{e:Isbd.1}) for $m$
[increase $c_6(m,\vec{n})'$ appropriately in order to turn
$\|x\|_{\infty}$ into $|\!|\!|x |\!|\!|$]. The proof is complete.
\end{pf*}

\subsection{We cannot do better than
$|J(x)| \leq c |x|^{-(d+2)}$\textup{:} An example}
\label{sub-counter-gauss}

We here present a ``counterexample,'' which mildly violates
the pointwise bound $|J(x)| \leq c |x|^{-(d+2)}$ and which does
not exhibit the Gaussian asymptotic form of (\ref{e:gauss-asmp.1}).
The pointwise bound is not a necessary condition, but the following
example shows that it is rather sharp for $d > 4$.

The author is grateful to K\^{o}hei Uchiyama concerning
the proof of Proposition \ref{prop-c-ex}.

\begin{prop}
\label{prop-c-ex}
Fix $d > 4$ and $0 < \varepsilon< (d-4)/4$,
and let $g(x)$ be a slowly varying, nonnegative,
${ {\mathbb Z}^d }$-symmetric function which
diverges as $|x| \rightarrow\infty$. Define
%
\begin{equation}
h(x) = g(x)^{-(1+\varepsilon)/d} ,
\end{equation}
and subsets of ${ {\mathbb Z}^d }$ as
%
\begin{eqnarray}
\mathcal{E} &:=& \{\pm l_n \bolds{e}_j |
1 \leq j \leq d, n \geq1 \} ,\nonumber\\[-8pt]\\[-8pt]
\tilde{\mathcal{E}} &:=& \{ y \in{ {\mathbb Z}^d } |
\exists x \in\mathcal{E}, |y-x| \leq h(x) |x|
\},\nonumber
\end{eqnarray}
where $\bolds{e}_j$ is the unit vector in the $j$th coordinate
axis.
Finally define
%
\begin{equation}
\label{e:J-conterexample-def}
J(x) := \frac{1-\delta}{2d} I[ |x|=1 ]
+ \frac{g(x)}{|x|^{d+2}} I[x \in\tilde{\mathcal{E}}],
\end{equation}
where $\delta$ is determined so that $\sum_x J(x) = 1$.
Then by choosing a sequence $l_n$ which diverges to infinity
sufficiently rapidly (depending on $g$), we can achieve
%
\begin{equation}
\limsup_{|x| \rightarrow\infty} |x|^{d-2} C(x) = \infty.
\end{equation}
That is, the model does \textit{not} exhibit the Gaussian
asymptotic form of Theorem \textup{\ref{th-gau}}.
\end{prop}

\begin{pf}
We choose $l_n$ which diverges sufficiently rapidly as
$n \rightarrow\infty$, so that (1) $J(x) \geq0$ for all
$x \in{\mathbb Z}^d$, and (2) $\sum_x |x|^2 J(x) < \infty$.
We prove, for $x \in\mathcal{E}$ with sufficiently large $|x|$,
%
\begin{equation}
\label{e:gauss-counter.11}
C(x) \geq\frac{c}{|x|^{d-2}} g(x) h(x)^{4}
\exp\{
- c' g(x) h(x)^{d}
\}
\end{equation}
with finite positive constants $c, c'$ which are independent of $x$.
This immediately implies
%
\begin{equation}\qquad
\label{e:gauss-counter.13}
\mathop{\mathop{\lim}_{|x| \rightarrow\infty}}_{x \in\mathcal{E}}
C(x) |x|^{d-2}
\geq\mathop{\mathop{\lim}_{|x| \rightarrow\infty}}_{x \in\mathcal{E}}
c g(x)^{(d-4-4\varepsilon)/d}
\exp\{
- c' g(x)^{-\varepsilon}
\}
= \infty,
\end{equation}
because of our choice of $g$ and $h$.
In the following, we explain how to get
(\ref{e:gauss-counter.11}).

First choose arbitrary but large $n$ and define $a = l_n$.
We prove (\ref{e:gauss-counter.11}) for $x = a \bolds{e}_1$,
which is sufficient.
Define
%
\begin{eqnarray}\qquad\quad
\label{e:gauss-counter.15}
q^a(y) &:=& J(y) \sum_{j=1}^d
\{ I[ |y + a \bolds{e}_j| \leq a h(a)]
+
I[ |y - a \bolds{e}_j| \leq a h(a)]
\}, \nonumber\\[-8pt]\\[-8pt]
p^{a} (y) &:=& J(y) - q^{a}(y).\nonumber
\end{eqnarray}
[With an abuse of notation, we write $g(a)$ and
$h(a)$ for $g(a \bolds{e}_j)$ and $h(a \bolds{e}_j)$.]
Because both $p^a$ and $q^a$ are nonnegative, we get a lower
bound on $C(x)$ by discarding some terms as
\begin{eqnarray}
\label{e:gauss-counter.17}
C(x) & = & \sum_{n=0}^\infty(p^a + q^a)^{(*n)} (x)\nonumber\\[-8pt]\\[-8pt]
&\geq&\sum_{n=0}^\infty n
\bigl( (p^a)^{(*(n-1))}*q^a \bigr) (x)
= ( q^a * C^a * C^a) (x)\nonumber
\end{eqnarray}
where we introduced locally
%
\begin{equation}
\label{e:gauss-counter.19}
C^a(y) := \sum_{n=0}^\infty(p^a)^{(*n)}(y) .
\end{equation}
We further get a lower bound of (\ref{e:gauss-counter.17}) by
restricting the sum arising from the convolution:
%
\begin{eqnarray}
\label{e:gauss-counter.23}
C(x)
&\geq&\sum_{y\dvtx |y-a\bolds{e}_1| \leq a h(a)}
q^a(y) ( C^a*C^a) (x-y)\nonumber\\[-8pt]\\[-8pt]
&\geq&\frac{g(a)}{2 a^{d+2}} \sum_{|z|\leq a h(a)}
(C^a * C^a) (z) ,\nonumber
\end{eqnarray}
where in the last step we used $g(y) \geq g(a)/2$,
because $g(x)$ is slowly varying.

To get a nice lower bound on $(C^a * C^a) (z)$,
we use the following integral representation
\begin{eqnarray}
\label{e:gauss-counter.25}
(C^a * C^a)(z)
& = &\int_0^\infty dt \,t \int_{[-\pi,\pi]^{d}}\frac{d^d k}{(2\pi)^d}
e^{ikz} e^{-t ( 1 - {\hat{p}}^a(k))}
\nonumber\\[-8pt]\\[-8pt]
&
\geq&
\int_{T_1}^{T_2} dt\, t \int_{[-\pi,\pi]^{d}}\frac{d^d k}{(2\pi)^d}
e^{ikz} e^{-t ( 1 - {\hat{p}}^a(k))},
\nonumber
\end{eqnarray}
where $T_1 := |z|^2, T_2 := 2 |z|^2$.
In the last step we used the fact that the integrand
[inverse Fourier transform of $e^{-t ( 1 - {\hat{p}}^a(k))}$] is
nonnegative. This fact can be seen by writing it as
\begin{eqnarray}\hspace*{40pt}
\int_{[-\pi,\pi]^{d}}\frac{d^d k}{(2\pi)^d}e^{ikz} e^{-t ( 1 -
{\hat{p}}^a(k))}
& = & e^{-t} \int_{[-\pi,\pi]^{d}}\frac{d^d k}{(2\pi)^d}e^{ikz} \sum
_{n=0}^\infty\frac{t^n}{n!}
({\hat{p}}^a(k))^n
\nonumber\\[-8pt]\\[-8pt]
&
= &e^{-t} \sum_{n=0}^\infty\frac{t^n}{n!} (p^a)^{(*n)}(z),\nonumber
\end{eqnarray}
and use the fact that $p^a(z)$ is nonnegative by its definition,
(\ref{e:gauss-counter.15}).
Because we have ${\hat{p}}^a(0) < 1$ now, we bound the right-hand side
of (\ref{e:gauss-counter.25}) as
\begin{eqnarray}\qquad\quad
\label{e:gauss.counter.27}
(C^a * C^a)(z) & \geq
\displaystyle\int_{T_1}^{T_2} dt\, t e^{-t ( 1 - {\hat{p}}^a(0))}
I_t^a (z)
\geq
e^{-T_2 ( 1 - {\hat{p}}^a(0))}
\int_{T_1}^{T_2} dt\, t I_t^a (z)
\end{eqnarray}
with
%
\begin{equation}
\label{e:gauss.counter.29}
I_t^a (z) := \int_{[-\pi,\pi]^{d}}\frac{d^d k}{(2\pi)^d}
e^{ikz} e^{-t ( {\hat{p}}^a(0) - {\hat{p}}^a(k))}
.
\end{equation}
The first exponent of (\ref{e:gauss.counter.27}) can be bounded as
\begin{eqnarray}
\label{e:gauss.counter.31}
T_2\bigl(1 - {\hat{p}}^a(0)\bigr)
& =& T_2 {\hat{q}}^a(0)
= T_2 \sum_{y\dvtx |y \pm a \bolds{e}_j| \leq a h(a)}
\frac{g(y)}{|y|^{d+2}}
\nonumber\\[-8pt]\\[-8pt]
&
\leq& c T_2 g(a) h(a)^d a^{-2}
\leq c g(a) h(a)^d,\nonumber
\end{eqnarray}
with some constant $c$,
where in the last step we used $|z| \leq a h(a) \leq a$.

The remaining integral of $I_t^a(z)$ can be estimated as we did in
Section \ref{sb-large-t}. By Lemma \ref{lem-large-t}, we have
%
\begin{equation}\quad
\label{e:gauss.counter.33}
I_t^a(z) = \biggl( \frac{d}{2 \pi K_1' t} \biggr)^{d/2}
\exp\biggl( - \frac{d |z|^2}{2 t K_1'} \biggr)
+ o(t^{-d/2}) + c_5 e^{-c_2/\varepsilon} t^{-d/2}
\end{equation}
with $K_1' \approx K_1$ for $\varepsilon>0$ and $t > 1/\varepsilon$.
For $\varepsilon$ sufficiently small and for $t \geq|z|^2$
sufficiently large depending on
$\varepsilon$, the first term of (\ref{e:gauss.counter.33})
dominates the rest. So we have
%
\begin{equation}\quad
\label{e:gauss.counter.35}
\int_{T_1}^{T_2} dt\, t
I_t^a(z)
\geq
\int_{T_1}^{T_2} dt\, t \frac{1}{2}
\biggl( \frac{d}{2 \pi K_1' t} \biggr)^{d/2}
\exp\biggl( - \frac{d |z|^2}{2 t K_1'} \biggr)
\geq
\frac{c''}{|z|^{d-4}}
\end{equation}
for sufficiently large $z$.

Combining (\ref{e:gauss.counter.27}), (\ref{e:gauss.counter.31})
and (\ref{e:gauss.counter.35}),
we have for sufficiently large $|z|$
%
\begin{equation}
\label{e:gauss.counter.37}
(C^a * C^a)(z) \geq c |z|^{4-d}
\exp\{ - c' g(a) h(a)^d \}
\end{equation}
with positive constants $c, c'$.
Going back to (\ref{e:gauss-counter.23}) yields, for
sufficiently large $a h(a)$,
\begin{eqnarray}
C(x)
& \geq& c \frac{g(a)}{a^{d+2}} \times
\sum_{L < |z|\leq a h(a)} c |z|^{4-d}
\exp\{ - c' g(a) h(a)^d \}
\nonumber\\[-8pt]\\[-8pt]
&
= &c \exp\{ - c' g(a) h(a)^d \}
\frac{g(a) h(a)^4}{a^{d-2}} .\nonumber
\end{eqnarray}
This proves (\ref{e:gauss-counter.11}).
\end{pf}

\subsection[Proof of (1.38)]{Proof of \textup{(\protect\ref{e:gauss-asmp.1a})}}
\label{sub-gau-err}

We here explain briefly how to prove (\ref{e:gauss-asmp.1a}).
The framework of the proof is the same as that of
(\ref{e:gauss-asmp.1}),
except that we choose different $T$ and that we use explicit
error bounds instead of Riemann--Lebesgue lemma.
Concretely, we proceed as follows.

First, instead of (\ref{e:T-choice}),
we now choose
%
\begin{equation}
\label{e:T-choice-err}
T := |x|^{2-(\rho\wedge2)/d},
\end{equation}
and use
the decomposition (\ref{e:G><-def}) and (\ref{e:int-rep.2}).

Improved bound (\ref{e:J-cnd.3'}) on $J(x)$ improves several
estimates concerning contributions from $t > T$.
First, the error term $\hat{R}_2(k)$ of
Lemma \ref{lem-Jhatbd.1} now obeys
$| \hat{R}_2(k) | \leq
\frac{K_2'}{2} |k|^{2+(\rho\wedge2)}$.
Taking $k_t = t^{-1/(2+\rho\wedge2)}$ and using this new bound on
$\hat{R}_2(k)$ improves Lemma \ref{lem-large-t}'s error bound as
$|R_3(t)| \leq c t^{-(d+\rho\wedge2)/2}$. This leads,
with the new choice of $T$, (\ref{e:T-choice-err}), to
%
\begin{equation}
\label{e:G>-err}
C_>(x) = \frac{a_d}{K_1} |x|^{2-d}
+ O\bigl(|x|^{-(d-2+(\rho\wedge2)/d)}\bigr)
.
\end{equation}

Not much is improved for $t<T$, and we use
Lemma \ref{lem-small-t} in its current form, that is,
$I_t(x) = O(|x|^{-d})$. Because of the new choice of $T$,
(\ref{e:T-choice-err}), this leads to slightly improved
%
\begin{equation}
\label{e:G<-err}
C_{<}(x) = O\bigl(|x|^{-(d-2+(\rho\wedge2)/d)}\bigr) .
\end{equation}

Combining (\ref{e:G>-err}) and (\ref{e:G<-err}) yields
(\ref{e:gauss-asmp.1a}), and completes the proof.

\section{Diagrammatic estimates}
\label{sc-diagram}

Here we prove several diagrammatic estimates,
Lemmas \ref{lem-Pihat} and \ref{lem-Pix} for $\Pi$
of the lace expansion.
These estimates are model dependent, and have to be proved
individually for each model.

\subsection{Brief notes on diagrammatic estimates}
\label{ss-diagram-notes}

We first introduce some graphical notation and briefly explain
basic techniques of diagrammatic estimates. These methods
have been extensively used in previous works.
Consult \cite{HS94,MS93,Slad04} for reviews on the lace expansion
and diagrammatic estimates involved.

\begin{figure}[b]

\includegraphics{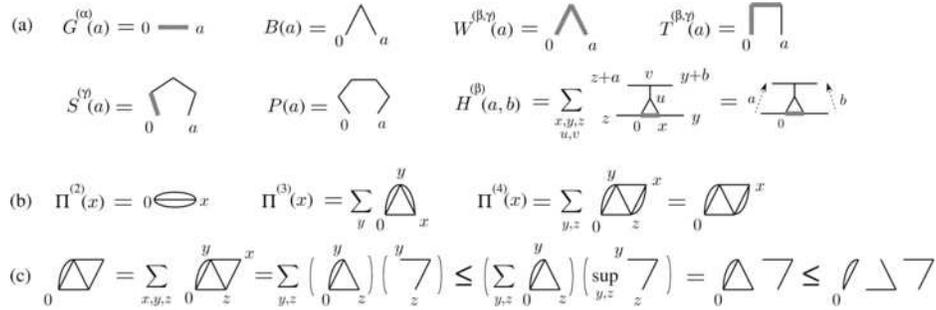}

\caption{
\textup{(a)} Diagrammatic representation of quantities defined in
(\protect\ref{e:Ba-def})--(\protect\ref{e:Ha-def}) for $a\neq0$.
Lines weighted with $|x|^\beta$ and $|x|^\gamma$ are represented
by thick shadowed lines. Two dashed arrows in $H^{(\beta)}(a,b)$
mean that we sum over these vertices, keeping
displacement vectors $a, b$ fixed.
\textup{(b)} Diagrams for $\Pi^{(n)}(x)$ ($n = 2, 3, 4$)
for self-avoiding walk.
\textup{(c)} Using our basic inequality, (\protect\ref{e:basic-ineq.1}).
All unlabeled vertices with degree $\geq2$ are
summed over.
}
\label{fig-diag1}
\end{figure}

For self-avoiding walk, $\Pi^{(0)}(x)$ is identically zero and
$\Pi^{(1)}(x)$ is nonzero only at $x = 0$.
Next few terms of $\Pi^{(n)}(x)$ are bounded as follows:
%
\begin{eqnarray}
&& \Pi^{(2)}(x) \leq G(x)^3,\qquad
\Pi^{(3)}(x) \leq\sum_{y\dvtx y \neq0, x}
G(y)^2 G(x-y)^2 G(x) ,
\nonumber\\[-8pt]\\[-8pt]
&& \Pi^{(4)} (x) \leq\mathop{\mathop{\sum}_{{y\dvtx y \neq0}}}_{z\dvtx z \neq x}
G(y)^2 G(x-y) G(z) G(x-z)^2 G(y-z) .\nonumber
\end{eqnarray}
We introduce diagrammatic expression to represent quantities
on the right-hand side.
In the diagram, a line connecting $x$ and $y$ represents $G(x-y)$,
and unlabeled vertices with degree $\geq2$ are summed
over.
Bounds on $\Pi^{(n)}(x)$ ($n = 2, 3, 4$) are thus represented as
%
\begin{eqnarray}
&& \Pi^{(2)}(x) \leq
\mbox{\protect
\includegraphics{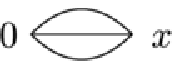}
}
,\qquad
\Pi^{(3)}(x) \leq
\mbox{\protect
\includegraphics{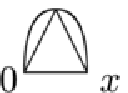}
}
, \nonumber\\[-8pt]\\[-8pt]
&& \Pi^{(4)}(x) \leq\sum_{y,z}
\mbox{\protect
\includegraphics{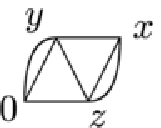}
}
=
\mbox{\protect
\includegraphics{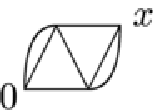}
}
.\nonumber
\end{eqnarray}
Diagrammatic representation for quantities defined
in (\ref{e:Gbardef})--(\ref{e:Ha-def}) are shown
in Figure \ref{fig-diag1}(a), using the above
convention.

%

Special care is required for vertices of degree one.
Vertices of degree one are not usually summed over, unless they
appear in a pair---we sometimes sum over two vertices $x$ and
$x+a$, while keeping $a$ fixed. Two examples appear in
the diagrammatic representation for $H^{(\beta)}(a,b)$ of
Figure \ref{fig-diag1}(a), where the constant vector $a$ and $b$
are represented by dashed arrows.

We next turn to our basic techniques in diagrammatic estimates,
which are used to estimate sums like $\sum_x \Pi^{(n)}(x)$
and $\sum_x |x|^2 \Pi^{(n)}(x)$. We perform this task by
breaking the sum into products of basic units,
using a simple inequality
%
\begin{equation}
\label{e:basic-ineq.1}
\sum_{x} f(x) g(x) \leq\biggl[ \sup_x f(x) \biggr]
\Biggl[ \sum_x g(x) \Biggr],
\end{equation}
which is valid for any nonnegative functions $f, g$.
Here $x$ could be a group of variables.

How to use this inequality in decomposing a diagram into
two small components, and finally into a product of (open)
bubbles, $B(a)$, is illustrated in Figure \ref{fig-diag1}(c).
[In these diagrams, all horizontal lines could be of length
zero; other (slant) lines' lengths are greater than zero.
Therefore, open bubbles are nothing but $B(a)$ with some $a$.]
Graphically, we can just ``peel off'' open bubbles from right
or left.

Arguing this way, we can bound $\sum_x \Pi^{(n)}(x)$
by a product of open bubbles, as
%
\begin{equation}
\label{e:diag-example.11}
\sum_x \Pi^{(n)}(x) \leq\biggl( \sup_{x \neq0} G(x) \biggr)
\biggl( \sup_a B(a) \biggr)^{n-1}
\leq\biggl( \sup_{x \neq0} G(x) \biggr) \bar{B}^{n-1}
.
\end{equation}
Estimates like these will be extensively used in what follows.

\subsection[Proof of Lemma 1.8 for self-avoiding walk]{Proof of Lemma \textup{\protect\ref{lem-Pihat}} for self-avoiding walk}
\label{ss-prf-PihatS}
We start from the proof of Lemma \ref{lem-Pihat} for self-avoiding
walk, which is the simplest of our diagrammatic estimates.
We will prove for $N \geq3$
%
\begin{equation}
\label{e:lem-Pihat-prf.1}
\sum_{x} |x|^{\alpha+\beta+\gamma} \Pi^{(N)}
\leq c N^{\alpha+ \beta+ \gamma+ 2} \lambda^{N-3}
\end{equation}
with a finite constant $c$ which is independent of $N$.
Summing this over $N \geq3$ (the sum converges as long as
$\lambda< 1$) and noting that lowest order ($N=2$)
is bounded by $\sum_x |x|^{\alpha+\beta+\gamma} G(x)^3 =
\sum_x G^{(\alpha)}(x) \times G^{(\beta)}(x) \times G^{(\gamma)}(x)
\leq\bar{G}^{(\alpha)} \bar{W}^{(\beta, \gamma)}= O(1)$
proves the lemma. In the following, we explain
how to prove (\ref{e:lem-Pihat-prf.1}).

\begin{figure}[b]

\includegraphics{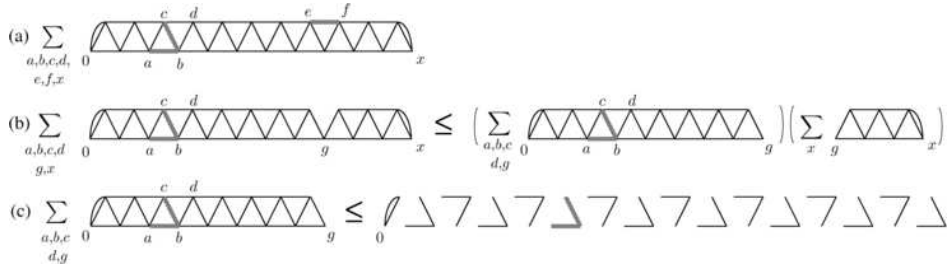}

\caption{\textup{(a)} A typical lace diagram for self-avoiding walk.
\textup{``}Long\textup{''} segments are indicated by thick shadowed lines.
\textup{(b)} After extracting $\bar{G}$ from the diagram \textup{(a)},
decompose at $g$.
\textup{(c)} How to decompose the first factor of \textup{(b)}
into little bubbles and
$W^{(\beta, \gamma)}$. Here all the unlabeled vertices
with degree $\geq2$ are summed over.
}
\label{fig-PihatS}
\end{figure}

\textit{Step \textup{1.}
Distributing the weight $|x|^{\alpha+\beta+\gamma}$.}
A typical lace expansion diagram for self-avoiding walk is shown in
Figure \ref{fig-PihatS}(a).
We want to multiply it with $|x|^{\alpha+\beta+\gamma}$
and sum over all the vertices (except $0$).
For this purpose,
we first distribute the weight $|x|^{\alpha+\beta+\gamma}$ over
suitable line segments of the diagram. Because there are
three distinct lines connecting $0$ and $x$
(the uppermost line, the lowermost line,
and the zigzag line), we pick a long segment out of
each line.

%

Concretely, we proceed as follows.
\begin{itemize}
\item
First pick the longest segment from the lowermost line
connecting $0$ and $x$. To be concrete, suppose
this is $ab$ in Figure \ref{fig-PihatS}(a).
Because the number of segments
of the lowermost line is $\lfloor N/2 \rfloor$,
this longest segment $ab$ is at least as long as
$|x|/\lfloor N/2 \rfloor\geq2|x|/N$.
\item
Next consider the triangle which contains this longest segment.
In Figure \ref{fig-PihatS}(a), this is triangle $abc$. Because
the edge $ab$ is longer than $2|x|/N$, at least one of $ac$
or $bc$ must be longer than $|x|/N$ (by the triangle
inequality).
Choose the longer one of $ac$ and $bc$ as our
second ``long'' segment.
(To be concrete, suppose this is $ac$.)
\item
Finally, choose the longest segment in the uppermost line
connecting $0$ and $x$. This is our third ``long segment.''
Because the number of segments of the uppermost line
is $\lceil N/2 \rceil$, the longest segment is at least as
long as $|x|/\lceil N/2 \rceil\geq|x|/N$.
For concreteness, suppose this is $ef$ in
Figure \ref{fig-PihatS}(a).
\end{itemize}
By the above choice, all three long segments are at least
as long as $|x|/N$. We use this relation to bound the factor
$|x|^{\alpha+\beta+\gamma}
= |x|^{\alpha} \cdot|x|^{\beta} \cdot|x|^{\gamma}$.
In our example, we have
%
\begin{equation}
\label{e:PihatS.51}
|x|^{\alpha+ \beta+ \gamma}
\leq N^{\alpha+ \beta+ \gamma} \times
|e-f|^{\alpha} |a-b|^{\beta} |a-c|^{\gamma} .
\end{equation}

\textit{Step \textup{2.} Decomposition of the diagram.}
Now we control the sum over all vertices of the diagram.
In this example of Figure \ref{fig-PihatS}(a),
we first peel off $\bar{G}^{(\alpha)}$ from
the edge $ef$. This just leaves the diagram with this edge
removed (and the summation over vertices are the same as before);
the result is the diagram on the left-hand side of
Figure \ref{fig-PihatS}(b).
This is further bounded as in Figure \ref{fig-PihatS}(b), by
decomposing it at vertex~$g$. Here, the right factor
looks like the one of Figure \ref{fig-diag1}(c) (with more loops),
and is bounded by a product of open bubbles as explained in
(\ref{e:diag-example.11}).
(For the right factor, we fix $g$ and sum over $x$.)

What remains is to bound the left factor, which is decomposed as
shown in Figure \ref{fig-PihatS}(c). As shown,
this is bounded by a product of open bubbles $B(a)$,
together with $W^{(\beta,\gamma)}(a)$.
There are $(N-2)$ open bubbles (each of which is bounded
by $\lambda$), so the example is bounded by
%
\begin{equation}
\lambda^{-(N-2)} \bar{G}^{(\alpha)}
\biggl( \sup_a W^{(\beta,\gamma)}(a)
\biggr)
= \lambda^{-(N-2)} \bar{G}^{(\alpha)}
\bar{W}^{(\beta, \gamma)}
\leq c \lambda^{-(N-2)} .
\end{equation}

Other diagrams occur, depending on which line segment
is the longest---even for the diagram in
Figure \ref{fig-PihatS}(a), we encounter
$\bar{W}^{(\beta, 0)} \bar{W}^{(0, \gamma)}$ instead of
$\bar{W}^{(\beta, \gamma)}$ alone,
if we pick $bc$ instead of $ac$.
These can be bounded in the same way, and all possible
cases are bounded by $c \lambda^{-(N-3)}$.
This is because each bubble is bounded by $\lambda$,
and there are at least $(N-3)$ of them. (The diagram consists of
$N$-loops, and at most three of them are used as
$\bar{G}^{(\alpha)}$ and $\bar{W}$'s.)

\textit{Step \textup{3.} Summary of the above.}
Each of the weighted $N$-loop $\Pi$ diagrams is bounded from above by
%
\begin{equation}
N^{\alpha+\beta+\gamma} \times\bigl[\bar{G}^{(\alpha)}
\bar{W}^{(\beta, \gamma)} \mbox{ or }
\bar{G}^{(\alpha)}\bar{W}^{(\beta,0)} \bar{W}^{(0,\gamma)}\bigr]
\times c \lambda^{-(N-3)}.
\end{equation}
The number of choices of long segments is bounded by
$\lfloor N/2 \rfloor\times2 \times\lceil N/2 \rceil\leq N^2$.
Thus, the $N$-loop contribution is bounded by
$c N^2 \times N^{\alpha+\beta+\gamma} \times\lambda^{(N-3)}$.
This proves (\ref{e:lem-Pihat-prf.1}), and proves the lemma.

\subsection[Proof of Lemma 1.7 for self-avoiding walk]{Proof of Lemma \textup{\protect\ref{lem-Pix}} for self-avoiding walk}
\label{ss-prf-PixS}

The proof proceeds in the same spirit as that of
Lemma \ref{lem-Pihat}. The diagrams look the same, but
different methods are required because we are now
\textit{fixing $x$}.

\begin{figure}[b]

\includegraphics{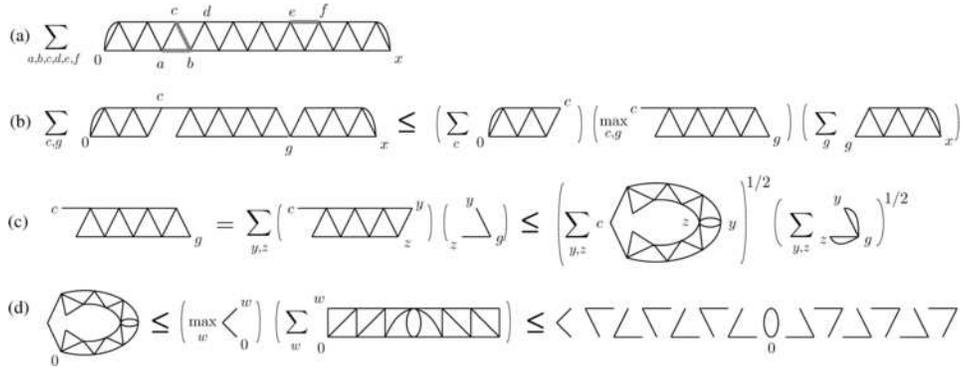}

\caption{\textup{(a)} A typical lace diagram for self-avoiding walk and
long segments.
\textup{(b)} Diagram of \textup{(a)} after extracting three long segments, and
how to bound it by decomposing into three factors.
\textup{(c)} How to use the Schwarz inequality: vertices $c, g$ are fixed.
\textup{(d)} How to decompose the first factor of \textup{(c)}.
Vertices of degree one are \textit{not} summed over.
}
\label{fig-PixS}
\end{figure}

\textit{Step \textup{1.} Picking and extracting
\textup{``}long\textup{''} segments.}
We illustrate by a typical diagram of Figure \ref{fig-PixS}(a).
We first pick and extract three ``long'' segments exactly as we
did in the proof of Lemma \ref{lem-Pihat}.
The result is that we get \textit{three} factors of
%
\begin{equation}
\label{e:Gbardef2}
G_{x,N} := \sup_{y\dvtx |y| \geq|x|/N} G(y)
\leq\beta\biggl( \frac{N}{|x|} \biggr)^\alpha,
\end{equation}
and a remaining diagram, which is shown on the left of
Figure \ref{fig-PixS}(b).
Our remaining task is to bound this diagram by $O(\lambda^{N-3})$.

%

\textit{Step \textup{2.} Decomposition of the diagram.}
We decompose the resulting diagram
as shown on the right of Figure \ref{fig-PixS}(b).
The left and right factors are further decomposed into open
bubbles easily (recall that we are now
fixing $0$ and $x$), and are
bounded by suitable powers of $\lambda$.
Our remaining task is to bound the middle factor.

\textit{Step \textup{3.} Bounding the middle factor.}
Consider the middle factor on the right of
Figure \ref{fig-PixS}(b) as a summation over
$y, z \in{ {\mathbb Z}^d }$ of the product of
two factors, and use the Schwarz inequality as
in Figure \ref{fig-PixS}(c).
The second factor on the right of (c) is just the bubble squared---to
be more precise, one of them has nonzero lines and
is bounded by $\lambda$, another is $O(1)$.

The first factor on the right of (c) is more complicated.
But here we \textit{fix only one} vertex of this diagram
and sum over all others.
Using translation invariance, we can move the fixed
vertex from $c$ to $0$ as shown on the left of
Figure \ref{fig-PixS}(d).
Having moved $c$, we can now decompose this into open
bubbles as shown. (Here we are using our convention that
no vertices of degree one are summed over.)

\textit{Step \textup{4.} Summary.}
We have seen that extracting three ``long'' lines
yields $(G_{x,N})^{3}$, while the remaining diagram
is bounded by $O(\lambda^{N-3})$.
We have to sum over all the possible choices of the long segments.
As shown in the proof of Lemma \ref{lem-Pihat},
the number of choices of long segments is bounded by $N^2$.
Using our assumption on the decay of $G$, we thus have
\begin{eqnarray}\qquad
\Pi^{(N)}(x)
& \leq& c N^2 \times\lambda^{N-3} \times
(G_{x,N})^{3}
\leq c
N^{2} \lambda^{N-3} \biggl( \frac{\beta}{(|x|/N)^{\alpha}}
\biggr)^3
\nonumber\\[-8pt]\\[-8pt]
&
= &c N^{2 + 3 \alpha} \lambda^{N-3}
\frac{\beta^3}{|x|^{3\alpha}}\nonumber
\end{eqnarray}
with a finite constant $c$.
Summing this over $N\geq3$ (the sum converges as long as
$\lambda< 1$), and noting that the lowest
order ($N=2$) is bounded by
$G(x)^{3} \leq\beta^3/|x|^{3\alpha}$ proves the lemma.

\subsection[Proof of Lemma 1.8 for percolation]{Proof of Lemma \textup{\protect\ref{lem-Pihat}} for percolation}
\label{ss-prf-lem-Pihat}

This is proven along the same line as for self-avoiding walk, but
we encounter more complicated percolation diagrams \cite{HS90a}.
Although we have to consider general $N$-loop diagrams, details are
explained by using 4-loop diagrams as examples.
General cases will be extrapolated rather easily.

\begin{figure}

\includegraphics{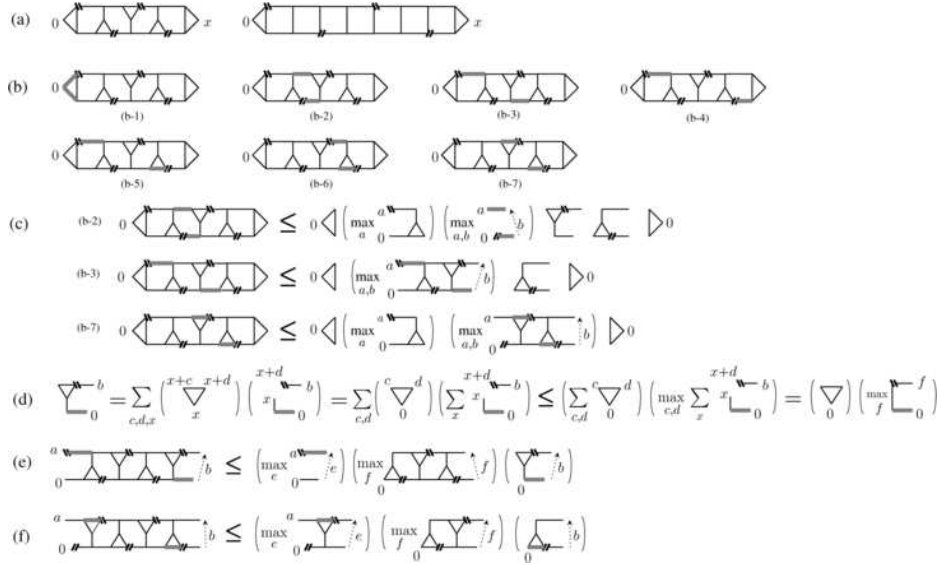}

\caption{
\textup{(a)} Typical four loop diagrams of $\Pi^{(4)}(x)$ for percolation.
\textup{(b)} Possible choices of \textup{``}long\textup{''} segments, indicated by
shadowed thick lines.
\textup{(c)} How to decompose some cases of \textup{(b)} into
simple components. Note that the left and right factors can be
further decomposed into (open) triangles, and produce powers of
$\lambda$.
\textup{(d)} How to decompose a factor appearing in \textup{(c)} into a triangle
and a weighted triangle, $\bar{T}^{(0,\gamma)}$.
The second equality follows from translation invariance.
The rightmost factor is
not exactly equal to, but is bounded by a constant multiple of,
$\bar{T}^{(0,\gamma)}$ as explained around (\protect\ref{e:Tpiv-bd.3}).
\textup{(e)} How to decompose the middle factor of \textup{(b-3)}
into basic components.
\textup{(f)} How to decompose the middle factor of \textup{(b-7)} into basic
components.
The leftmost diagram is new and is bounded by a
constant multiple of $\bar{H}^{(\beta)}$.
}
\label{fig-PihatP}
\end{figure}

Diagrams of $\Pi^{(4)}$ look like
those of Figure \ref{fig-PihatP}(a), plus 14 others.
[In general, there are $2^N$ diagrams for $\Pi^{(N)}$.]
Dealing with the right diagram (and 14 others) is easier,
and we only explain how to deal with the left one.

Before going into details we explain about a special feature of
percolation diagrams. In percolation diagrams, we encounter
\begin{picture}(33,6)
\put(4,0){$x$}
\put(10,-0.5){\line(0,1){4.5}}
\put(10.2,-0.5){\line(0,1){4.5}}
\put(10.4,-0.5){\line(0,1){4.5}}
\put(12,-0.5){\line(0,1){4.5}}
\put(12.2,-0.5){\line(0,1){4.5}}
\put(12.4,-0.5){\line(0,1){4.5}}
\put(12,1.5){\line(1,0){15}}
\put(29,0){$y$}
\end{picture}
,
which represents $2 d p (D*G)(y-x)$
\cite{HS90a}. This is almost the same as $G(y-x)$ for large
$|y-x|$, because
%
\begin{eqnarray}
\label{e:Tpiv-bd.1}
2d p (D*G)(y-x) &=& 2dp \sum_{z\dvtx |z-x|=1} \frac{1}{2d} G(y-z)\nonumber\\[-8pt]\\[-8pt]
&\leq&(1 + c_4 \lambda)
\sum_{z\dvtx |z-x|=1} \frac{1}{2d} G(y-z) .\nonumber
\end{eqnarray}
Some care is needed when $|y-x|=1$ can happen. For example,
the rightmost factor of Figure \ref{fig-PihatP}(d) is
\begin{eqnarray}
\label{e:Tpiv-bd.2}
&&2 d p \sum_{|u|=1} \frac{1}{2d} \bigl(G^{(\gamma)}*G*G\bigr)(f-u)\nonumber\\
&&\qquad= p \sum_{|u|=1} \bigl(G*G^{(\gamma)}*G\bigr)(f-u)
\\
&&\qquad
= p \sum_{|u|=1} \bigl\{
T^{(0,\gamma)}(f-u) + \delta_{\gamma,0}\delta_{f-u, 0} \bigr\}.\nonumber
\end{eqnarray}
When $|f| \neq1$, the above is bounded by
$2dp \bar{T}^{(0,\gamma)}
\leq(1+ c_4 \lambda) \bar{T}^{(0,\gamma)}$.
But when $|f| =1$, $f-u$ can be zero for one $u$. In this case
we get
\begin{eqnarray}
\label{e:Tpiv-bd.3}
&&
2 d p \sum_{|u|=1} \frac{1}{2d} \bigl(G^{(\gamma)}*G*G\bigr)
(f-u)\nonumber\\
&&\qquad\leq p \bigl[ (2d-1) \bar{T}^{(0,\gamma)}
+ \bigl(1 + \bar{T}^{(0,\gamma)}\bigr)
\bigr]
\\
&& \qquad
\leq(1+ c_4 \lambda) \biggl[ \bar{T}^{(0,\gamma)}
+ \frac{1}{2d}
\biggr] .\nonumber
\end{eqnarray}
We can thus conclude that the rightmost factor of
Figure \ref{fig-PihatP}(d) is bounded by (\ref{e:Tpiv-bd.3}).

%

\textit{Step \textup{1.}
Distributing the weight $|x|^{\beta+\gamma}$.}
To deal with $|x|^{\beta+\gamma}$, we first note that
there are \textit{two} (upper and lower) disjoint paths which
connect $0$ and $x$.
Out of each line, we pick up the longest segment, as we did for
self-avoiding walk. Because there are at most $(2N+1)$ segments
for each of the upper and lower lines of a $N$-loop diagram,
these ``long'' segments are not shorter than $|x|/(2N+1)$.
Various choices of these elements are illustrated as
Figure \ref{fig-PihatP}(b), where long segments are indicated
by thick shadowed lines. Suppose for concreteness that
$|x|^\beta$ is on the upper line, and $|x|^\gamma$ is on the
lower line.

\textit{Step \textup{2.} Decomposition of the diagram.}
Next we control the sum over all vertices of the diagram.
This procedure is illustrated in Figure \ref{fig-PihatP}(c).
We can peel off (open) triangles from left and right, leaving
$|x|^{\beta}$-, $|x|^{\gamma}$-weighted parts in the middle.

For (b-2), the middle factor is nothing but
$W^{(\beta,\gamma)}$, which is assumed to be finite,
and we are done.

The case (b-3) is explained in Figure \ref{fig-PihatP}(e).
(We have increased the number of loops in the middle, to illustrate
more general $N$-loop diagrams.)
As shown, we can peel off $W^{(\beta, 0)}$ from the left, decompose
the middle part into triangles, and are left with the right factor.
The right factor itself is decomposed as in
Figure \ref{fig-PihatP}(d), and is bounded by the product of a
triangle and $T^{(0,\gamma)}$.

The case (b-7) is more complicated. Decomposing as before,
we encounter the leftmost component of Figure \ref{fig-PihatP}(f).
[Other parts can be decomposed into triangles and $T^{(0,\gamma)}$,
and are controlled well.]
This is nothing but $H^{(\beta)}(a, b)$ of (\ref{e:Ha-def}),
and is finite by the assumption of the lemma.

\textit{Step \textup{3.} Summary.}
Proceeding this way, we see all the cases of
weighted $N$-loop $\Pi$ diagrams are bounded above by
%
\begin{equation}\qquad\quad
(2N+1)^{\beta+\gamma} \times\bigl[
\bar{W}^{(\beta, \gamma)} \mbox{ or }
\bar{W}^{(\beta,0)} \bar{T}^{(0,\gamma)} \mbox{ or }
\bar{H}^{(\beta)} \bar{T}^{(0,\gamma)}\bigr]
\times(\mbox{triangles}).
\end{equation}
The diagram consists of $N$ nontrivial loops,
and at most two of them are used as
$\bar{W}^{(\beta, \gamma)}$, $\bar{W}^{(\beta, 0)}$,
$\bar{H}^{(\beta)}$ and/or $\bar{T}^{(0,\gamma)}$.
So there are at least $(N-2)$ open triangles,
each of which is bounded by $2 d p (\lambda+ \frac{1}{2d})$
($\leq2 \lambda$ for sufficiently large $d$ and small $\lambda$).
In addition, we have small triangles which are bounded by
$(1+\lambda)$ \cite{HS90a}.
So the triangles contribute $(c\lambda)^{N-2}$ with some $c$.

The number of choices of ``long'' segments are bounded by
$(2N+1)^2$, because there are at most $(2N+1)$ segments for
upper and lower lines. Also, there are $2^N$ diagrams
for $\Pi^{(N)}$. The
$N$-loop contribution is thus bounded by
%
\begin{equation}
c 2^N \times N^2 \times(2N+1)^{\beta+\gamma}
\times(c'\lambda)^{N-2}
= c N^{2+\beta+\gamma} (2c'\lambda)^{N-2} .
\end{equation}
Summing this over $N \geq2$ and taking care of $N= 0, 1$
separately proves (\ref{e:lem-Pihat-prf.1}) and the lemma.
(The cases of $N=0, 1$ are rather simple,
and the details are omitted.)

\subsection[Proof of Lemma 1.7 for percolation]{Proof of Lemma \textup{\protect\ref{lem-Pix}} for percolation}
\label{ss-prf-lem-Pix}

The basic idea is the same as for the self-avoiding walk.
We extract \textit{two} (cf. \textit{three} for
self-avoiding walk) factors of long $G$ from the upper
and lower lines connecting $0$ and $x$, and bound the
rest by $(c\lambda)^{N-3}$.

\begin{figure}

\includegraphics{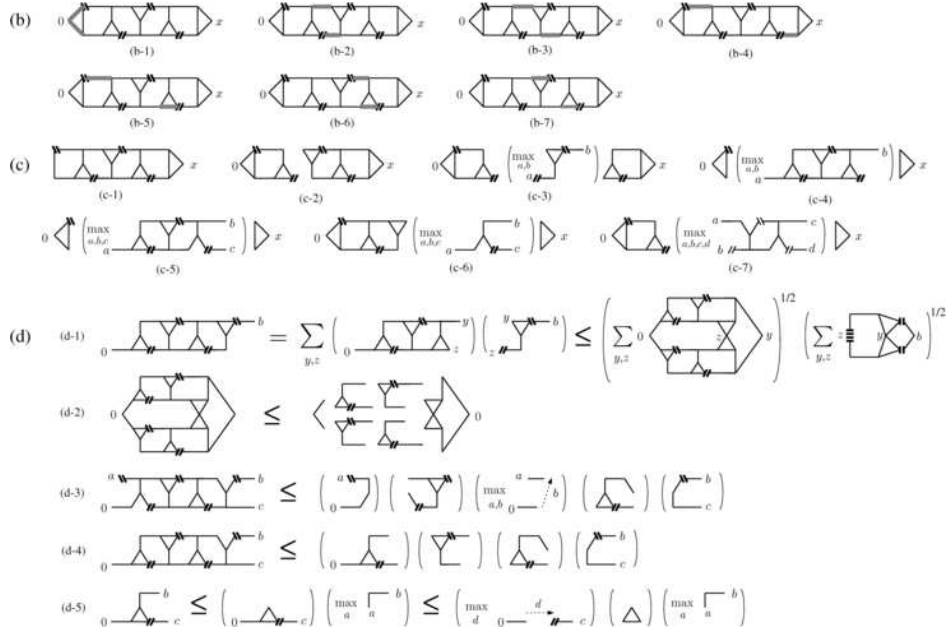}

\caption{
\textup{(b)} Several cases of \textup{``}long\textup{''} lines for the diagram of
Figure \protect\ref{fig-PihatP}\textup{(a)} on the left.
\textup{(c)} Diagrams of \textup{(b)}, after extracting two long lines, and how
to decompose them into smaller components.
\textup{(d)} Typical nasty diagrams of case 2. \textup{(d-1)} is bounded by
the Schwarz inequality, and the result is further decomposed
as shown in \textup{(d-2)}.
\textup{(d-3)} is decomposed into triangles, squares, and two $G$\textup{'}s.
\textup{(d-4)} is decomposed into triangles, squares, and
a factor of \textup{(d-5)}, which is further decomposed into a triangle
and bubbles.
}
\label{fig-PixP}
\end{figure}

As for self-avoiding walk, we first pick ``long'' segments.
We have a segment of length $\geq|x|/(2N+1)$ on the
upper and lower sides of the diagram connecting
$0$ and $x$. These long segments can be any lines which lie
on the upper and lower sides of the diagram,
so the total number of choices are bounded by $(2N+1)^2$.
Several different cases are shown in Figure \ref{fig-PixP}(b),
where the shaded thick lines represent these long segments.
These cases are grouped into two.

%

\textit{Case \textup{1.}}
This is when (i) we have these ``long'' segments
on two lines on a rectangle (or triangle) facing each other,
like Figure \ref{fig-PixP}(b-1) and (b-2),
or (ii) we have long lines on adjacent rectangles,
like Figure \ref{fig-PixP}(b-3).
In either case, we just bound the diagram by extracting
two factors of
%
\begin{equation}
\label{e:longline.bd1}
G_{x,N} := \sup_{y\dvtx |y| \geq|x|/(2N+1)} G(y)
\leq\frac{\beta(2N+2)^{\alpha}}{|x|^{\alpha}} ,
\end{equation}
where the second inequality follows from our assumption
(\ref{e:PixS-ass}) of the lemma on $G(x)$.

The effect of extracting these $G$'s is nothing but
erasing these two lines in the diagram, so the case (b-2) is bounded
by (c-2), after extracting two factors of $G_{x, N}$.
The remaining components in (c-2) as well as in (c-1)
are easily bounded in terms of
triangles, because $0$ and $x$ are now fixed.

The case (b-3) is similar. By peeling off from
left and right, we get (c-3).
Now the factor in the middle is easily seen to be bounded by
two triangles (just extract the small triangle first).

To summarize,\vadjust{\goodbreak} case 1 can be bounded by
$G_{x,N}^{2}$ times convergent diagrams,
which are bounded by some powers of
triangles. By counting the number of nontrivial loops, we see
that these triangles are bounded by a $O((c\lambda)^{N-2})$,
just as in the proof of Lemma \ref{lem-Pihat}.

\textit{Case \textup{2.}}
There remain more complicated cases, but the basic idea is the
same. As shown in Figure \ref{fig-PixP}(c-4) through
Figure \ref{fig-PixP}(c-7), we decompose into the component
in the middle which is hard to deal with, and the left and the
right components which can be easily decomposed into
triangles. We now concentrate on the component in the middle.

There are essentially three kinds of these, which are shown in
Figure \ref{fig-PixP}(d), as (d-1), (d-3) and (d-4).
The case (d-3) has two lines at each end of the diagram,
while the case (d-1) has only one line.
The case (d-4) is a kind of ``cross-term'' of these two.

\textit{This is where we have to impose the restriction $d > 8$}
even if we assume \mbox{$\lambda\ll1$.}
It is natural that the lemma holds for $d > 6$, but currently we
cannot control the middle factor in $6 < d \leq8$.
In $d>8$, we can bound the middle factor by decomposing it into
open triangles and a square $\bar{S}^{(0)}$, as shown in
Figure \ref{fig-PixP}(d). The infrared bound
(\ref{e:tr-Ghat-90}) guarantees that $\bar{S}^{(0)}$
is finite in $d > 8$.

The factor of Figure \ref{fig-PixP}(d-1) is bounded by the
Schwarz inequality as has been done for self-avoiding walk diagrams,
as shown in the right-hand side of Figure \ref{fig-PixP}(d-1).
The resulting components are further bounded by $\bar{B}, \bar{T}$
and $\bar{S}$ as shown in Figure \ref{fig-PixP}(d-2).
The factor of Figure \ref{fig-PixP}(d-3) is bounded
as shown, in terms of open triangles, squares, and two $G$'s.
The factor of Figure \ref{fig-PixP}(d-4) is decomposed
as shown, and its first factor is further decomposed as in (d-5).

In all these cases, we can collect at least $(N-3)$ factors of
$c\lambda$ and two factors of $G_{x,N}$ for
each diagram. Multiplying by the number of different choices of
``long'' segments [which is $O(N^2)$], and summing over $N$
proves the lemma.

\subsection[Proof of Lemma 1.8 for lattice trees
and animals]{Proof of Lemma \textup{\protect\ref{lem-Pihat}} for lattice trees
and animals}
\label{ss-prf-lem-PihatLTLA}

The proof proceeds along the same line as for
self-avoiding walk and
percolation, and we will be brief.

\begin{figure}

\includegraphics{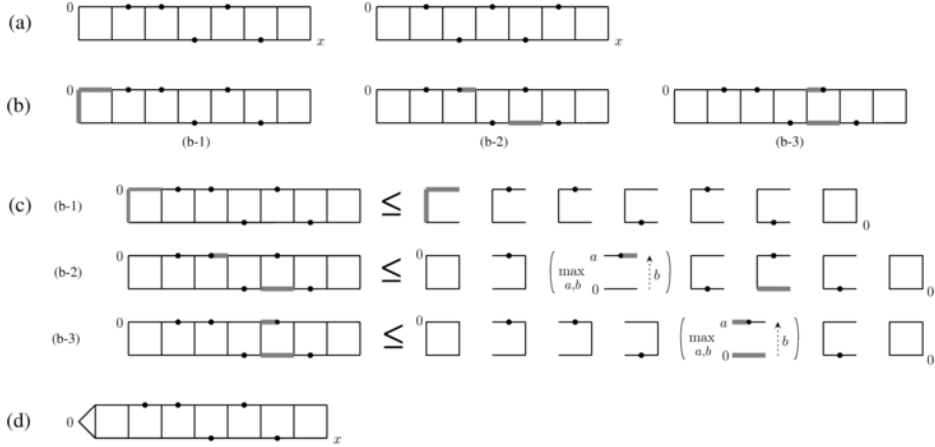}

\caption{
\textup{(a)} Two examples of diagrams of $\Pi^{(7)}(x)$
for lattice trees.
\textup{(b)} Possible choices of \textup{``}long\textup{''} segments, represented by
shadowed thick lines, for the left diagram of \textup{(a)}.
\textup{(c)} How to decompose three cases of \textup{(b)} into
basic components.
\textup{(d)} A typical diagram of $\Pi^{(7)}(x)$ for lattice animals.
The only difference between lattice trees and animals is that
we have an extra triangle on the left (at $0$).
}
\label{fig-PihatLTLA}
\end{figure}

Typical diagrams for $\Pi^{(7)}(x)$ of lattice trees are given in
Figure \ref{fig-PihatLTLA}(a).
In general, diagrams for $\Pi^{(N)}(x)$ consist of $N$ small squares,
with an extra vertex on each inner square.
These inner extra vertices (and $x$ itself) can appear on
either (upper and lower) side of the diagram, and there
are $2^{N-1}$ diagrams for $\Pi^{(N)}(x)$.

%

As in the percolation diagrams, there are two
(upper and lower) disjoint paths which connect $0$ and $x$.
Out of each line, we pick the longest segment, as we did for
self-avoiding walk. Because there are at most $2N$ segments
for each of the upper and lower lines of a $N$-loop diagram,
these ``long'' segments are not shorter than $|x|/(2N)$.
Several choices of these segments are illustrated in
Figure \ref{fig-PihatLTLA}(b), where long segments are
represented by thick shadowed lines. Suppose for concreteness
that $|x|^\beta$ is on the upper line, and $|x|^\gamma$ is
on the lower line.

Next we control the sum over all vertices of the diagram.
This procedure is illustrated in Figure \ref{fig-PihatLTLA}(c).
We peel off $S^{(0)}(a)$, $S^{(\gamma)}(a)$,
or $T^{(\beta, \gamma)}(a)$ from right to left.
For (b-1), we peel off open squares from the right,
and the remaining leftmost factor is bounded by
$\bar{T}^{(\beta, \gamma)}$.
For (b-2), we proceed similarly, but encounter
$\bar{S}^{(\gamma)}$ and $\bar{T}^{(\beta, 0)}$ in the process.
For (b-3), we encounter $\bar{T}^{(\beta, \gamma)}$.

Proceeding this way, we see that all the cases of
weighted $N$-loop $\Pi$ diagrams are bounded above by
%
\begin{equation}
(2N)^{\beta+\gamma} \times\bigl[
\bar{T}^{(\beta, \gamma)} \mbox{ or }
\bar{T}^{(\beta,0)} \bar{S}^{(\gamma)} \bigr]
\times(\mbox{squares}).
\end{equation}
The diagram consists of $N$ nontrivial loops,
and at most two of them are used as
$\bar{T}^{(\beta, \gamma)}$, $\bar{T}^{(\beta, 0)}$,
and/or $\bar{S}^{(\gamma)}$.
So there are at least $(N-2)$ open squares,
each of which is bounded by $\lambda$.

The number of choices of ``long'' segments are bounded by
$(2N)^2$, because there are at most $2N$ segments for
upper and lower lines. Also, there are $2^N$ diagrams
for $\Pi^{(N)}$. The
$N$-loop contribution is thus bounded by
%
\begin{equation}
c 2^N \times(2N)^2 \times(2N)^{\beta+\gamma}
\times(\lambda)^{N-2}
= c N^{2+\beta+\gamma} (2\lambda)^{N-2} .
\end{equation}
Summing this over $N \geq2$ and considering $N= 1$
separately proves (\ref{e:lem-Pihat-prf.1}) and the lemma
for lattice trees.

Diagrams for lattice animals are almost the same as those for
lattice trees, except that there is an extra triangle at $0$,
as shown in Figure \ref{fig-PihatLTLA}(d). (Diagrams in
Figure \ref{fig-PihatLTLA}(d) incorporate an
improvement achieved in \cite{HHS03}
over the analysis in \cite{HS90b}.)
These can be handled in the same way as for lattice trees.

\subsection[Proof of Lemma 1.7 for lattice trees
and animals]{Proof of Lemma \textup{\protect\ref{lem-Pix}} for lattice trees
and animals}
\label{ss-prf-lem-PixLTLA}

\begin{figure}

\includegraphics{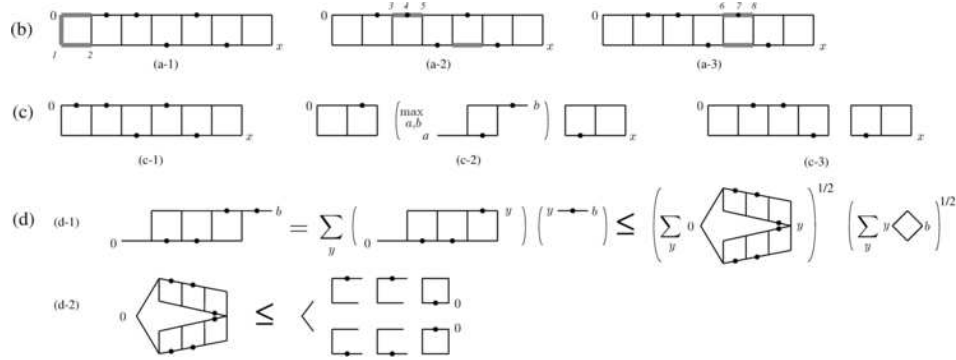}

\caption{
\textup{(b)} Several cases of \textup{``}long\textup{''} segments for a lattice tree
diagram. Vertices 1 through 8 are summed over; they are here
just for the explanation in the main text.
\textup{(c)} Diagrams of \textup{(b)}, after extracting two long segments.
\textup{(d)} How to decompose the middle factor (with more loops) of \textup{(c-2)}.
}
\label{fig-PixLTLA}
\end{figure}

The basic idea is again the same as for self-avoiding walk
and percolation, and we will be brief.
We illustrate for a typical example of
the left of Figure \ref{fig-PihatLTLA}(a).
As for percolation, we extract two factors of long segments
from the upper and lower lines connecting $0$ and $x$,
and bound the rest by $(c\lambda)^{N-3}$.
However, we now consider
the convolution $G*G$ appearing in the diagram as one segment.
Three typical choices of long segments are shown in
Figure \ref{fig-PixLTLA}(b). Here each of 0-1-2, 3-4-5, and
6-7-8 is considered to be a single segment.

%

Because of our modified definition of line segments, there are
$N$ segments for upper and lower lines connecting $0$ and $x$.
Therefore, each long segment is at least as long as $|x|/N$,
and there are $N^2$ choices of these long segments.

We now extract contributions
of ``long'' segments from upper and lower lines.
Because of our modified definition of
line segments, factors extracted will be either
$G_{x,N} := \sup_{|y| \geq|x|/N} G(y)$ (as before), or
%
\begin{equation}
\sup_{|y| \geq|x|/N} (G*G)(y) \leq c \beta^2
\biggl( \frac{N}{|x|} \biggr)^{2\alpha-d}\qquad
(\mbox{if } 0 < 2\alpha- d < d)
\end{equation}
(which is new), where the inequality comes from our assumption
(\ref{e:PixS-ass}) and a basic property of convolution,
Lemma \ref{lem-conv} (a). Contributions from
two long segments are thus bounded by
%
\begin{equation}
\label{e:LTLA-bd.3}
c \beta^2 \biggl( \frac{N}{|x|} \biggr)^{2\alpha}
\vee
c \beta^4 \biggl( \frac{N}{|x|} \biggr)^{4\alpha-2d}
.
\end{equation}

Examples of remaining factors are shown in
Figure \ref{fig-PixLTLA}(c). Of these, (c-1) and \mbox{(c-3)}
are easily decomposed into (open) squares, and pose no problem.
The middle factor of (c-2) is more complicated, like several
middle factors of Figure \ref{fig-PixP}(c) for percolation.

For this (and similar middle factors) we use the Schwarz inequality
as we did for percolation. Concretely (we increased the number
of loops to illustrate more complicated typical cases), we proceed
as in Figure \ref{fig-PixLTLA}(d). We first use the Schwarz
inequality to get two diagrams on the right of (d-1).
The second factor is
bounded by $\bar{S}^{(0)}+1$. The first factor is bounded by
decomposing it into open squares and pentagons $\bar{P}$,
as shown in (d-2).
Existing bound (\ref{e:tr-Ghat-90})
guarantees that $\bar{P}$ is finite in $d > 10$.

In all these cases, we can collect at least $(N-3)$ factors of
$c\lambda$ and a factor of (\ref{e:LTLA-bd.3}) for
each diagram. Multiplying by the number of different choices of
``long'' segments (which is $N^2$) and the number of $N$-loop
diagrams (which is $2^N$), and summing over $N$ proves the lemma.

The proof for lattice animals proceeds similarly and is omitted.

\section[Proof of Lemma 1.9]{Proof of Lemma \protect\ref{lem-PiToGbarWT}}
\label{sb-apr-saw}

In this section, we prove Lemma \ref{lem-PiToGbarWT}.
The basic idea of the proof is simple, but technical details
can be complicated (especially for noninteger exponents).
Therefore, we first explain the framework of the proof
in Section \ref{sb-lemPiToG-framework}, and give details in
later sections.

\subsection[Framework of the proof of Lemma 1.9]{Framework of the proof of Lemma
\textup{\protect\ref{lem-PiToGbarWT}}}
\label{sb-lemPiToG-framework}

\subsubsection{Reduction of the proof to certain integrability
conditions}
\label{subsub-lemPiToG-integrability}

Our goal is to prove that $G^{(\alpha)}(a)$,
$W^{(\beta, \gamma)}(a)$, \ldots, $H^{(\beta)}(a, b)$ are
finite uniformly in $a, b \in{ {\mathbb Z}^d }$.
However, it is cumbersome to deal with
$|x|^{\alpha}$-weighted quantities, especially when $\alpha$ is
not an even integer. We thus define $(j, l = 1, 2, \ldots, d$)
%
\begin{equation}
G^{(\alpha)}_j(a) := |a_j|^{\alpha} G(a) ,\qquad
W_{jl}^{(\beta, \gamma)}(a)
:= \bigl(G_{j}^{(\beta)} * G_{l}^{(\gamma)}\bigr)(a)
,
\end{equation}
%
%
\begin{eqnarray}
T_{jl}^{(\beta,\gamma)}(a) &:=&
\bigl( G_{j}^{(\beta)} * G_{l}^{(\gamma)} * G \bigr) (a),
\nonumber\\[-8pt]\\[-8pt]
S_{j}^{(\gamma)}(a) &:=&
\bigl( G_{j}^{(\gamma)} * G*G*G \bigr) (a)\nonumber
,
\end{eqnarray}
and similarly $H_j^{(\beta)}(a, b)$.
In view of an elementary inequality
%
\begin{eqnarray}
|x|^\alpha&=& \Biggl( \sum_{j=1}^d x_j^2 \Biggr)^{\alpha/2}
\leq c_{\alpha} \sum_{j=1}^d |x_j|^{\alpha}
\qquad\mbox{with}\nonumber\\[-8pt]\\[-8pt]
c_{\alpha} &=&
\cases{ d^{\alpha-1}, & \quad$(\alpha\geq1),$
\cr
d^{\alpha}, &\quad $(0 < \alpha< 1)$
}\nonumber
\end{eqnarray}
it suffices to prove that
$G_{j}^{(\alpha)}(a)$, $W_{jl}^{(\beta, \gamma)}(a), \ldots,
H_{j}^{(\beta)}(a, b)$ are finite uniformly
in $a, b\in{ {\mathbb Z}^d }$.

Now, these quantities are represented in Fourier space as
\begin{eqnarray}
 G_j^{(\alpha)}(a) &=& \int_{[-\pi,\pi]^{d}}\frac{d^d k}{(2\pi)^d}e^{ika}
{\hat{G}}_j^{(\alpha)}(k) ,
\label{e:Wij-exp.0}\\
W_{jl} ^{(\beta, \gamma)}(a)
&=& \int_{[-\pi,\pi]^{d}}\frac{d^d k}{(2\pi)^d}e^{i k a}
{\hat{G}}_j^{(\beta)}(k)
{\hat{G}}_l^{(\gamma)}(k) ,
\label{e:Wij-exp.0a}\\
T_{jl}^{(\beta, \gamma)}(a) &=& \int_{[-\pi,\pi]^{d}}\frac{d^d
k}{(2\pi)^d}e^{i k a}
{\hat{G}}_j^{(\beta)} (k) {\hat{G}}_l^{(\gamma)} (k) {\hat{G}}(k),
\label{e:Wij-exp.1}\\
S_{j}^{(\gamma)}(a) &=& \int_{[-\pi,\pi]^{d}}\frac{d^d k}{(2\pi
)^d}e^{i k a}
{\hat{G}}_j^{(\gamma)} (k) {\hat{G}}(k)^3,
\label{e:Wij-exp.1a}\\
H_{j} ^{(\beta)}(a, b)
&=& \int\!\!\int\!\!\int_{[-\pi, \pi]^{3d}}
\frac{d^d k}{(2\pi)^d} \frac{d^d l}{(2\pi)^d}
\frac{d^d p}{(2\pi)^d}
e^{i (k a + l b)}\nonumber\\
&&\hspace*{58.1pt}{}\times{\hat{G}}_j^{(\beta)}(p)
{\hat{G}}(k)^2 {\hat{G}}(l)^2
\label{e:Hj-exp.2}\\
&&\hspace*{58.1pt}{}
\times
{\hat{G}}(p-k) {\hat{G}}(p+l) {\hat{G}}(k+l) .\nonumber
\end{eqnarray}
Therefore, if we have a good control over ${\hat{G}}_j^{(\alpha)}(k)$,
${\hat{G}}_j^{(\beta)}(k), \ldots,$
so that we can prove integrability of ${\hat{G}}_j^{(\alpha)}$,
${\hat{G}}_j^{(\beta)}{\hat{G}}_l^{(\gamma)}, \ldots\,,$ we are done.

\subsubsection[Proof of Lemma 1.9 for
even integer exponents]{Proof of Lemma \textup{\protect\ref{lem-PiToGbarWT}} for
even integer exponents}
\label{subsub-lemPiToG-even-integer}

The above scenario works perfectly
when $\alpha, \beta, \gamma$ are even integers, because
when $n$ is a positive integer
($\partial_j := \frac{\partial}{\partial k_j}$),
%
\begin{eqnarray}
\label{e:Gj-der-exp.1}
G_j^{(2n)}(x) &&= |x_j|^{2n} G(x) = (-1)^n
\int_{[-\pi,\pi]^{d}}\frac{d^d k}{(2\pi)^d}e^{ik\cdot x} (\partial
_j )^{2n}
{\hat{G}}(k)
\nonumber\\[-8pt]\\[-8pt]
&&\Longrightarrow\quad
{\hat{G}}_j^{(2n)}(k) = (-1)^n (\partial_j)^{2n} {\hat{G}}(k)
, \nonumber
\end{eqnarray}
and a good bound on the derivative is given by the following lemma.
%
\begin{lemma}
\label{lem-m-derivative}
Suppose we have
%
\begin{equation}
\label{e:mder.1}
\sum_x |x|^M |\Pi(x)| < \infty
\end{equation}
for a positive integer $M$. Then, ${\hat{G}}(k)$ of
(\ref{e:Gsawx.1})--(\ref{e:Pihat-def.1}) satisfies,
for all $1 \leq m \leq M$ and $j = 1, 2, \ldots, d$
%
\begin{equation}
\label{e:mder.2}
\biggl| \frac{\partial^m}{\partial k_j^m} {\hat{G}}(k)
\biggr| \leq
\frac{c}{|k|^{2+m}}
\end{equation}
with a possibly $m$-dependent constant $c$.
\end{lemma}

\begin{pf*}{Proof of Lemma \protect\ref{lem-PiToGbarWT} when
$\alpha, \beta, \gamma$ are even integers, assuming
Lemma \ref{lem-m-derivative}}
Lemma \ref{lem-PiToGbarWT} for even integer exponents can now
be proved, by counting powers of $k$ and checking integrability.
When $\alpha$ (resp. $\beta$, $\gamma$) is an even positive integer
which satisfies $\alpha\leq\phi$ (resp.
$\beta\leq\lfloor\phi\rfloor$,
$\gamma\leq\lfloor\phi\rfloor$),
the assumption of the lemma (\ref{e:PiToGbarWT.1}) guarantees
that (\ref{e:mder.1}) holds with $M = \alpha$
(resp. $\beta$, $\gamma$). This allows us to use
(\ref{e:Gj-der-exp.1}) and (\ref{e:mder.2}) to get
$|{\hat{G}}_j^{(m)}(k)| \leq c |k|^{-2-m}$ ($m= \alpha$, or $\beta$,
or $\gamma$).
This implies
%
\begin{equation}
G_j^{(\alpha)}(a) \leq\int_{[-\pi,\pi]^{d}}\frac{d^d k}{(2\pi)^d}
\bigl| {\hat{G}}_j^{(\alpha)}(k)\bigr |
\leq
\int_{[-\pi,\pi]^{d}}\frac{d^d k}{(2\pi)^d}\frac{c}{|k|^{2+\alpha
}} ,
\end{equation}
which is finite for $2 + \alpha< d$. Similarly, by
(\ref{e:Wij-exp.0}) and (\ref{e:Wij-exp.1}),
%
\begin{eqnarray}
W_{jl} ^{(\beta, \gamma)}(a)
&\leq&\int_{[-\pi,\pi]^{d}}\frac{d^d k}{(2\pi)^d}\frac
{c}{|k|^{2+\beta} |k|^{2+\gamma}}
,\nonumber\\[-8pt]\\[-8pt]
S_{j}^{(\gamma)}(a)
&\leq&\int_{[-\pi,\pi]^{d}}\frac{d^d k}{(2\pi)^d}
\frac{c}{|k|^{2+\gamma} |k|^6} .\nonumber
\end{eqnarray}
The first integral is finite if $4 + \beta+ \gamma< d$.
The second integral is finite if $2+\gamma+6 < d$.
$T_{jl}^{(\beta,\gamma)}(a)$ is handled in exactly the same way.

$H_{j}^{(\beta)}(a,b)$ requires more care. Using
(\ref{e:Hj-exp.2}), we have
\begin{eqnarray}\quad
&&H_{j} ^{(\beta)}(a, b)
\leq\int\!\!\int\!\!\int_{[-\pi, \pi]^{3d}}
\frac{d^d k}{(2\pi)^d} \frac{d^d l}{(2\pi)^d}
\frac{d^d p}{(2\pi)^d} \nonumber\\[-8pt]\\[-8pt]
&&\hspace*{115.3pt}{}\times\frac{c}{|p|^{2+\beta}
|k|^{4} |l|^4 |p-k|^{2}
|p+l|^{2} |k+l|^{2} } .\nonumber
\end{eqnarray}
This $3d$-dimensional integral is seen to be finite by
elementary power counting. In short, these integrals are finite,
as long as singularities at the origin are
integrable when some (or all) integral variables are sent to
zero simultaneously (see \cite{Reis88b,Reis88a} for details).
In our case, this is satisfied if $2+\beta<d$, $2+\beta+4+2 < 2d$,
$2+\beta+ 14 < 3d$. These conditions are satisfied when $d > 6$
and $4+\beta< d$.
\end{pf*}

\subsubsection[Proof of Lemma 1.9 for
noninteger exponents\textup{,} $\alpha< \lfloor\phi\rfloor$]{Proof of Lemma \textup{\protect\ref{lem-PiToGbarWT}} for
noninteger exponents\textup{,} $\alpha< \lfloor\phi\rfloor$}
\label{subsub-lemPiToG-non-integer}

When $\alpha, \beta, \gamma$ are not
even integers, the Fourier transform of
$G_j^{(\alpha)}(x) = |x_j|^{\alpha} G(x)$ is not a simple
derivative of ${\hat{G}}(k)$.
[When $\alpha$ is an odd integer, Fourier transform of
$(x_j)^{\alpha} G(x)$ is given by a simple derivative;
this is not true for $|x_j|^{\alpha} G(x)$.]
The answer is given in terms of \textit{fractional derivatives},
which is explained in Section \ref{sec-fourier-frac-power}. As
a result, we get:

\begin{lemma}
\label{lem-Ghatalphabd}
Suppose we have
%
\begin{equation}
\label{e:mder.11}
\sum_x |x|^M | \Pi(x) | < \infty
\end{equation}
for a positive integer $M$. Then $G$ of
(\ref{e:Gsawx.1})--(\ref{e:Pihat-def.1}) satisfies,
for any integer $n \in[1, M \wedge(d-2)]$
and for $0 < \varepsilon< 1$,
%
\begin{equation}
\label{e:G1-ne.rep1}
|x_1|^{n-\varepsilon} G(x)
= G_1^{(n-\varepsilon)}(x) = \int_{[-\pi,\pi]^{d}}\frac{d^d
k}{(2\pi)^d}e^{ikx}
{\hat{G}}_{1}^{(n - \varepsilon)} (k)
\end{equation}
with
%
\begin{equation}
\label{e:Ghat-n-epsilon.5}
\bigl| {\hat{G}}_{1}^{(n - \varepsilon)} (k_1, \vec{k})
\bigr|
\leq
\frac{c}{|k_1|^{1-\varepsilon} |\vec{k}|^n |k|},
\end{equation}
where $c$ may depend on $\varepsilon$.
Here $k = (k_1, \vec{k})$ and
$|k| := (|k_1|^2 + |\vec{k}|^2)^{1/2}$.
\end{lemma}

\begin{pf*}{Proof of Lemma \protect\ref{lem-PiToGbarWT} for
noninteger exponents, assuming Lemma \protect\ref{lem-Ghatalphabd}}
Thanks to Lemma \ref{lem-Ghatalphabd},
we have the bound (\ref{e:Ghat-n-epsilon.5})
for $1 \leq n \leq\lfloor\phi\rfloor\wedge(d-2)$.
We estimate our quantities of interest one by one, using the
above bound.

We start from $\bar{G}^{(\alpha)}$ when
$\alpha< \lfloor\phi\rfloor$.
In this case, $1 \leq n \leq\lfloor\phi\rfloor$ is
satisfied if we write $\alpha= n -\varepsilon$ with
$\varepsilon\in(0,1)$.
Therefore, the bound (\ref{e:Ghat-n-epsilon.5})
allows us to conclude
%
\begin{equation}
G_1^{(n-\varepsilon)}(a) \leq\int_{[-\pi,\pi]^{d}}\frac{d^d
k}{(2\pi)^d}
\bigl| {\hat{G}}_1^{(n-\varepsilon)}(k) \bigr|
\leq
\int_{[-\pi,\pi]^{d}}\frac{d^d k}{(2\pi)^d}\frac
{c}{|k_1|^{1-\varepsilon} |\vec{k}|^n
|k|}
.\hspace*{-35pt}
\end{equation}
Dividing the integration region according to $|k_1| > |\vec{k}|$
or not, we can easily see that the above integral is finite
as long as $1-\varepsilon< 1$, $n < d-1$, and $2+n-\varepsilon<d$.
(These conditions are equivalent to $2+\alpha< d$.)
This proves the lemma for $\bar{G}^{(\alpha)}$, for
$\alpha< \lfloor\phi\rfloor\wedge(d-2)$.
We need a separate argument to deal with $\bar{G}^{(\alpha)}$ for
$\alpha> \lfloor\phi\rfloor$, to which we will come back later.

Controlling $\bar{S}^{(\gamma)}$ and $\bar{H}^{(\beta)}$ is similar.
By ${ {\mathbb Z}^d }$-symmetry, it suffices to show that
$\bar{S}_{1}^{(\gamma)}$ and $\bar{H}_{1}^{(\beta)}$ are finite.
$1 \leq n \leq\lfloor\phi\rfloor$ is satisfied for noninteger
$\gamma$ satisfying $\gamma\leq\lfloor\phi\rfloor$,
if we write $\gamma= n - \varepsilon$ with $\varepsilon\in(0,1)$.
Using (\ref{e:Ghat-n-epsilon.5}) and the Fourier representation
(\ref{e:Wij-exp.1}), we have
\begin{eqnarray}
S_1^{(\gamma)}(a)
& \leq&
\int_{[-\pi,\pi]^{d}}\frac{d^d k}{(2\pi)^d}\bigl| {\hat
{G}}_1^{(n-\varepsilon)} (k) {\hat{G}}(k)^3
\bigr|\nonumber\\[-8pt]\\[-8pt]
&\leq&
\int_{[-\pi,\pi]^{d}}\frac{d^d k}{(2\pi)^d}
\frac{c}{|k_1|^{1-\varepsilon} |\vec{k}|^{n} |k|}
\times\frac{1}{|k|^6}
.\nonumber
\end{eqnarray}
The integral on the right is finite as long as
$n< d-1$ and $n+8-\varepsilon< d$, or equivalently, $\gamma+ 8< d$.
Similarly, writing $\beta= m - \delta$, using
(\ref{e:Ghat-n-epsilon.5}) and the Fourier representation
(\ref{e:Hj-exp.2}), we have
\begin{eqnarray}
H_{1} ^{(\beta)}(a, b)
&
\leq&
\int\!\!\int\!\!\int_{[-\pi, \pi]^{3d}}
\frac{d^d k}{(2\pi)^d} \frac{d^d l}{(2\pi)^d}
\frac{d^d p}{(2\pi)^d}
\frac{c}{|p_1|^{1-\delta} |\vec{p}|^m |p|}
\nonumber\\[-8pt]\\[-8pt]
&&\hspace*{55.3pt}{}
\times\frac{c}{|k|^4 |l|^4 |p-k|^2 |p+l|^2 |k+l|^2}
.\nonumber
\end{eqnarray}
This integral is finite as long as $d>6$, $m-\delta+4 < d$, and
$m < d-1$. This condition is satisfied if $\beta< d-4$ and
$d > 6$.
These prove Lemma \ref{lem-PiToGbarWT} for $\bar{S}^{(\gamma)}$
and $\bar{H}^{(\beta)}$.

Next we move on to $W^{(\beta, \gamma)}$. By ${ {\mathbb Z}^d }$-symmetry,
it suffices to show that $W_{11}^{(\beta, \gamma)}$ and
$W_{12}^{(\beta, \gamma)}$ are finite.
Writing $\beta= m-\delta$ and $\gamma= n - \varepsilon$,
using (\ref{e:Ghat-n-epsilon.5}) and
the Fourier representation (\ref{e:Wij-exp.0}),
%
\begin{equation}
W_{11}^{(\beta, \gamma)} (a)
\leq\int_{[-\pi,\pi]^{d}}\frac{d^d k}{(2\pi)^d}
\frac{c}{|k_1|^{1-\delta} |\vec{k}|^{m} |k|}
\times
\frac{c}{|k_1|^{1-\varepsilon} |\vec{k}|^{n} |k|}
.
\end{equation}
This integral is finite as long as $n+m < d-1$,
$2-\varepsilon-\delta<1$, and
$(n+m+2)+(2-\varepsilon-\delta) < d$. These conditions are
equivalent to $\beta+ \gamma< d-4$ and
$\beta+\gamma-(\lfloor\beta\rfloor+ \lfloor\gamma\rfloor) <1$.

$W_{12}^{(\beta, \gamma)}$ is similar.
By ${ {\mathbb Z}^d }$-symmetry, ${\hat{G}}_{2}^{(\gamma)}$ obeys the same
bound as
${\hat{G}}_{1}^{(\gamma)}$, if we interchange $k_1$ and $k_2$.
Writing $\bolds{k}$ for $k_3, k_4, \ldots, k_d$, and using
the Fourier representation (\ref{e:Wij-exp.0}), we have
\begin{eqnarray}
&&W_{12}^{(\beta, \gamma)} (a)
\leq\int_{[-\pi,\pi]^d} \frac{d^d k}{(2\pi)^d}
\frac{c}{|k_1|^{1-\delta} (|k_2|^2+|\bolds{k}|^2)^{m/2} |k|}
\nonumber\\[-8pt]\\[-8pt]
&&\hspace*{93.8pt}{}\times
\frac{c}{|k_2|^{1-\varepsilon} (|k_1|^2+|\bolds{k}|^2)^{n/2} |k|}
.\nonumber
\end{eqnarray}
This integral is seen to be finite if $\beta+ \gamma< d-4$, by
exhausting six cases depending on the lengths of
$|k_1|$, $|k_2|$, and $|\bolds{k}|$.
We have thus proved Lemma \ref{lem-PiToGbarWT} for
$\bar{W}^{(\beta, \gamma)}$. Proof for $\bar{T}^{(\beta, \gamma)}$
proceeds in exactly the same way and is omitted.
\end{pf*}

\subsubsection[Proof of Lemma 1.9 for
noninteger $\alpha> \lfloor\phi\rfloor$]{Proof of Lemma \textup{\protect\ref{lem-PiToGbarWT}} for
noninteger $\alpha> \lfloor\phi\rfloor$}
\label{subsub-lemPiToG-non-integer-alpha}

Finally, we control $\bar{G}^{(\alpha)}$ for
$\alpha> \lfloor\phi\rfloor$. This is the most complicated
of all the cases relevant for Lemma \ref{lem-PiToGbarWT}.
In this case, $n$ exceeds $\lfloor\phi\rfloor$ if we write
$\alpha= n-\varepsilon$. Thus the assumption
of the lemma is not sufficient to guarantee the finiteness of
$\partial_1^n {\hat{G}}(k)$, and we cannot rely on the bound
(\ref{e:Ghat-n-epsilon.5}), which has been so useful in
previous cases.

To overcome this difficulty, we proceed as follows.
Instead of directly controlling the Fourier transform of
$G^{(n-\varepsilon)}_1(a)$, we treat this quantity
by considering it as a product of $|a_1|^{1-\varepsilon}$ and
$(a_1)^{n-1} G(a)$. When $n$ is even, this product has a different
sign from $|a_1|^{n-\varepsilon}$, but this suffices for our purpose.

The Fourier transform of $(a_1)^{n-1} G(a)$
is given by $(i \partial_1)^{n-1} {\hat{G}}(k)$.
Using the explicit differentiation formula [i.e., (\ref{e:tau.8'})
in the proof of Lemma \ref{lem-m-derivative}], we see that
terms in $(i\partial_1)^{n-1} {\hat{G}}(k)$ can be grouped
into two: (1) terms with $(n-1)$ derivatives on a single
function [i.e., terms with $\partial_1^{n-1}{\hat{J}}(k)$
or $\partial_1^{n-1}{\hat{g}}(k)$],
and (2) terms which contain lower order
derivatives of ${\hat{J}}$ and ${\hat{g}}$. We call the first group
$\hat{P}(k)$, and the second $\hat{Q}(k)$. Explicitly,
%
\begin{eqnarray}
{\hat{P}}(k) &=&
\frac{{\hat{g}}(k) (i\partial_1)^{n-1}{\hat{J}}(k)}
{\{ 1 - {\hat{J}}(k)\}^{2}}
+ \frac{(i\partial_1)^{n-1}{\hat{g}}(k)}{1 - {\hat{J}}(k)}
,\nonumber\\[-8pt]\\[-8pt]
{\hat{Q}}(k) &=& (i\partial_1)^{n-1} {\hat{G}}(k) - {\hat{P}}(k) .\nonumber
\end{eqnarray}
We denote their inverse Fourier transforms
by $P(a)$ and $Q(a)$, so that
%
\begin{eqnarray}
G(a) &=& P(a) + Q(a)\quad \mbox{and}\nonumber\\[-8pt]\\[-8pt]
|a_1|^{1-\varepsilon} G(a) &=&
|a_1|^{1-\varepsilon} P(a) + |a_1|^{1-\varepsilon} Q(a) .\nonumber
\end{eqnarray}
Our task is to show that two quantities on the right
are finite uniformly in $a$.

We begin with $|a_1|^{1-\varepsilon} P(a)$.
We introduce $\hat{P}_1(k)$, $\hat{P}_2(k)$,
$\hat{\psi}_1$ and $\hat{\psi}_2$ as
%
\begin{eqnarray}
{\hat{P}}_1(k) &:= &\frac{{\hat{g}}(k) (i\partial_1)^{n-1}{\hat{J}}(k)}
{\{ 1 - {\hat{J}}(k)\}^{2}}
:=
\hat{\psi}_1(k) (i\partial_1)^{n-1}{\hat{J}}(k),
\nonumber\\[-8pt]\\[-8pt]
{\hat{P}}_2(k) &:=&
\frac{(i\partial_1)^{n-1}{\hat{g}}(k)}{1 - {\hat{J}}(k)}
:=
\hat{\psi}_2(k) (i\partial_1)^{n-1}{\hat{g}}(k)
.\nonumber
\end{eqnarray}
We only consider $P_1$, because dealing with $P_2$ is similar
and easier.

Consider $P_1$ in $x$-space, which reads
%
\begin{equation}
P_1(a) =
\sum_y (y_1)^{n-1} J(y) \psi_1(a-y)
.
\end{equation}
Multiply both sides by $|a_1|^{1-\varepsilon}$,
and on the right-hand side use
$|a_1|^{1-\varepsilon} \leq c (|a_1-y_1|^{1-\varepsilon}
+ |y_1|^{1-\varepsilon})$ with some constant $c$.
As a result, we get two terms:
\begin{eqnarray}\quad
\label{e:d=gen-Pbd.11}
|a_1|^{1-\varepsilon} | P_1(a) |
& \leq &
c \sum_y |y_1|^{n-\varepsilon} |J(y)| \times| \psi_1(a-y) |
\nonumber\\
&&{}
+
c \sum_y |y_1|^{n-1} |J(y)| \times
|a_1-y_1|^{1-\varepsilon} | \psi_1(a-y) |
\nonumber\\[-8pt]\\[-8pt]
&
\leq& c \Biggl[
\sum_{y} |y_1|^{n-\varepsilon} |J(y)| \Biggr]
\biggl[ \sup_x |\psi_1(x)| \biggr]
\nonumber\\
&&{}
+
c \Biggl[ \sum_{y} |y_1|^{n-1} |J(y)| \Biggr]
\biggl[ \sup_x |x_1|^{1-\varepsilon} |\psi_1(x)| \biggr] .\nonumber
\end{eqnarray}
Our task is to show that the four factors are all finite.

First $\sum_{y} |y_1|^{n-\varepsilon} |J(y)|$ is finite, because of
our assumption
(recall $n-\varepsilon= \alpha\leq\phi$).
This also shows that $\sum_{y} |y_1|^{n-1} |J(y)|$ is finite.

To prove that $\sup_x |x_1|^{1-\varepsilon} |\psi_1(x)|$ is finite,
we use the following fact which is proved in
Section \ref{sb-bounds-on-frac-Ghat} using fractional derivatives:
%
\begin{eqnarray}
\label{e:d=gen-Pbd.13aa}
|x_1|^{1-\varepsilon} \psi_1(x) &=&
\int_{[-\pi,\pi]^{d}}\frac{d^d k}{(2\pi)^d}e^{ikx} \hat\psi
_1^{(1-\varepsilon)}(k) \qquad
\mbox{with}\nonumber\\[-8pt]\\[-8pt]
\bigl| \hat{\psi}_1^{(1-\varepsilon)}(k) \bigr|
&\leq&\frac{c}{|k_1|^{1-\varepsilon} |\vec{k}|^3 |k|}
.\nonumber
\end{eqnarray}
$\hat{\psi}_1^{(1-\varepsilon)}(k)$ is
integrable if $d > 5-\varepsilon$, and
$\sup_x |x_1|^{1-\varepsilon} |\psi_1(x)|$
and $\sup_x |\psi_1(x)|$ are finite [$\psi_1(0)$ is easily
seen to be finite in $d >4$ by (\ref{e:d=gen-Pbd.15})].
We have therefore shown that (\ref{e:d=gen-Pbd.11}) is finite
uniformly in $a$.

We now turn to $|a_1|^{1-\varepsilon} Q(a) = Q_1^{(1-\varepsilon)}(a)$.
We will prove in Section \ref{sb-bounds-on-frac-Ghat}:
%
\begin{eqnarray}
\label{e:Q1-frac-bd.11}
|x_1|^{1-\varepsilon} Q_1(x) &=&
\int_{[-\pi,\pi]^{d}}\frac{d^d k}{(2\pi)^d}e^{ikx} {\hat
{Q}}_1^{(1-\varepsilon)}(k)
\qquad
\mbox{with}\nonumber\\[-8pt]\\[-8pt]
\bigl| {\hat{Q}}_1^{(1-\varepsilon)} (k) \bigr|
&\leq&\frac{c}{|k_1|^{1-\varepsilon} |\vec{k}|^{n} |k|}
.\nonumber
\end{eqnarray}
${\hat{Q}}_1^{(1-\varepsilon)} (k)$ is integrable in $k$,
and thus $|a_1|^{1-\varepsilon} Q(a)$ is
finite, as long as $2+n-\varepsilon< d$, or $\alpha+2 < d$.

We have thus shown that both $|a|^{1-\varepsilon}P(a)$
and $|a|^{1-\varepsilon}Q(a)$ are finite uniformly in $a$, and
the proof is complete.

In the following, we prove Lemma \ref{lem-m-derivative},
Lemma \ref{lem-Ghatalphabd}, (\ref{e:d=gen-Pbd.13aa}),
and (\ref{e:Q1-frac-bd.11}), one by one.

\subsection[Proof of Lemma 4.1]{Proof of Lemma \textup{\protect\ref{lem-m-derivative}}}
\label{sb-lemm-derivative}

By ${\mathbb Z}^d$-symmetry,
it suffices to prove (\ref{e:mder.2}) for \mbox{$j=1$,}
and we abbreviate $\partial^{m}$ for
$\frac{\partial^{m}}{\partial k_{1}^{m}}$.
We first note that (\ref{e:mder.1}) implies
%
\begin{equation}
\label{e:tau.7}
| \partial^{m} {\hat{g}}(k) |,
| \partial^{m} {\hat{J}}(k) |< \infty
\qquad\mbox{for all } m \leq M.
\end{equation}
By explicit differentiation, we see from (\ref{e:Gsawx.1})
that
\begin{eqnarray}\qquad\quad
\label{e:tau.8'}
\partial^{m} {\hat{G}}(k)
&=&
\partial^{m} \biggl( \frac{{\hat{g}}(k)}{1 - {\hat{J}}(k)} \biggr)
= \sum_{p = 0}^{m} \pmatrix{m \cr p}
( \partial^{m-p} {\hat{g}}(k) )
\partial^{p} \biggl( \frac{1}{1 - {\hat{J}}(k)} \biggr)
\nonumber\\[-8pt]\\[-8pt]
&=& \partial^m {\hat{g}}(k)
+ \sum_{p=1}^{m} \pmatrix{m \cr p}
( \partial^{m-p} {\hat{g}}(k) )
\sum_{q=2}^{p+1} \sum_{\vec{r}}
C_{\vec{r}}
\frac{\prod_{\ell\geq1}
[ \partial^{\ell} {\hat{J}}(k) ]^{r_{\ell}}}
{\{1 - {\hat{J}}(k)\}^{q}} .\nonumber
\end{eqnarray}
In the above, $\vec{r} = \{ r_{\ell} \}_{\ell\geq1}$
is a vector of nonnegative integers,
$C_{\vec{r}}$ is a coefficient which depends on $\vec{r}$.
The vector $\vec{r}$ satisfies
\begin{subequation}
\begin{eqnarray}
\label{e:tau.9a}
\sum_{\ell\geq1} \ell r_{\ell} & = & p ,
\qquad
\sum_{\ell\geq1} r_{\ell} = q-1 ,
\\
\label{e:tau.9c}
r_{\ell} & = & 0 \hspace*{7.6pt}\qquad\mbox{for } \ell+ q \geq p+3
.
\end{eqnarray}
\end{subequation}

Because $\partial^{\ell} {\hat{J}}(k)$ are all finite (for
relevant values of $\ell$) and
because ${\hat{J}}(k)$ is even in $k_{j}$, we
have for \textit{odd} $\ell$ satisfying $\ell+1 \leq M$
%
\begin{equation}
\label{e:tau.10}
| \partial^{\ell} {\hat{J}}(k) |
\leq\sup_{k}
| \partial^{\ell+1} {\hat{J}}(k) | |k|
= c |k| .
\end{equation}

By (\ref{e:tau.9a}),
%
\begin{equation}
\sum_{\ell\geq2} (\ell-1) r_{\ell}
= \sum_{\ell\geq1} \ell r_{\ell}
- \sum_{\ell\geq1} r_{\ell}
= p - q + 1
\end{equation}
and we have
%
\begin{eqnarray}
\label{e:tau.12}
r_{1} &=& \sum_{\ell\geq1} r_{\ell}
- \sum_{\ell\geq2} r_{\ell}
\geq\sum_{\ell\geq1} r_{\ell}
- \sum_{\ell\geq2} (\ell- 1) r_{\ell}\nonumber\\[-8pt]\\[-8pt]
&=& q-1 - (p-q+1) = 2 q - p - 2.\nonumber
\end{eqnarray}

We now combine (\ref{e:tau.10}) and (\ref{e:tau.12}), dividing
into two cases:

(i) For $q$ sufficiently large such that
$2q > p+2$, we have $r_{1} > 0$ by (\ref{e:tau.12}). We have at
least $r_{1}$ powers of $|k|$ in the numerator, and
the terms in (\ref{e:tau.8'}) with $2q > p+2$ are bounded as
%
\begin{equation}
\label{e:tau.13}
| \mbox{case (i) of } (\ref{e:tau.8'}) |
\leq c\frac{|k|^{r_{1}}}{|k|^{2q}}
= c |k|^{2q-p-2 - 2q} = c |k|^{-p-2} .
\end{equation}

(ii) For $2q \leq p+2$, it may happen that there is
no first derivative in the numerator.
But we at least know that the numerator is finite.
We simply bound these terms as
%
\begin{equation}
\label{e:tau.14}
| \mbox{case (ii) of } (\ref{e:tau.8'}) |
\leq\frac{c}{|k|^{2q}} \leq\frac{c} {|k|^{p+2}}.
\end{equation}

Combining these two cases, we get (for $m \leq M$)
%
\begin{equation}
\label{e:tau.15}
| \partial^{m} {\hat{G}}(k) |
\leq c \sum_{p=0}^{m}
[ |k|^{-p-2} + |k|^{-p-2} ]
\leq\frac{c}{|k|^{m+2}} .
\end{equation}
This proves (\ref{e:mder.2}).
\qed

\subsection{Fourier analysis of fractional powers}
\label{sec-fourier-frac-power}

One way to prove that a given function $f(x)$ decays at least as
fast as $|x|^{-n}$ when $|x| \rightarrow\infty$,
where $n$ is a positive integer, is to
show that the $n$th-derivative of its Fourier
transform $\hat{f}(k)$ is integrable. However, there are cases
where the $n$th-derivative is not integrable, whereas
suitably defined $(n-\varepsilon)$th-derivative
is, for some $0 < \varepsilon< 1$.
We then expect that $f(x)$ decays
at least as fast as $|x|^{-(n-\varepsilon)}$. In this subsection, we
summarize results which will be useful in such cases.
The subject is closely related to fractional derivatives
and can be considered as a special case of \textit{Weyl fractional
derivatives} (\cite{SKM93}, Section 19) if we consider
$f(x)$ as the ``Fourier coefficient'' of ${\hat{f}}(k)$.

In this subsection, $f(x)\dvtx { {\mathbb Z}^d } \rightarrow{\mathbb R}$
always denotes a
${ {\mathbb Z}^d }$-symmetric function, which is represented as
%
\begin{equation}
\label{e:f-fourier.1}
f(x) = \int_{[-\pi,\pi]^{d}}\frac{d^d k}{(2\pi)^d}e^{ikx} {\hat
{f}}(k)\qquad \mbox{with }
{\hat{f}}(k) \in L^1([-\pi, \pi]^d)
\end{equation}
and ${\hat{f}}(k)$ is periodic in each $k_j$
($k=1, 2, \ldots, d$) with period $2\pi$.
We treat the first component $k_1$ of $k$
differently from $k_2, k_3, \ldots, k_d$, and write
$k = (k_1, \vec{k})$. Also, we write $\partial_1 {\hat{f}}(k)$
for the partial derivative with respect to the first
argument of ${\hat{f}}$.
We define
%
\begin{equation}
\operatorname{sgn} x_1 =
\cases{
1, &\quad $( x_1 > 0)$, \cr
0, & \quad$( x_1 = 0)$, \cr
-1, &\quad $(x_1 < 0)$,
}
\end{equation}
and ($\alpha\in{\mathbb R}$)
%
\begin{equation}\qquad
f_1^{(\alpha)}(x)
:= |x_1|^\alpha I[x_1\neq0] f(x) ,\qquad
{f_1'}^{(\alpha)}(x) := |x_1|^\alpha
(\operatorname{sgn} x_1) f(x) .
\end{equation}
Note that the prime on ${f_1'}^{(\alpha)}(x)$ does
\textit{not} mean a derivative.
We introduce for $0 < \varepsilon< 1$
%
\begin{eqnarray}
\label{e:Lodd-def.1}
L_{o, \varepsilon}(p_1)
& :=& \frac{1}{2 \pi i \Gamma(\varepsilon)}
\int_0^\infty dt\, t^{\varepsilon-1} \frac{\sin p_1 }
{\cosh t - \cos p_1} , \nonumber\\[-8pt]\\[-8pt]
L_{e, \varepsilon}(p_1)
& :=& \frac{1}{2 \pi\Gamma(\varepsilon)}
\int_0^\infty dt\, t^{\varepsilon-1} \frac{\cos p_1 - e^{-t} }
{\cosh t - \cos p_1} .\nonumber
\end{eqnarray}
These are Fourier transforms of $|x_1|^{-\varepsilon} (\operatorname{sgn}
x_1)$ and
$|x_1|^{-\varepsilon} I[ x_1 \neq0]$ respectively, in the sense that
%
\begin{eqnarray}
\label{e:Loe-Fourier.1}
|x_1|^{-\varepsilon} ( \operatorname{sgn} x_1 ) &=&
\int_{-\pi}^\pi dp_1 e^{i p_1 x_1} L_{o, \varepsilon}(p_1)
, \nonumber\\[-8pt]\\[-8pt]
|x_1|^{-\varepsilon} I[x_1 \neq0] &=&
\int_{-\pi}^\pi dp_1 e^{i p_1 x_1} L_{e, \varepsilon}(p_1)\nonumber
\end{eqnarray}
hold.
These identities can be proved, for example, by interchanging the order
of $t$ and $p_1$ integrations and using residue calculus.

We begin with the following proposition which represents
$f_1^{(-\varepsilon)}(x)$ and ${f_1'}^{(-\varepsilon)}(x)$ in terms
of Fourier transforms for $0 < \varepsilon< 1$.
The proposition looks almost obvious in view of
(\ref{e:Loe-Fourier.1}); it is a special case of a well-known
fact that the Fourier transform of $f(x)g(x)$ is
given by ${\hat{f}}*{\hat{g}}$.
%
\begin{prop}
\label{prop-f-alpha}
Suppose $f(x)$ is represented by (\ref{e:f-fourier.1}), and define
for $0 < \varepsilon< 1$
%
\begin{eqnarray}
{\hat{f}}_1^{(-\varepsilon)} (k) &:=& \int_{-\pi}^\pi
L_{e, \varepsilon} (p_1) {\hat{f}}(k_1 - p_1, \vec{k})
\,d p_1
,
\nonumber\\[-8pt]\\[-8pt]
{\hat{f'}_1}^{(-\varepsilon)} (k)
&:=& \int_{-\pi}^\pi
L_{o, \varepsilon} (p_1) {\hat{f}}(k_1 - p_1, \vec{k})
\,d p_1 .\nonumber
\end{eqnarray}
Then we have
%
\begin{eqnarray}
\label{e:f-alpha.rep1}
f_1^{(-\varepsilon)}(x)
& =& \int_{[-\pi,\pi]^{d}}\frac{d^d k}{(2\pi)^d}e^{ikx} {\hat
{f}}_1^{(-\varepsilon)} (k)
,
\nonumber\\[-8pt]\\[-8pt]
{f'_1}^{(-\varepsilon)}(x)
& =& \int_{[-\pi,\pi]^{d}}\frac{d^d k}{(2\pi)^d}e^{ikx}
\mbox{${\hat f'_1}$}^{(-\varepsilon)} (k)
.\nonumber
\end{eqnarray}
\end{prop}

For $f_1^{(\alpha)}(x)$ with $\alpha> 0$, which is of our main
interest, several representations
with differing conditions of applicability can be obtained.
Of these, the following will be useful for our analysis.

\begin{prop}
\label{prop-f-n-alpha}
Suppose $f(x)$ is represented by (\ref{e:f-fourier.1}).
Let $m$ be a positive integer, $0 < \varepsilon< 1$,
and assume $(\partial_1)^m {\hat{f}}(k)$ is integrable in $k_1$
for each $\vec{k} \neq\vec{0}$. Define
%
\begin{equation}
\label{e:F1hat-alpha.rep1}
{\hat{f}}_1^{(m-\varepsilon)} (k_1, \vec{k}) := \int_{-\pi}^{\pi}
L_{*, \varepsilon}(p_1) \{ (i \partial_1)^m
{\hat{f}}(k_1-p_1, \vec{k}) \}
\,dp_1
,
\end{equation}
where $L_{*, \varepsilon}= L_{o, \varepsilon}$ if $m$ is odd,
and $L_{*, \varepsilon} = L_{e, \varepsilon}$ if $m$ is even.
Then,
%
\begin{equation}
\label{e:F1-alpha.rep0}
f_1^{(m-\varepsilon)}(x) = \lim_{\delta\downarrow0}
\mathop{{\int}_{{[-\pi,\pi]^{d-1}}}}_{|\vec{k}| > \delta}
\frac{d^{d-1} \vec{k}}{(2\pi)^{d-1}}
\int_{-\pi}^{\pi} \frac{d k_1}{2\pi}
e^{ikx} {{\hat{f}}_1}^{(m-\varepsilon)} (k) .
\end{equation}
If we further assume
${\hat{f}}_1^{(m-\varepsilon)} \in L^1([-\pi, \pi]^d)$, then
%
\begin{equation}
\label{e:F1-alpha.rep1}
f_1^{(m-\varepsilon)}(x) = \int_{[-\pi,\pi]^{d}}\frac{d^d k}{(2\pi)^d}
e^{ikx} {{\hat{f}}_1}^{(m-\varepsilon)} (k) .
\end{equation}
\end{prop}

As for the integral kernels, we have:
\begin{prop}
\label{prop-Lbd}
Fix $0 < \varepsilon< 1$.
$L_{o, \varepsilon}(p_1)$ is pure imaginary, odd in $p_1$,
and satisfies for $p_1 \in[-\pi, \pi]$
%
\begin{equation}
\label{e:L2bd.1}
| L_{o, \varepsilon}(p_1) |
\leq\tfrac{1}{2} |p_1|^{\varepsilon-1} ,\qquad
| \partial L_{o, \varepsilon}(p_1) | \leq|p_1|^{\varepsilon-2}
.
\end{equation}
$L_{e, \varepsilon}(p_1)$ is real-valued, even in $p_1$, and
satisfies for $p_1 \in[-\pi, \pi]$
%
\begin{equation}
\label{e:L4bd.1}
- \frac{\log2}{\pi} \leq L_{e, \varepsilon}(p_1)
\leq\frac{|p_1|^{\varepsilon-1}}{\pi(1-\varepsilon)} ,\qquad
| \partial L_{e, \varepsilon}(p_1) |
\leq\frac{|p_1|^{\varepsilon-2}}{\pi} .
\end{equation}
\end{prop}

The above bounds on the derivatives of $L_{o, \varepsilon}$ and $L_{e,
\varepsilon}$
are not optimal, in the sense that the coefficients on the right-hand
side can be multiplied by $1-\varepsilon$. However, the current bounds
suffice for our purpose.

We in the following briefly prove these propositions, in this order.
Because Proposition \ref{prop-Lbd} can be proved independently
of the rest, we use its result (especially the integrability
of $L_{*, \varepsilon}$) in the proofs of
Propositions \ref{prop-f-alpha} and \ref{prop-f-n-alpha}.

\begin{pf*}{Sketch of the proof of
Proposition \textup{\ref{prop-f-alpha}}}
The proof is almost identical for $f_1^{(-\varepsilon)}$ and
${f'_1}^{(-\varepsilon)}$, and we only treat $f_1^{(-\varepsilon)}$.
Note first that ${\hat{f}}_1^{(-\varepsilon)}$ is well defined
and is integrable, thanks to the integrability of $L_{e, \varepsilon}$
and ${\hat{f}}$.
We calculate (\ref{e:f-alpha.rep1}) using the
definition of $L_{e, \varepsilon}$.
Starting from
\[
\int_{[-\pi,\pi]^{d}}\frac{d^d k}{(2\pi)^d}e^{ikx} {\hat
{f}}_1^{(-\varepsilon)} (k)
= \int_{[-\pi,\pi]^{d}}\frac{d^d k}{(2\pi)^d}e^{ikx} \int_{-\pi
}^\pi dp_1
L_{e, \varepsilon} (p_1) {\hat{f}}(k_1 - p_1, \vec{k})
,
\]
we write the $d$-dimensional integral as an iterated integral
(guaranteed by the integrability of $L_{e, \varepsilon}$ and ${\hat{f}}$)
and then change variables from $k_1, p_1$ to
$k_1' = k_1 - p_1, p_1' = p_1$.
By the periodicity of ${\hat{f}}$, we get
\begin{eqnarray}\qquad
&&
\int_{[-\pi, \pi]^{d-1}}
\frac{d^{d-1}\vec{k}}{(2\pi)^{d-1}} e^{i \vec{k}\vec{x}}
\int_{-\pi}^\pi\frac{d k_1}{2\pi} e^{i k_1 x_1}
\int_{-\pi}^\pi d p_1
L_{e, \varepsilon} (p_1) {\hat{f}}(k_1 - p_1, \vec{k})
\nonumber\\
&&\qquad
= \int_{[-\pi, \pi]^{d-1}}
\frac{d^{d-1}\vec{k}}{(2\pi)^{d-1}} e^{i \vec{k}\vec{x}}
\int_{-\pi}^\pi\frac{d k_1'}{2\pi}\nonumber\\
&&\qquad\quad{}\times
\int_{-\pi}^\pi d p_1'
e^{i (k_1'+p_1') x_1}
L_{e, \varepsilon} (p_1')
{\hat{f}}(k_1', \vec{k})
\\
&&\qquad
= \int_{[-\pi,\pi]^{d}}\frac{d^d k}{(2\pi)^d}e^{ikx} {\hat
{f}}(k_1, \vec{k})
\int_{-\pi}^\pi dp_1 e^{i p_1 x_1} L_{e, \varepsilon}(p_1)
\nonumber\\
&&\qquad
= f(x)
\int_{-\pi}^\pi dp_1 e^{i p_1 x_1} L_{e, \varepsilon}(p_1)
.\nonumber
\end{eqnarray}
Now the last integral is $|x_1|^{-\varepsilon} I[x_1 \neq0]$
by (\ref{e:Loe-Fourier.1}).
\end{pf*}

\begin{pf*}{Sketch of the proof of
Proposition \textup{\ref{prop-f-n-alpha}}}
We deal with even $m$ only---odd $m$ can be treated in the
same way.
Because $L_{e, \varepsilon}$ is integrable and
$(i\partial_1)^m {\hat{f}}(k)$ is integrable in $k_1$
for $\vec{k} \neq\vec{0}$, we can interchange
the $p_1$-integral and $(i\partial_1)^m$ to get
%
\begin{eqnarray}
{\hat{f}}_1^{(m-\varepsilon)} (k_1, \vec{k})
&=& (i \partial_1)^m \int_{-\pi}^{\pi}
L_{e, \varepsilon}(p_1) {\hat{f}}(k_1-p_1, \vec{k})\,
dp_1\nonumber\\[-8pt]\\[-8pt]
&=& (i \partial_1)^m {\hat{f}}_1^{(-\varepsilon)}(k_1, \vec{k})\nonumber
\end{eqnarray}
for $\vec{k} \neq\vec{0}$.
Using this, the right-hand side of
(\ref{e:F1-alpha.rep0}) can be calculated as
\begin{eqnarray}
&&\lim_{\delta\downarrow0}
\mathop{{\int}_{{[-\pi,\pi]^{d-1}}}}_{|\vec{k}| > \delta}
\frac{d^{d-1} \vec{k}}{(2\pi)^{d-1}}
\int_{-\pi}^{\pi} \frac{d k_1}{2\pi}
e^{ikx} {{\hat{f}}_1}^{(m-\varepsilon)} (k)
\nonumber\\
&&\qquad=
\lim_{\delta\downarrow0}
\mathop{{\int}_{{[-\pi,\pi]^{d-1}}}}_{|\vec{k}| > \delta}
\frac{d^{d-1} \vec{k}}{(2\pi)^{d-1}}
\int_{-\pi}^{\pi} \frac{d k_1}{2\pi}
e^{ikx}
\bigl\{ (i \partial_1)^m {\hat{f}}_1^{(-\varepsilon)}(k_1, \vec{k})
\bigr\}
\nonumber\\[-8pt]\\[-8pt]
&&\qquad= (x_1)^m \lim_{\delta\downarrow0}
\mathop{{\int}_{{[-\pi,\pi]^{d-1}}}}_{|\vec{k}| > \delta}
\frac{d^{d-1} \vec{k}}{(2\pi)^{d-1}}
\int_{-\pi}^{\pi} \frac{d k_1}{2\pi}
e^{ikx}
{\hat{f}}_1^{(-\varepsilon)}(k_1, \vec{k})
\nonumber\\
&&\qquad= (x_1)^m \int_{[-\pi,\pi]^{d}}\frac{d^d k}{(2\pi)^d}
e^{ikx}
{\hat{f}}_1^{(-\varepsilon)}(k_1, \vec{k})
.\nonumber
\end{eqnarray}
Here the second equality follows from integration by parts
with respect to $k_1$, and the last equality follows because
${\hat{f}}_1^{(-\varepsilon)}$ is integrable in $k$.
In view of Proposition~\ref{prop-f-alpha}, the integral on
the far right is nothing but $f_1^{(-\varepsilon)}(x)$
for even $m$, and we get (\ref{e:F1-alpha.rep0}).
Now (\ref{e:F1-alpha.rep1}) follows trivially.
\end{pf*}

\begin{pf*}{Sketch of the proof of
Proposition \ref{prop-Lbd}}
That $i L_{o, \varepsilon}(p_1)$ is odd in $p_1$ and is positive
for $p_1 >0$ is easily seen from its integral representation
(\ref{e:Lodd-def.1}). The integral
can be performed exactly by residue calculus,
and we get for $0 < p_1 < \pi$
%
\begin{eqnarray}\qquad\quad
\label{e:Ltwobd.51}
&&i L_{o, \varepsilon}(p_1)\nonumber\\[-8pt]\\[-8pt]
&&\qquad= \frac{\csc(({\pi}/{2})\varepsilon)}{2\Gamma(\varepsilon)}
\Biggl[
p_1^{\varepsilon-1} - \sum_{n=1}^{\infty}
\{
(2\pi n - p_1)^{\varepsilon-1} - (2\pi n + p_1)^{\varepsilon-1}
\}
\Biggr]
.\nonumber
\end{eqnarray}
As the summand is positive, we immediately get for $0 < p_1 < \pi$
%
\begin{eqnarray}
\label{e:Ltbd.53}
0 \leq i L_{o, \varepsilon}(p_1) \leq
\frac{\csc(({\pi}/{2})\varepsilon)}{2\Gamma(\varepsilon)}
p_1^{\varepsilon-1} = C_{o, \varepsilon} p_1^{\varepsilon-1}\nonumber\\[-8pt]\\[-8pt]
\eqntext{\mbox{with }
C_{o, \varepsilon} :=
\dfrac{\csc(({\pi}/{2})\varepsilon)}{2 \Gamma(\varepsilon) }
.}
\end{eqnarray}
The coefficient $C_{o, \varepsilon}$ is increasing in $\varepsilon$
for $0 < \varepsilon\leq1$ and is bounded by its value
at $\varepsilon=1$: $C_{o, \varepsilon}\leq1/2$.
We thus get the first half of
(\ref{e:L2bd.1}).

Differentiating (\ref{e:Ltwobd.51}), we get
%
\begin{eqnarray}\qquad\quad
\label{e:Ltbd.57}
&&i \partial L_{o, \varepsilon}(p_1)\nonumber\\[-8pt]\\[-8pt]
&&\qquad = - C_{o,\varepsilon}
(1-\varepsilon)
\Biggl[
p_1^{\varepsilon-2} + \sum_{n=1}^{\infty}
\{
(2\pi n - p_1)^{\varepsilon-2} + (2\pi n + p_1)^{\varepsilon-2}
\}
\Biggr]
.\nonumber
\end{eqnarray}
This is even in $p_1$ and is obviously negative. To get its lower
bound, we bound the summation by its value at $p_1=\pi$ (because
the summand is increasing in $|p_1|$), to get
\begin{eqnarray}
\label{e:Ltbd.59}
&&\sum_{n=1}^{\infty}
\{
(2\pi n - p_1)^{\varepsilon-2} + (2\pi n + p_1)^{\varepsilon-2}
\}\nonumber\\
&&\qquad \leq
\sum_{n=1}^{\infty}
\{
(2\pi n - \pi)^{\varepsilon-2} + (2\pi n + \pi)^{\varepsilon-2}
\}
\\
&&\qquad
= \pi^{\varepsilon-2} \Biggl[
1 + 2 \sum_{n=1}^\infty(2n+1)^{\varepsilon-2}
\Biggr].
\nonumber
\end{eqnarray}
Because $(2x+1)^{\varepsilon-2}$ is convex, we can bound the sum as
%
\begin{equation}
\label{e:Ltbd.61}
\sum_{n=1}^\infty(2n+1)^{\varepsilon-2}
\leq\int_{1/2}^{\infty} (2x +1)^{\varepsilon-2} \,dx
= \frac{2^{\varepsilon-1}}{2(1-\varepsilon)}
= \frac{2^{\varepsilon-2}}{1-\varepsilon}
.
\end{equation}
As a result, we get
%
\begin{eqnarray}
\label{e:Ltbd.61a}
- i \partial L_{o, \varepsilon}(p_1)
&\leq&
C_{o,\varepsilon} (1-\varepsilon) p_1^{\varepsilon-2}
+ C_{o,\varepsilon} \pi^{\varepsilon-2} (
1-\varepsilon+ 2^{\varepsilon-1} )\nonumber\\[-8pt]\\[-8pt]
&\leq&\frac{1-\varepsilon}{2} p_1^{\varepsilon-2}
+ \frac{1}{2\pi} .\nonumber
\end{eqnarray}
In the last step above, we used $C_{o, \varepsilon}\leq1/2$ and
the fact that $\pi^{\varepsilon-2}(1-\varepsilon+ 2^{\varepsilon-1})$
is increasing in $\varepsilon$ and is bounded by its value
at $\varepsilon=1$, that is, by $1/\pi$.

As for $L_{e, \varepsilon}(p_1)$, we start from an integral
representation of its derivative:
\begin{eqnarray}
\label{e:Lfbd.51}
\partial L_{e, \varepsilon}(p_1)
&=& - \frac{\sin p_1}{2\pi\Gamma(\varepsilon)}
\int_0^\infty dt \frac{t^{\varepsilon-1} \sinh t}
{(\cosh t - \cos p_1)^2}
\nonumber\\
&=& - \frac{\sin p_1}{2\pi\Gamma(\varepsilon)}
\lim_{\delta\rightarrow0}
\biggl[ \frac{\delta^{\varepsilon-1}}{\cosh\delta- \cos p_1}\\
&&\hspace*{67.9pt}{}+ (\varepsilon-1) \int_{\delta}^\infty dt
\frac{t^{\varepsilon-2}}{\cosh t - \cos p_1}
\biggr] ,\nonumber
\end{eqnarray}
where in the second step we integrated by parts. The first line
of (\ref{e:Lfbd.51}) shows that $\partial L_{e, \varepsilon}(p_1)$ is
negative. The integral on the second line
can be performed by residue calculus, and leads to the following
representation:
%
\begin{eqnarray}\qquad
\label{e:Lfbd.53}
&&\partial L_{e, \varepsilon}(p_1)
= - C_{e, \varepsilon}
\Biggl[
p_1^{\varepsilon-2} - \sum_{n=1}^\infty
\{
(2\pi n - p_1)^{\varepsilon-2}\nonumber \\[-8pt]\\[-8pt]
&&\hspace*{130.5pt}{}- (2\pi n + p_1)^{\varepsilon-2}
\}
\Biggr] ,\nonumber
\end{eqnarray}
with
$C_{e, \varepsilon}
= (1-\varepsilon) \sec(\frac{\pi}{2}\varepsilon)/\{2\Gamma
(\varepsilon)\}$.
The coefficient $C_{e, \varepsilon}$ is increasing in $\varepsilon$
for $0 < \varepsilon< 1$ and is bounded by its limiting value
at $\varepsilon= 1^{-}$:
$C_{e, \varepsilon} \leq1/\pi$.
Because the summand in (\ref{e:Lfbd.53}) is positive, we get
%
\begin{equation}
\label{e:Lfbd.55}
\partial L_{e, \varepsilon}(p_1)
\geq- C_{e, \varepsilon}
p_1^{\varepsilon-2}
\geq- \frac{1}{\pi} p_1^{\varepsilon-2} .
\end{equation}
Finally, we turn to $L_{e, \varepsilon}(p_1)$.
Because $\partial L_{e, \varepsilon}(p_1)$ is negative, we can get
a lower bound on $L_{e, \varepsilon}(p_1)$ as
%
\begin{eqnarray}
\label{e:Lfbd.57f}
L_{e, \varepsilon}(p_1) &\geq& L_{e, \varepsilon}(\pi)
= - \frac{1}{\pi\Gamma(\varepsilon)} \int_0^\infty
\frac{t^{\varepsilon-1}}{e^t+1}\, dt\nonumber\\[-8pt]\\[-8pt]
&=&
\frac{1}{\pi} \sum_{n=1}^\infty\frac{(-1)^n}{n^\varepsilon}
\geq-\frac{\log2}{\pi} ,\nonumber
\end{eqnarray}
where in the last step we bounded the sum by its value
at $\varepsilon= 1^{-}$.
To get an upper bound, we integrate
(\ref{e:Lfbd.53}) from $p_1$ to $\pi$, to get
\begin{eqnarray}\quad\quad\quad
\label{e:Lfbd.57}
&&L_{e, \varepsilon}(p_1)
=
\frac{C_{e, \varepsilon}}{1-\varepsilon}
\Biggl[
p_1^{\varepsilon-1} - \pi^{\varepsilon-1}\nonumber\\
&&\hspace*{83pt}{}+ \sum_{n=1}^\infty
\{
(2\pi n - p_1)^{\varepsilon-1} + (2\pi n + p_1)^{\varepsilon-1}
\\
&&\hspace*{105pt}{}
- (2\pi n - \pi)^{\varepsilon-1} - (2\pi n + \pi)^{\varepsilon-1}
\}
\Biggr]
+ L_{e, \varepsilon}(\pi)
.\nonumber
\end{eqnarray}
The sum in the above is negative because
$(2\pi n-p)^{\varepsilon-1}+(2\pi n + p)^{\varepsilon-1}$ is
increasing in $p$. Because $L_{e, \varepsilon}(\pi) \leq0$
as is seen in (\ref{e:Lfbd.57f}), we get an upper bound
%
\begin{equation}\qquad
L_{e, \varepsilon}(p_1)
\leq
\frac{C_{e, \varepsilon}}{1-\varepsilon}
(p_1^{\varepsilon-1}-\pi^{\varepsilon-1})
\leq
\frac{C_{e, \varepsilon}}{1-\varepsilon}
p_1^{\varepsilon-1}
\leq\frac{1}{\pi(1-\varepsilon)}
p_1^{\varepsilon-1} .
\end{equation}
\upqed\end{pf*}

\subsection{Bounds on the Fourier transform of
\textup{``}fractionally weighted\textup{''} two-point functions and related quantities}
\label{sb-bounds-on-frac-Ghat}

In this subsection, we make use of the representation obtained
in the previous subsection and
prove bounds on the Fourier transform
of $G_{j}^{(\alpha)}(x)$, Lemma \ref{lem-Ghatalphabd}, and
two bounds (\ref{e:d=gen-Pbd.13aa}) and (\ref{e:Q1-frac-bd.11}).
We start from the following simple lemma
concerning one-dimensional convolution, whose proof is postponed
to the end of this section.
%
\begin{lemma}
\label{lem-Ghatbd-convolution}
Fix $0 < \varepsilon< 1$ and $\rho> 1$, and suppose
%
\begin{equation}
\label{e:Ghatbd-conv.1}
| f(k_1, \vec{k}) |
\leq\frac{1}{|k|^{\rho}},
\qquad
| \partial_1 f(k_1, \vec{k}) |
\leq\frac{1}{|k|^{\rho+1}}
\end{equation}
and
%
\begin{equation}
\label{e:Ghatbd-conv.3}
| g(p_1) | \leq|p_1|^{\varepsilon-1},\qquad
| \partial_1 g(p_1) | \leq|p_1|^{\varepsilon-2}
.
\end{equation}
Then the $1$-dimensional convolution
%
\begin{equation}
\label{e:Ghatbd-conv.5}
(\partial_1 f*g)(k_1, \vec{k}) = \int_{-\pi}^\pi
g(p_1) \partial_1 f(k_1-p_1, \vec{k})\, dp_1
\end{equation}
obeys
%
\begin{eqnarray}
\label{e:Ghatbd-conv.9}
| (\partial_1 f*g) (k_1, \vec{k})
|
&\leq&
c \times
\left\{\matrix{
|k_1|^{\varepsilon-1} |\vec{k}|^{-\rho}
& (\mbox{for } |k_1| \leq|\vec{k}|)
\cr
|k_1|^{\varepsilon-2} |\vec{k}|^{1-\rho}
& (\mbox{for } |k_1| \geq|\vec{k}|)
}
\right\}\nonumber\\[-8pt]\\[-8pt]
&\approx&
\frac{c}{|k_1|^{1-\varepsilon} |\vec{k}|^{\rho-1} |k|}\nonumber
\end{eqnarray}
with a possibly $\varepsilon$-dependent constant $c$.
\end{lemma}

\begin{pf*}{Proof of Lemma \protect\ref{lem-Ghatalphabd}}
We consider only positive $k_1$; bounds on negative
$k_1$ follow by ${ {\mathbb Z}^d }$-symmetry.
We start from the following expression for the Fourier
transform suggested by (\ref{e:F1hat-alpha.rep1}):
%
\begin{equation}
\label{e:Ghat-n-epsilon.1}
{\hat{G}}_{1}^{(n - \varepsilon)} (k)
=
\int_{-\pi}^{\pi} L_{*, \varepsilon} (p_1) \{
(i \partial_1)^{n} {\hat{G}}(k_1 - p_1, \vec{k})
\}\,
d p_1
,
\end{equation}
where $L_{*, \varepsilon}= L_{o, \varepsilon}$ if $n$ is odd,
and $L_{*, \varepsilon} = L_{e, \varepsilon}$ if $n$ is even.
[We will soon check that the above ${\hat{G}}_{1}^{(n - \varepsilon)} (k)$
in fact satisfies (\ref{e:G1-ne.rep1}).]
Lemma \ref{lem-m-derivative} gives
%
\begin{equation}
\label{e:Ghat-n-epsilon.2}
| \partial_1^{n}
{\hat{G}}(k) |
\leq c |k|^{-2-n}
\qquad\mbox{for } n \leq M
,
\end{equation}
and Proposition \ref{prop-Lbd} gives
%
\begin{equation}
| L_{*, \varepsilon}(p) |
\leq c |p|^{\varepsilon-1},
\qquad
| \partial L_{*, \varepsilon}(p) |
\leq c |p|^{\varepsilon-2} .
\end{equation}
We combine these and estimate (\ref{e:Ghat-n-epsilon.1})
by the following lemma,
by setting $f(k) = \partial_1^{n-1} {\hat{G}}$ and
$\rho= n+1$. (We can apply the lemma, because $\rho>1$
thanks to $n \geq1$).
The result turns out to be (\ref{e:Ghat-n-epsilon.5}).

The proof is complete if we show that ${\hat{G}}_1^{(n-\varepsilon)}(k)$
of (\ref{e:Ghat-n-epsilon.1}) does satisfy (\ref{e:G1-ne.rep1})
for $1 \leq n \leq M \wedge(d-2)$.
For this, note that (\ref{e:Ghat-n-epsilon.2}) guarantees
the integrability of $\partial_1^n {\hat{G}}(k)$ in $k_1$ for fixed
$\vec{k} \neq\vec{0}$. Also, (\ref{e:Ghat-n-epsilon.5}) means
${\hat{G}}_1^{(n-\varepsilon)}(k)$ is
integrable in $k$, for $n$ under consideration.
Therefore Proposition \ref{prop-f-n-alpha} can be applied
and (\ref{e:F1-alpha.rep1}) holds for $G$,
which is nothing but (\ref{e:G1-ne.rep1}).
\end{pf*}

\begin{pf*}{Proof of (\protect\ref{e:d=gen-Pbd.13aa})}
As we did for $G_1^{(n-\varepsilon)}$, we start from
the following expression for the Fourier
transform suggested by (\ref{e:F1hat-alpha.rep1}):
%
\begin{equation}
\label{e:d=gen-Pbd.13}
\hat\psi_1^{(1-\varepsilon)} (k) = \int_{-\pi}^\pi
L_{o, \varepsilon}(k_1-p_1) \partial_1 \hat{\psi}_1(p_1, \vec{k})
\,dp_1
.
\end{equation}
By ${\hat{J}}_{p_c}(0)=1$, the estimate (\ref{e:tr-Pi-90}), and
the fact that ${\hat{J}}(k)$ is even in $k_1$, it is easily
seen that $\hat{\psi}_1(k)$ obeys the bound
%
\begin{equation}
\label{e:d=gen-Pbd.15}
| \hat{\psi}_1(k) | \leq c |k|^{-4} ,\qquad
|\partial_1 \hat{\psi}_1(k) | \leq c |k|^{-5}
.
\end{equation}
So Lemma \ref{lem-Ghatbd-convolution} implies the desired bound,
%
\begin{equation}
\label{e:d=gen-Pbd.17}
| \hat{\psi}_1^{(1-\varepsilon)}(k) |
\leq\frac{c}{|k_1|^{1-\varepsilon} |\vec{k}|^3 |k|}
.
\end{equation}
The proof is complete if we show that the inverse Fourier
transform of $\hat\psi_1^{(1-\varepsilon)}(k)$ is equal to
$|x_1|^{1-\varepsilon} \psi_1(x)$, but this can be done in
exactly the same way as for ${\hat{G}}^{(n-\varepsilon)}$.
\end{pf*}

\begin{pf*}{Proof of (\protect\ref{e:Q1-frac-bd.11})}
The proof for $|a_1|^{1-\varepsilon} Q(a) = Q_1^{(1-\varepsilon)}(a)$
is similar. We start from
%
\begin{equation}
\label{e:Ghat-expression.15}
{\hat{Q}}_1^{(1-\varepsilon)} (k) = \int_{-\pi}^\pi
L_{o, \varepsilon}(k_1-p_1) \partial_1 Q(p_1, \vec{k})
\,dp_1 .
\end{equation}
We first note that the total
number of differentiations appearing in each term of
${\hat{Q}}(k)$ is $n-1$, and that of $\partial_1{\hat{Q}}(k)$ is $n$.
Second, we note that all the derivatives appearing in the expression
of $\partial_1 {\hat{Q}}(k)$
are finite; this is because the highest order of differentiation in
${\hat{Q}}(k)$ is $n-2$, and thus the highest order of differentiation
in $\partial_1 {\hat{Q}}(k)$ is $n-1 \leq\lfloor\phi\rfloor$.
So the expression for $\partial_1 {\hat{Q}}(k)$ now reads
[cf. (\ref{e:tau.8'})]:
\begin{equation}
\label{e:tau.8''}
\partial_1 {\hat{Q}}(k)
 = \sum_{p=1}^{n} \pmatrix{n\cr p}
\bigl( \partial_1^{n-p} {\hat{g}}(k) \bigr)
\sum_{q=2}^{p+1} \sum_{\vec{r}}
C_{\vec{r}}
\frac{\prod_{\ell\geq1}
[ \partial_1^{\ell} {\hat{J}}(k) ]^{r_{\ell}}}
{\{1 - {\hat{J}}(k)\}^{q}} ,
\end{equation}
but $C_{\vec{r}}=0$ if $p=n$ and $q=2$ (i.e., only $\ell< n$ is
allowed in the numerator).
Arguing as in the proof of Lemma \ref{lem-m-derivative}
and counting powers of $|k|$ of each term, we get
%
\begin{equation}
| \hat{Q}(k) | \leq c |k|^{-(1+n)},
\qquad
| \partial_1 \hat{Q}(k) | \leq c |k|^{-(2+n)}.
\end{equation}
Therefore, estimating (\ref{e:Ghat-expression.15})
using Lemma \ref{lem-Ghatbd-convolution} leads to
%
\begin{equation}
\bigl| {\hat{Q}}_1^{(1-\varepsilon)} (k) \bigr|
\leq\frac{c}{|k_1|^{1-\varepsilon} |\vec{k}|^{n} |k|}
.
\end{equation}
Finally, we can show that the inverse Fourier transform of
the above ${\hat{Q}}_1^{(1-\varepsilon)}(k)$ is in fact
$|x_1|^{1-\varepsilon}Q(x)$, just as we did for
${\hat{G}}_1^{(n-\varepsilon)}(k)$.
\end{pf*}

\begin{pf*}{Proof of Lemma \protect\ref{lem-Ghatbd-convolution}}
We first rewrite (\ref{e:Ghatbd-conv.5}), dividing the
integration region and integrating by parts as follows.
To simplify notation, we write $a := k_1$ and
$b := |\vec{k}|$:
\begin{eqnarray}
\label{e:Ghatbd-conv.11}
(\partial_1 f*g)(k)
&=&
\int_{|p_1|< a/2}
g (a-p_1) \partial_1 f(p_1, \vec{k})\, d p_1\nonumber\\
&&{}+
\int_{a/2< |p_1| <\pi} g (a-p_1)
\partial_1 f(p_1, \vec{k}) \,d p_1
\nonumber\\
&=&
[ g(a-p_1) f(p_1, \vec{k})
]_{-a/2}^{a/2}\\
&&{}+
\int_{|p_1|< a/2} \partial_1 g (a-p_1)
f(p_1, \vec{k})\, d p_1
\nonumber\\
&&{}+
\int_{a/2< |p_1| <\pi} g (a-p_1)
\partial_1 f(p_1, \vec{k})\, d p_1
.\nonumber
\end{eqnarray}
The integration by parts was done only for the interval
$[-a/2, a/2]$---this is justified because there is no
singularity of $g(p_1)$ in this interval.
We estimate these terms one by one.

The first and second terms are simple. For the
first term, we have
\begin{eqnarray}\qquad
\label{e:Ghat-n-epsilon.11}
| [
g (a-p_1) f(p_1, \vec{k})
]_{-a/2}^{a/2} |
& \leq&
\biggl| g\biggl(\frac{a}{2}\biggr) f\biggl(\frac{a}{2}, \vec{k}\biggr)
\biggr|
+ \biggl|
g\biggl(\frac{3a}{2}\biggr) f\biggl(-\frac{a}{2}, \vec{k}\biggr) \biggr|
\nonumber\\[-8pt]\\[-8pt]
&
\leq &c a^{\varepsilon-1} \times(a^2 + b^2)^{-\rho/2} .
\nonumber
\end{eqnarray}
For the second term, we have
\begin{eqnarray}
\label{e:Ghat-n-epsilon.13}
&&\biggl|
\int_{|p_1|< a/2} \partial g(a-p_1)
f(p_1, \vec{k}) \,d p_1 \biggr|\nonumber\\
&&\qquad\leq c
\int_{|p_1|< a/2} |a-p_1|^{\varepsilon-2}
(p_1^2 + b^2)^{-\rho/2} \,dp_1
\nonumber\\[-8pt]\\[-8pt]
&&\qquad\leq
c a^{\varepsilon-2} \int_{-a/2}^{a/2}
(p_1^2 + b^2)^{-\rho/2} \,dp_1\nonumber\\
&&\qquad\leq c a^{\varepsilon-2} \times(a \wedge b) b^{-\rho}
= c
a^{\varepsilon-1} b^{1-\rho} (a\vee b)^{-1} .\nonumber
\end{eqnarray}

The third term is bounded and divided as
\begin{eqnarray}
\label{e:Ghat-n-epsilon.15}
&&\biggl|
\int_{a/2 < |p_1|< \pi} g (a-p_1)
\partial_1 f(p_1, \vec{k})\, d p_1 \biggr|\nonumber\\
&&\qquad\leq c
\int_{a/2 < |p_1|< \pi} |a-p_1|^{\varepsilon-1}
(p_1^2 + b^2)^{-(\rho+1)/2} \,dp_1
\nonumber\\[-8pt]\\[-8pt]
&&\qquad= \int_{-\pi}^{-a/2} (\cdots)\, dp_1
+ \int_{a/2}^{3a/2} (\cdots) \,dp_1
+ \int_{3a/2}^{\pi} (\cdots)\, dp_1\nonumber\\
:\!\!\!\!\!\!\hspace*{-18.09pt}&&\qquad=
(I) + (\mathit{II}) + (\mathit{III})
.\nonumber
\end{eqnarray}
In (I) and (III), we have $|a-p_1|\geq|p_1|/3$ and
$|a-p_1|^{\varepsilon-1} \leq3^{1-\varepsilon} p_1^{\varepsilon-1}$.
Therefore, we can bound them as
%
\begin{equation}
\label{e:Ghat-n-epsilon.21}
(I) + (\mathit{III}) \leq c
\int_{a/2}^\infty
(p_1^2 + b^2)^{-(\rho+1)/2} p_1^{\varepsilon-1}\, dp_1
.
\end{equation}
This integral can be bounded in two ways. First, by neglecting
$b^2$ in the integrand,
%
\begin{eqnarray}
\label{e:Ghat-n-epsilon.23}
(I)+(\mathit{III}) &\leq& c \int_{a/2}^\infty
(p_1^2 + b^2)^{-(\rho+1)/2} p_1^{\varepsilon-1} \,dp_1\nonumber\\[-8pt]\\[-8pt]
&\leq&\int_{a/2}^\infty
p_1^{\varepsilon-1-\rho-1} \,dp_1
= c a^{\varepsilon-\rho-1}
.\nonumber
\end{eqnarray}
Also, extending the integration region to $p_1 \geq0$ and
changing the variable to $q_1 = p_1/b$,
\begin{eqnarray}
\label{e:Ghat-n-epsilon.25}
(I)+(\mathit{III})
& \leq& c \int_{a/2}^\infty
(p_1^2 + b^2)^{-(\rho+1)/2} p_1^{\varepsilon-1}\, dp_1
\nonumber\\[-8pt]\\[-8pt]
& \leq& b^{\varepsilon-\rho-1} \int_0^\infty
(1+q_1^2)^{-(\rho+1)/2} q_1^{\varepsilon-1}\, dq_1
.
\nonumber
\end{eqnarray}
For $0 < \varepsilon< \rho+1$, the last integral is finite. We have\vadjust{\goodbreak}
thus shown $(I)+(\mathit{III}) \leq c (a\vee b)^{\varepsilon-\rho-1}$.
Now in (II), $p_1^2 + b^2$ is of the same order as $a^2+b^2$.
Thus
%
\begin{eqnarray}
\label{e:Ghat-n-epsilon.27}
(\mathit{II}) &\leq& c (a^2+b^2)^{-(\rho+1)/2}
\int_{a/2}^{3a/2} |a-p_1|^{\varepsilon-1} \,dp_1\nonumber\\[-8pt]\\[-8pt]
&=& c (a^2+b^2)^{-(\rho+1)/2} a^{\varepsilon}
.\nonumber
\end{eqnarray}

We can thus conclude
%
\begin{eqnarray}
\label{e:Ghat-n-epsilon.29}
(I) + (\mathit{II}) + (\mathit{III}) &\leq& c (a \vee b)^{\varepsilon-\rho-1}
+ c (a^2+b^2)^{-(\rho+1)/2} a^{\varepsilon}\nonumber\\[-8pt]\\[-8pt]
&\leq& c (a \vee b)^{\varepsilon-\rho-1}.\nonumber
\end{eqnarray}
Combining (\ref{e:Ghat-n-epsilon.11}), (\ref{e:Ghat-n-epsilon.13})
and (\ref{e:Ghat-n-epsilon.29}), we get
\begin{eqnarray}
\label{e:Ghat-n-epsilon.29a}
&&| (\partial_1 f*g) (k_1, \vec{k})
|\nonumber\\
&&\qquad
\leq c [ a^{\varepsilon-1}(a\vee b)^{-\rho}
+ a^{\varepsilon-1} b^{1-\rho} (a\vee b)^{-1}
+ (a \vee b)^{\varepsilon-\rho-1} ]
\\
&&\qquad
\leq c
\cases{
a^{\varepsilon-1} b^{-\rho} &\quad $(a < b)$, \cr
a^{\varepsilon-2} b^{1-\rho} &\quad $( a \geq b)$,
}
\nonumber
\end{eqnarray}
which proves the lemma [note that
$(a\vee b) \leq|k| \leq2 (a\vee b)$].
\end{pf*}

\begin{appendix}
\section{Quantities at $p=p_c$}
\label{sb-ap-Jcont}

Proposition \ref{thm-TW-90a} is a slightly
improved version of corresponding results obtained
in previous works: \cite{HS92b,HS92a} (SAW),
\cite{HS90a} (percolation), and \cite{HS90b} (LTLA).
It is slightly improved, in the sense
that original works mainly dealt with quantities for
$p < p_c$ (although all the estimates were uniform in $p$).
More precisely, estimates (\ref{e:tr-Pi-90}) and
(\ref{e:tr-GB-90})--(\ref{e:tr-LTLA-90aNN}),
the Fourier representation (\ref{e:Gsawx.1}),
and the bound (\ref{e:tr-Ghat-90}) are proved for $p < p_c$;
the critical point $p_c$ is characterized by
%
\begin{equation}
\label{e:crit-char.1ap}
\lim_{p \uparrow p_c} \hat{J}_{p}(0) = 1
\end{equation}
instead of (\ref{e:crit-char.1}).
We in this Appendix show how to extend these results to $p= p_c$,
so that we have Proposition \ref{thm-TW-90a}.

1. We first explain how to extend estimates
(\ref{e:tr-Pi-90}) and
(\ref{e:tr-GB-90})--(\ref{e:tr-LTLA-90aNN}) to $p= p_c$.
Note that $G_p(x)$ is left continuous and increasing in $p$;
this is because $G_p(x)$
can be realized as an increasing limit (finite sum$/$volume
approximation) of a function which is continuous and increasing
in $p$.
[In fact $G_p(x)$ for percolation is continuous in $p$ for all $p$
(\cite{Grim99}, page 203), although we do not need this fact.]
The left continuity of $G_p(x)$ in $p$ and the dominated
convergence theorem establish
(\ref{e:tr-GB-90})--(\ref{e:tr-LTLA-90aNN}) at $p=p_c$.
Diagrammatic bounds of the lace expansion and the dominated
convergence theorem now guarantee absolute convergence
of the sums over $x$ and $n$ defining ${\hat{\Pi}}_p(k)$ at
$p= p_c$. [We are not using continuity of ${\hat{\Pi}}_p(k)$ here;
we just bound each term of ${\hat{\Pi}}_{p_c}(k)$ in terms of
quantities appearing in
(\ref{e:tr-GB-90})--(\ref{e:tr-LTLA-90aNN}) at $p=p_c$.]
Therefore (\ref{e:tr-Pi-90}) holds even at $p=p_c$.

2.
Moreover, (\ref{e:tr-Pi-90aa}) is now proved, where $h^{(n)}(x)$
is obtained by bounding diagrams for $\Pi_{p_c}^{(n)}(x)$ in
terms of (products of) critical two-point function $G_{p_c}$.
Finally, both equations of (\ref{e:Fourier-def.1})
hold even at $p=p_c$ for
$f(x) = \Pi_{p_c}(x), g_{p_c}(x), J_{p_c}(x)$.

3. We will, later in this Appendix,
show that ${\hat{\Pi}}_p(k)$ and ${\hat{g}}_p(k)$
are left continuous at $p = p_c$. Equation
(\ref{e:crit-char.1}) now follows from (\ref{e:crit-char.1ap}) by
the left-continuity.

4. Finally, we deal with (\ref{e:Gsawx.1}) and
(\ref{e:tr-Ghat-90}) at $p=p_c$. This is rather subtle,
because $G_{p_c}(x)$ is not summable as Theorem \ref{thm-saw}
implies. To make sense of (\ref{e:Gsawx.1}),
let $p \uparrow p_c$ on both sides of (\ref{e:Gsawx.1}).
By the left continuity of $G_p(x)$ in $p$,
the left-hand side of (\ref{e:Gsawx.1}) goes to $G_{p_c}(x)$.
On the right-hand side,
the integrand is integrable in $k$ uniformly in
$p < p_c$, thanks to the infrared bound (\ref{e:tr-Ghat-90})
(recall that we are considering $d >2$).

Therefore, by the dominated convergence theorem and the
left-continuity of ${\hat{\Pi}}_p(k)$ and ${\hat{g}}_p(k)$ stated above,
\begin{eqnarray}
\label{e:Gpc.1}
G_{p_c}(x)
&=& \lim_{p \uparrow p_c} G_p(x)
= \lim_{p \uparrow p_c}
\int_{[-\pi,\pi]^{d}}\frac{d^d k}{(2\pi)^d}e^{ikx} \frac{{\hat
{g}}_p(k)}{1 - {\hat{J}}_p(k)}
\nonumber\\
&=&
\int_{[-\pi,\pi]^{d}}\frac{d^d k}{(2\pi)^d}e^{ikx}
\biggl( \lim_{p \uparrow p_c}
\frac{{\hat{g}}_p(k)}{1 - {\hat{J}}_p(k)}
\biggr)\\
&=& \int_{[-\pi,\pi]^{d}}\frac{d^d k}{(2\pi)^d}e^{ikx}
\frac{{\hat{g}}_{p_c}(k)}{1 - {\hat{J}}_{p_c}(k)}
.\nonumber
\end{eqnarray}
Thus, we still have the Fourier representation (\ref{e:Gsawx.1})
and the infrared bound (\ref{e:tr-Ghat-90}) at $p=p_c$,
with the understanding that ${\hat{G}}_{p_c}(k)$ is defined
by the integrand of the right-hand side of (\ref{e:Gpc.1}).

5.
Our remaining task is to prove the left-continuity of
$\hat{J}_p(k)$ and $\hat{g}_p(k)$ at \mbox{$p=p_c$.}
The proof is based on the following lemma:

\begin{lemma}
\label{lem-Picont.ap.2}
Consider SAW, percolation, or LTLA for which
Proposition \textup{\ref{thm-TW-90a}} holds. $\Pi_p^{(n)}(x)$ is
continuous in $p$ for $p < p_c$, and is
left-continuous at $p = p_c$.
\end{lemma}

\begin{pf*}{Proof that $\hat{\Pi}_p(k)$
is continuous in $p$ for $p \leq p_c$,
assuming Lemma \protect\ref{lem-Picont.ap.2}}
Equation
(\ref{e:tr-Pi-90aa}) implies the double sum
$\sum_{x} \sum_{n}(-1)^n \Pi_p^{(n)}(x) e^{-ikx}$
is absolutely and uniformly convergent
for $p \leq p_c$, and Lemma \ref{lem-Picont.ap.2} claims
the summand is continuous in $p$ for $p < p_c$ and is
left-continuous at $p=p_c$.
Because the uniform convergent limit of a continuous
function is continuous, $\hat{\Pi}_p(k)$ is
continuous in $p$ for $p < p_c$, and is
left-continuous at $p = p_c$.
$\hat{J}_p(k)$ and $\hat{g}_p(k)$ are now left-continuous
at $p=p_c$, as is easily seen from their definition.
\end{pf*}

In the rest of this Appendix, we prove
Lemma \ref{lem-Picont.ap.2} above.\vadjust{\goodbreak}

\subsection{\texorpdfstring{Proof of Lemma \protect\ref{lem-Picont.ap.2}
for SAW and LTLA}{Proof of Lemma A.1 for SAW and LTLA}}
\label{sb-ap-Jcont-saw-LTLA}

For self-avoiding walk and lattice trees$/$animals,
Lemma \ref{lem-Picont.ap.2} follows rather easily.
$\Pi_p^{(n)}(x)$ of these models are, by definition,
power series in $p$ with positive coefficients:
$\Pi_p^{(n)}(x) = \sum_{m=0}^\infty a_m(n,x) p^m$. Equation
(\ref{e:tr-Pi-90aa}) guarantees that the radius of
convergence of this series is at least $p_c$, and that
the series converges absolutely at $p=p_c$.
The finite sum $f_M(p) := \sum_{m=0}^M a_m(n,x) p^m$ is of course
continuous in $p$.
Therefore, $\Pi_p^{(n)}(x)$, being a uniformly convergent limit of
a continuous function $f_M(p)$, is thus
continuous in $p$ for $p < p_c$ and is left-continuous at $p=p_c$.

\subsection{\texorpdfstring{Proof of Lemma \protect\ref{lem-Picont.ap.2} for
percolation}{Proof of Lemma A.1 for percolation}}
\label{sb-ap-Jcont-percolation}

Proving Lemma \ref{lem-Picont.ap.2} for percolation
is more subtle, because
finite volume approximation to $\Pi_p^{(n)}(x)$ does not
seem to be either increasing or decreasing in $p$.
To overcome this difficulty, we decompose $\Pi_p(x)$ further,
and express it in terms of increasing/decreasing
events.

\textit{Step} 1.
We begin by recalling the definitions of $\Pi_p^{(n)}(x)$.
Because the following description is very brief,
the reader is advised to consult \cite{HS90a,HS94,Slad04} for details.
\begin{itemize}
\item
Given a set of sites $A \in{\mathbb Z}^d$ and a bond configuration,
two sites $x$ and $y$ are \textit{connected in $A$} if there is
a path of occupied bonds from $x$ to $y$
having all of its sites in $A$, or if $x=y \in A$.
The set of all sites which are connected to $x$ is denoted by $C(x)$.
[This $C(x)$ has nothing to do with $C(x)$ of Theorem \ref{th-gau}.]
\item
Given a set of sites $A \in{\mathbb Z}^d$ and a bond configuration,
two sites $x$ and $y$ are \textit{connected through $A$} if they are
connected but they are \textit{not} connected in
${\mathbb Z}^d\backslash A$.
\item
Given a bond configuration and a
bond $\{u, v\}$, we define $\tilde{C}^{\{u,v\}}(x)$ to be
the set of sites which remain connected to $x$ in the
new configuration obtained by setting $\{u, v\}$ to be
vacant.
\end{itemize}
For $x, y \in{\mathbb Z}^d$ and $A \subset{\mathbb Z}^d$,
let $E_0(x, y)$ be the event that $x$ and $y$
are doubly connected, and let $E_2(x, y; A)$ be the
event that $x$ is connected to $y$ through $A$ and
there is no pivotal bond for the connection
from $x$ to $y$ whose first endpoint is connected to
$x$ through $A$.
We now define
%
\begin{equation}
\Pi_p^{(0)}(x) := {\mathbb E}[I[E_0(0, x)]]
\end{equation}
and for $n \geq1$
%
\begin{equation}
\label{e:Pidef-decmp.1}
\Pi_p^{(n)}(x) := \sum_{(y_1, y_1')} p
\sum_{(y_2, y_2')} p \cdots\sum_{(y_n, y_n')}
p
\Pi_p^{(n)}(x; y_1, y_1', y_2, y_2', \dots, y_n, y_n')
.\hspace*{-30pt}
\end{equation}
Here the sums are over all directed pairs of nearest neighbor
sites, and
%
\begin{eqnarray}
&&\Pi_p^{(n)}(x; y_1, y_1', \dots, y_n, y_n')\nonumber\\[-8pt]\\[-8pt]
&&\qquad:=
{\mathbb E}_0\bigl[I_0 {\mathbb E}_1\bigl[I_1
{\mathbb E}_2\bigl[I_2 \cdots{\mathbb E}_{n-1}\bigl[I_{n-1}
{\mathbb E}_n[ I_n] \bigr] \cdots\bigr] \bigr] \bigr],\nonumber
\end{eqnarray}
where we abbreviated $I_0 = I[E_0(0, y_1)]$,
$I_j = I[E(y_{j}', y_{j+1}; \tilde{C}_{j-1})]$
with $\tilde{C}_{j-1} = \tilde{C}^{\{y_j, y_j'\}}(y_{j-1}')$
and $y_{n+1}= x$, and ${\mathbb E}_j$'s represent nested
expectations.

\textit{Step} 2.
We in the following prove that $\Pi_p^{(0)}(x)$ and
$\Pi_p^{(n)}(x; y_1, y_1', \dots, y_n, y_n')$
are continuous in $p$ for $p < p_c$ and are left continuous
at $p=p_c$. Once this is done,
continuity of $\Pi_p^{(n)}(x)$ follows easily as follows.
In the proof of Proposition \ref{thm-TW-90a}, one proves
that there is a function
$h^{(n)}(x; y_1, y_1', \dots, y_n, y_n')$ which
satisfies
%
\begin{equation}\qquad
\Pi_p^{(n)}(x; y_1, y_1', \dots, y_n, y_n')
\leq h^{(n)}(x; y_1, y_1', \dots, y_n, y_n')\qquad
(p \leq p_c),
\end{equation}
and
%
\begin{equation}
\sum_{(y_1, y_1')} p
\sum_{(y_2, y_2')} p \dots\sum_{(y_n, y_n')}
p
h^{(n)}(x; y_1, y_1', \dots, y_n, y_n')
< \infty.
\end{equation}
This bound guaranties that the sum over $(y_j, y_j')$
($j = 1, 2, \dots, n$) in (\ref{e:Pidef-decmp.1})
converges uniformly and absolutely
for $p \leq p_c$. This implies the limit $\Pi_p^{(n)}(x)$
is continuous for $p < p_c$ and is left continuous at $p=p_c$.

Our task has thus been reduced to proving continuity of
$\Pi_p^{(0)}(x)$ and $\Pi_p^{(n)}(x; y_1, y_1', \dots, y_n, y_n')$.
We consider
$\Pi_p^{(n)}(x; y_1, y_1', \dots, y_n, y_n')$
only, because $\Pi_p^{(0)}(x)$ is (much) easier.

\textit{Step} 3.
To prove continuity
of $\Pi_p^{(n)}(x; y_1, y_1', \dots, y_n, y_n')$,
we express it in terms of increasing$/$decreasing events.
For $A \subset{\mathbb Z}^d$, we define:
\begin{itemize}
\item
$F_1(x, y;A)$: the event that $x$ and $y$ are connected
in ${\mathbb Z}^d \backslash A$.
\item
$F_2(x, y; A)$: the event that $x$ and $y$ are connected, and
$x$ and $z$ are connected in ${\mathbb Z}^d \backslash A$,
where $z$ is the first endpoint of the last pivotal bond
for the connection from $x$ to $y$. (When there is no
pivotal bond for the connection from $x$ to $y$, $F_2$ is
simply the event that $x$ and $y$ are connected.)
\end{itemize}
Note that $F_1(x, y;A) \subset F_2(x, y;A)$ and
$E_2(x, y; A) = F_2(x, y;A) \backslash F_1(x, y;A)$.
So, abbreviating
$I_j^{(\varepsilon)} = I[F_{\varepsilon}(y_{j}', y_{j+1};
\tilde{C}_{j-1})]$ (for $\varepsilon= 1, 2$), we have
$I_j = I_j^{(2)} - I_j^{(1)}$.
Using this, we can decompose as
%
\begin{equation}
\Pi_p^{(n)}(x; y_1, y_1', \dots, y_n, y_n')
= \sum_{\vec{\varepsilon}} (-1)^{\sum_{j=1}^n\varepsilon_j}
\Pi_p^{(n, \vec{\varepsilon})}
(x; y_1, y_1', \dots, y_n, y_n'),\hspace*{-30pt}
\end{equation}
where $\vec{\varepsilon}$ stands for $(\varepsilon_1, \varepsilon_2,
\ldots, \varepsilon_n)$, the sum over
$\vec{\varepsilon}$ runs over all choices of $\varepsilon_j=1, 2$,
and
\begin{eqnarray}
&&\Pi_p^{(n, \vec{\varepsilon})}
(x; y_1, y_1', \dots, y_n, y_n')
\nonumber\\[-8pt]\\[-8pt]
&&\qquad:=
{\mathbb E}_0\bigl[I_0 {\mathbb E}_1\bigl[I_1^{(\varepsilon_1)}
{\mathbb E}_2\bigl[I_2^{(\varepsilon_2)} \cdots
{\mathbb E}_{n-1}\bigl[I_{n-1}^{(\varepsilon_{n-1})}
{\mathbb E}_n\bigl[ I_n^{(\varepsilon_n)}\bigr] \bigr] \cdots\bigr] \bigr] \bigr]
.\nonumber
\end{eqnarray}
We in the following prove that
$\Pi_p^{(n, \vec{\varepsilon})}$
is continuous in $p$ for every choice of $\varepsilon_j=1,2$
($1 \leq j \leq n$).
Continuity of $\Pi_p^{(n)}$ in $p$ immediately follows from this.

\textit{Step} 4.
Continuity of $\Pi_p^{(n, \vec{\varepsilon})}$ is proved by considering
its finite volume approximations which are increasing$/$decreasing
in the volume. Let $\Lambda$ be a finite set of
sites in ${\mathbb Z}^d$, and write $\partial\Lambda$ for
its boundary sites. We define for $A \subset\Lambda$ and for
$\varepsilon= 1, 2$:
\begin{itemize}
\item
$\tilde{E}_{0, \Lambda}(x,y) := E_0(x, y) \cap
\{ C(x) \cap\partial\Lambda= \varnothing\}$,
\item
$\tilde{\tilde{E}}_{0, \Lambda}(x,y) := E_0(x, y) \cup
\{ C(x) \cap\partial\Lambda\neq\varnothing\}$,
\item
$\tilde{F}_{\varepsilon, \Lambda} (x, y; A) :=
F_{\varepsilon}(x, y ; A) \cap
\{ C(x) \cap\partial\Lambda= \varnothing\}$,
\item
$\tilde{\tilde{F}}_{\varepsilon, \Lambda} (x, y; A) :=
F_{\varepsilon}(x, y ; A) \cup
\{ C(x) \cap\partial\Lambda\neq\varnothing\}$.
\end{itemize}
Now define $\tilde{\Pi}_{p, \Lambda}^{(n, \vec{\varepsilon})}$
by replacing $E_0$ and $F_{\varepsilon}$
by $\tilde{E}_{0, \Lambda}$
and $\tilde{F}_{\varepsilon, \Lambda}$
in the definition of $\Pi_p^{(n, \vec{\varepsilon})}$;
also define
$\tilde{\tilde{\Pi}}{}_{p, \Lambda}^{(n, \vec{\varepsilon})}$ by replacing
$E_0$ and $F_{\varepsilon}$ by $\tilde{\tilde{E}}_{0, \Lambda}$ and
$\tilde{\tilde{F}}_{\varepsilon, \Lambda}$.

$\tilde{\Pi}_{p, \Lambda}^{(n, \vec{\varepsilon})}$ and
$\tilde{\tilde{\Pi}}{}_{p, \Lambda}^{(n, \vec{\varepsilon})}$
converge to $\Pi_p^{(n, \vec{\varepsilon})}$ in the infinite
volume limit as long as the percolation density is zero
(which has been proven to be the case for $p \leq p_c$
in high dimensions).
Thanks to their definition,
$\tilde{\Pi}{}_{p, \Lambda}^{(n, \vec{\varepsilon})}$ is
increasing in the volume, while
$\tilde{\tilde{\Pi}}{}_{p, \Lambda}^{(n, \vec{\varepsilon})}$ is
decreasing in the volume.
Moreover, these functions are continuous in $p$ for $p \leq p_c$;
this is because events with tildes and double tildes are
essentially finite volume events.

The (increasing) infinite volume limit of a continuous function
$\tilde{\Pi}_{p, \Lambda}^{(n, \vec{\varepsilon})}$
is lower semicontinuous in $p$ for $p \leq p_c$.
The (decreasing) infinite volume limit of a continuous function
$\tilde{\tilde{\Pi}}{}_{p, \Lambda}^{(n, \vec{\varepsilon})}$
is upper semicontinuous in $p$ for $p \leq p_c$.
Their common limit, $\Pi_p^{(n, \vec{\varepsilon})}$, is
thus continuous in $p$ for $p < p_c$
and is left-continuous at $p=p_c$.

\section{Basic properties of convolution}
\label{sb-ap-conv}
\begin{lemma}
\label{lem-conv}
\textup{(i)} Let $f, g$ be functions on ${ {\mathbb Z}^d }$ which satisfy
%
\begin{equation}
\label{e:fg.1}
| f(x) | \leq\frac{1}{|\!|\!|x |\!|\!|^{\alpha}},
\qquad
| g(x) | \leq\frac{1}{|\!|\!|x |\!|\!|^{\beta}},
\end{equation}
with $\alpha, \beta>0$. Then there exists a constant $C$
depending on $\alpha, \beta, d$ such that
%
\begin{equation}
\label{e:fg.2}
| (f*g)(x) | \leq
\cases{
C |\!|\!|x |\!|\!|^{-(\alpha\land\beta)}, &
\quad($\alpha> d$ or  $\beta> d$), \cr
C |\!|\!|x |\!|\!|^{-(\alpha+ \beta-d)}, &
\quad($\alpha, \beta< d$ and
$\alpha+ \beta> d$).
}
\end{equation}

\textup{(ii)} Let $d > 2$ and let $f, g$ be ${ {\mathbb Z}^d }$-symmetric
functions on ${ {\mathbb Z}^d }$, which satisfy
%
\begin{equation}
f(x) = \frac{A}{|\!|\!|x |\!|\!|^{d-2}} + O \biggl(
\frac{B}{|\!|\!|x |\!|\!|^{d-2+\rho}} \biggr) ,\qquad
| g(x) | \leq\frac{C}{|\!|\!|x |\!|\!|^{d+\rho}}
\end{equation}
with positive $A, B, C$ and $0 < \rho< 2$. Then
%
\begin{equation}
(f*g)(x) = \frac{A \sum_y g(y)}{|\!|\!|x |\!|\!|^{d-2}} + O \biggl(
\frac{C(A+B)}{|\!|\!|x |\!|\!|^{d-2+\rho}} \biggr) ,
\end{equation}
where the constants in the error term depend on $d$ and $\rho$.

\textup{(iii)}
Let $f, g$ be functions on ${ {\mathbb Z}^d }$ which satisfy
%
\begin{equation}
\label{e:GH1a}
|f(x)| \leq\frac{1}{|\!|\!|x|\!|\!|^{\alpha}},\qquad
\sum_{x} | g(x) | = K ,\qquad
| g(x) |\leq\frac{K}{|x|^{d}}
\end{equation}
with positive $K$ and $0 < \alpha< d$.
Then, there exists a constant $C$ depending on $\alpha, d$
such that
%
\begin{equation}
\label{e:GH4a}
|(f*g) (x)| \leq\frac{C K}{|x|^{\alpha}} .
\end{equation}

\textup{(iv)}
Let $f, g$ be functions on ${ {\mathbb Z}^d }$ which satisfy
%
\begin{eqnarray}
\label{e:GH1}
f(x) &\sim&\frac{A}{|x|^{\alpha}}, \qquad(|x|
\uparrow\infty),\nonumber\\[-8pt]\\[-8pt]
\sum_{x} | g(x) | &=& K ,\qquad
| g(x) |\leq\frac{K}{|x|^{d}}\nonumber
\end{eqnarray}
with positive $A, K$ and $0 < \alpha< d$.
Then,
%
\begin{equation}
\label{e:GH4}
(f*g) (x) \sim\frac{A \sum_y g(y)}{|x|^{\alpha}}\qquad
(|x| \uparrow\infty) .
\end{equation}
\end{lemma}

\begin{pf}
Parts (i) and (ii) are Proposition 1.7 of
\cite{HHS03}, and their proofs are omitted.
We now prove (iii) and (iv). The case $x=0$ is trivial, so
we only consider $x\neq0$.

(iii) Divide the sum defining $f*g$ into two, $(f*g)(x) = T_1 + T_2$,
with
%
\begin{equation}
\label{e:Hrep.GH4a-prf.1}
T_1 := \sum_{y\dvtx |y| < |x|/2} f(y) g(x-y),
\qquad
T_2 := \sum_{y\dvtx |y| \geq|x|/2} f(y) g(x-y).
\end{equation}
For $T_1$, note that $|x-y| \geq|x|/2$. Therefore,
%
\begin{eqnarray}
|T_1| &\leq&\sum_{y\dvtx |y| < |x|/2}
\frac{1}{|\!|\!|y |\!|\!|^\alpha}
\frac{K}{|x-y|^d}\nonumber\\[-8pt]\\[-8pt]
&\leq&\frac{K}{(|x|/2)^d} \sum_{y\dvtx |y| < |x|/2}
\frac{1}{|\!|\!|y |\!|\!|^\alpha}
\leq
C' K 2^{d} |x|^{-\alpha}\nonumber
\end{eqnarray}
with some constant $C'$. On the other hand, using $|y|\geq|x|/2$
for $T_2$, we have
%
\begin{eqnarray}
|T_2| &\leq&\sum_{y\dvtx |y| \geq|x|/2}
\frac{1}{|\!|\!|y |\!|\!|^\alpha} |g(x-y)|\nonumber\\[-8pt]\\[-8pt]
&\leq&\frac{1}{(|x|/2)^\alpha} \sum_{y} |g(x-y)|
\leq2^d |x|^{-\alpha} \times K .\nonumber
\end{eqnarray}
Combining the above two proves (iii).

(iv) Fix $0 < \varepsilon\ll1$, and divide the sum defining $f*g$
into three parts,
$( f * g ) (x) = S_1 + S_2 + S_3$, with
\begin{eqnarray}
\label{e:Hrep.02a}
 S_1 &:=& \sum_{y\dvtx |y| < \varepsilon|x|} f(x-y) g(y),
\nonumber\\
S_2 &:=& \sum_{y\dvtx |x-y| < \varepsilon|x|} f(x-y) g(y),
\\
S_3 &:=& (f*g)(x) - S_1 - S_2 .\nonumber
\end{eqnarray}
In the following, we prove
\begin{eqnarray}
\label{e:GH11}
S_{1} & =& \frac{A}{|x|^{\alpha}} \sum_y g(y) + o(|x|^{-\alpha})
+ \varepsilon O(|x|^{-\alpha}) ,
\nonumber\\[-8pt]\\[-8pt]
S_{2} & =& \varepsilon^{d-\alpha} O(|x|^{-\alpha}) ,\qquad
S_{3} = \varepsilon^{-\alpha} o(|x|^{-\alpha}) ,\nonumber
\end{eqnarray}
where $O(|x|^{-\alpha})$ does not, but
$o(|x|^{-\alpha})$ may, depend on $\varepsilon$.
These give, for fixed $0 < \varepsilon\ll1$,
\begin{eqnarray}
\label{e:GH12}
\limsup_{|x|\rightarrow\infty} |x|^{\alpha} (f*g)(x)
& \leq &A \sum_y g(y)
+
[
C' \varepsilon+ C'' \varepsilon^{d-\alpha}
] ,
\\
\liminf_{|x|\rightarrow\infty} |x|^{\alpha} (f*g)(x)
& \geq &A \sum_y g(y)
- [
C' \varepsilon+ C'' \varepsilon^{d-\alpha}
]
\end{eqnarray}
with some $C', C''$. Letting $\varepsilon\downarrow0$
yields
$\lim_{|x|\rightarrow\infty} |x|^{\alpha} (f*g)(x) = A \sum_y g(y)$
for $\alpha< d$, and proves the lemma.
In the following, we prove (\ref{e:GH11}).

We first note that the asymptotic condition (\ref{e:GH1}) implies
%
\begin{equation}
\label{e:GH17}
\forall\varepsilon> 0\ \exists M > 0 \mbox{ s.t.}
\qquad
\frac{A-\varepsilon}{|x|^\alpha} \leq f(x)
\leq\frac{A + \varepsilon}{|x|^{\alpha}}
\qquad\mbox{for }|x| \geq M .
\end{equation}
We also note that by taking $M'$ large, we
can have a uniform bound for all $x \in{ {\mathbb Z}^d }$:
%
\begin{equation}
\label{e:GH18}
| f(x) | \leq\frac{M'}{|\!|\!|x |\!|\!|^{\alpha}} .
\end{equation}
We fix $0 < \varepsilon\ll1$, and choose $M, M'$ as above.
We only consider sufficiently large $|x|$ depending on
$M, M'$ and $\varepsilon$.

\textit{Dealing with $S_{1}$.}
We further divide $S_1$ as
\begin{eqnarray}\qquad
\label{e:GH19}
S_{1}
& =& \sum_{y\dvtx |y| < \varepsilon|x|} \frac{A}{|x|^{\alpha}} g(y)
+ \sum_{y\dvtx |y| < \varepsilon|x|}
\biggl[ \frac{A}{|x|^{\alpha}} - \frac{A}{|x-y|^{\alpha}}
\biggr] g(y)
\nonumber\\[-8pt]\\[-8pt]
&&{}
+ \sum_{y\dvtx |y| < \varepsilon|x|}
\biggl[ f(x-y) - \frac{A}{|x-y|^{\alpha}} \biggr] g(y)
=: S_{11} + S_{12} + S_{13}.
\nonumber
\end{eqnarray}
For $S_{11}$, we have
\begin{eqnarray}\qquad
S_{11}
& :=& \sum_{y\dvtx |y| < \varepsilon|x|}
\frac{A}{|x|^{\alpha}} g(y)
=
\Biggl[ \sum_{y\in{\mathbb Z}^d} g(y) \Biggr]
\frac{A}{|x|^{\alpha}}
- \Biggl[ \sum_{|y| \geq\varepsilon|x|} g(y) \Biggr]
\frac{A}{|x|^{\alpha}}
\nonumber\\[-8pt]\\[-8pt]
&
= &\frac{A}{|x|^{\alpha}} \sum_{y} g(y) + o(|x|^{-\alpha})
,
\nonumber
\end{eqnarray}
where we used the fact that $\sum_{|y| \geq\varepsilon|x|} |g(y)|$
goes to zero as $|x| \rightarrow\infty$,
because $\sum_{y} |g(y)| < \infty$.
Here $o(|x|^{-\alpha})$ may depend on $\varepsilon$.

For $S_{12}$, note that
$|x-y|^{-\alpha} = (1+O(\varepsilon)) |x|^{-\alpha}$
for $|y| \leq\varepsilon|x|$.
So, for $0 < \varepsilon\ll1$,
\begin{eqnarray}
\label{e:GH26}
| S_{12} |
& \leq &
\sum_{y\dvtx |y| < \varepsilon|x|}
\biggl| \frac{A}{|x|^{\alpha}} - \frac{A}{|x-y|^{\alpha}}
\biggr|
| g(y) |\nonumber\\[-8pt]\\[-8pt]
&\leq&
O(\varepsilon) \frac{A}{|x|^{\alpha}}
\sum_{y\dvtx |y| < \varepsilon|x|} | g(y) |
= \varepsilon O(|x|^{-\alpha}).\nonumber
\end{eqnarray}

For $S_{13}$, we use (\ref{e:GH17}) to conclude
\begin{eqnarray}
\label{e:GH28}
| S_{13} |
& :=&
\Biggl|
\sum_{y\dvtx |y| < \varepsilon|x|}
\Biggl[ f(x-y) - \frac{A}{|x-y|^{\alpha}} \Biggr] g(y)
\Biggr|\nonumber\\
&\leq&
\sum_{y\dvtx |y| < \varepsilon|x|}
\frac{\varepsilon}{|x-y|^{\alpha}} | g(y) |
\\
& \leq&
\frac{\varepsilon}{(1-\varepsilon)^{\alpha} |x|^{\alpha}}
\sum_{y\dvtx |y| < \varepsilon|x|} | g(y) |
\leq
\frac{\varepsilon}{(1-\varepsilon)^{\alpha} |x|^{\alpha}}
\times K .\nonumber
\end{eqnarray}
As a result, we have
%
\begin{equation}
\label{e:GH29}
S_{1}(x) = \frac{A}{|x|^{\alpha}} \sum_y g(y)
+ o(|x|^{-\alpha}) + \varepsilon O(|x|^{-\alpha}) .
\end{equation}

\textit{Dealing with $S_{2}$.}
Note first that $|y| \geq(1-\varepsilon) |x|$ if
$|x-y| < \varepsilon|x|$. So using (\ref{e:GH18}) to bound $f$ and
the pointwise bound on $g$ of (\ref{e:GH1}), we get
\begin{eqnarray}
\label{e:GH33}
| S_{2}(x) |
& :=&
\Biggl|
\sum_{y\dvtx |x-y| < \varepsilon|x|} f(x-y) g(y)
\Biggr|
\nonumber\\
&
\leq&
\Biggl( \sum_{y\dvtx |x-y| < \varepsilon|x|} | f(x-y) |
\Biggr)
\times
\biggl( \sup_{y\dvtx |x-y| < \varepsilon|x|} | g(y) |
\biggr)
\\
& \leq&
\Biggl( \sum_{y\dvtx |x-y| < \varepsilon|x|}
\frac{M'}{|\!|\!|x-y |\!|\!|^{\alpha}} \Biggr)
\times\frac{C'}{|x|^{d}}
= \varepsilon^{d-\alpha} O(|x|^{-\alpha})
.\nonumber
\end{eqnarray}

\textit{Dealing with $S_{3}$.}
We use (\ref{e:GH17}) (we only consider those $x$'s
satisfying $\varepsilon|x| \geq M$) and calculate as
\begin{eqnarray}\qquad
\label{e:GH41}
| S_{3}(x) |
&:=&
\Biggl| \mathop{\mathop{\sum}_{{
y\dvtx |x-y| \geq\varepsilon|x|}}}_{|y| \geq\varepsilon|x|}
f(x-y) g(y)\Biggr |
\leq
\mathop{\mathop{\sum}_{{
y\dvtx |x-y| \geq\varepsilon|x|}}}_{|y| \geq\varepsilon|x|}
\frac{A+\varepsilon}{|x-y|^{\alpha}} | g(y) |
\nonumber\\
&\leq&
\mathop{\mathop{\sum}_{{
y\dvtx |x-y| \geq\varepsilon|x|}}}_{|y| \geq\varepsilon|x|}
\frac{A+\varepsilon}{(\varepsilon|x|)^{\alpha}}
| g(y) |\\
&\leq&
\frac{A+\varepsilon}{\varepsilon^{\alpha} |x|^{\alpha}} \times
\sum_{|y| \geq\varepsilon|x|} | g(y) |
= \varepsilon^{-\alpha} o(|x|^{-\alpha}) .\nonumber
\end{eqnarray}
The last equality follows again because $|g(y)|$ is summable,
and $o(|x|^{-\alpha})$ may depend on $\varepsilon$.
\end{pf}
\end{appendix}

\section*{Acknowledgments}
This work is a continuation of \cite{HHS03} which
started when I was staying at the Fields Institute in the fall of
1998, whose hospitality is gratefully
acknowledged. I am very much indebted to Remco van der Hofstad
and Gordon Slade for stimulating discussions in the early
stages of this work, and for enjoyable collaborations which led
to \cite{HHS03}.
I am indebted to K\^{o}hei Uchiyama and Gregory Lawler for
valuable discussions and letting me know of their unpublished
works. I also thank Akira Sakai, Remco van der Hofstad,
Gordon Slade, and Tetsuya Hattori for constructive comments
on previous versions of the manuscript.

Parts of the results reported in this paper were obtained while
I was at the Department of Mathematics, Tokyo Institute of Technology
(1998--2000), and the rest was obtained at the Faculty of Mathematics,
Kyushu University (2004--). I am deeply indebted to everyone
who expressed his/her concern and encouragements during the
four years while I did not work on this problem.

\printaddresses


\begin{thebibliography}{99}

\bibitem{Aize97}
\textsc{Aizenman, M.} (1997).
On the number of incipient spanning clusters.
\textit{Nuclear. Phys. B} \textbf{485} 551--582.
\MR{1431856}

\bibitem{AB87}
\textsc{Aizenman, M.} and \textsc{Barsky, D. J.} (1987).
Sharpness of the phase transition in percolation models.
\textit{Comm. Math. Phys.} \textbf{108} 489--526.
\MR{0874906}

\bibitem{BS85}
\textsc{Brydges, D. C.} and \textsc{Spencer, T.} (1985).
Self-avoiding walk in 5 or more dimensions.
\textit{Comm. Math. Phys.} \textbf{97} 125--148.
\MR{0782962}

\bibitem{Grim99}
\textsc{Grimmett, G.} (1999).
\textit{Percolation}, 2nd ed.
Springer, Berlin.
\MR{1707339}

\bibitem{HM54}
\textsc{Hammersley, J. M.} and \textsc{Morton, K. W.} (1954).
Poor man's {Monte} {Carlo}.
\textit{J. Roy. Statist. Soc. Ser. B} \textbf{16} 23--38.
\MR{0064475}

\bibitem{Hara07b}
\textsc{Hara, T.} Paper in preparation.

\bibitem{HHS03}
\textsc{Hara, T.}, \textsc{van der Hofstad, R.}
and \textsc{Slade, G.} (2003).
Critical two-point functions and the lace expansion for spread-out
high-dimensional percolation and related models.
\textit{Ann. Prob.} \textbf{31} 349--408.
\MR{1959796}

\bibitem{HS90a}
\textsc{Hara, T.} and \textsc{Slade, G.} (1990).
Mean-field critical behaviour for percolation in high dimensions.
\textit{Comm. Math. Phys.} \textbf{128} 333--391.
\MR{1043524}

\bibitem{HS90b}
\textsc{Hara, T.} and \textsc{Slade, G.} (1990).
On the upper critical dimension of lattice trees and lattice animals.
\textit{J. Statist. Phys.} \textbf{59} 1469--1510.
\MR{1063208}

\bibitem{HS92b}
\textsc{Hara, T.} and \textsc{Slade, G.} (1992).
The lace expansion for self-avoiding walk in five or more dimensions.
\textit{Rev. Math. Phys.} \textbf{4} 235--327.
\MR{1174248}

\bibitem{HS92a}
\textsc{Hara, T.} and \textsc{Slade, G.} (1992).
Self-avoiding walk in five or more dimensions. {I.}
{The} critical behaviour.
\textit{Commun.\ Math.\ Phys.} \textbf{147} 101--136.
\MR{1171762}

\bibitem{HS94}
\textsc{Hara, T.} and \textsc{Slade, G.} (1994).
Mean-field behaviour and the lace expansion.
In \textit{Probability and Phase Transition} (G. Grimmett, ed.)
87--122.
Kluwer, Dordrecht.
\MR{1283177}

\bibitem{Hugh95}
\textsc{Hughes, B. D.} (1995).
\textit{Random Walks and Random Environments}. \textbf{1}. \textit{Random Walks}.
Oxford Univ. Press.
\MR{1341369}

\bibitem{Klar67}
\textsc{Klarner, D. A.} (1967).
Cell growth problems.
\textit{Canad. J. Math.} \textbf{19} 851--863.
\MR{0214489}

\bibitem{Klei81}
\textsc{Klein, D. J.} (1981).
Rigorous results for branched polymer models with excluded volume.
\textit{J. Chem. Phys.} \textbf{75} 5186--5189.

\bibitem{Lawl91}
\textsc{Lawler, G. F.} (1991).
\textit{Intersections of Random Walks}.
Birkh\"{a}user, Boston.
\MR{1117680}

\bibitem{Lawl94a}
\textsc{Lawler, G. F.} (1994).
A note on {G}reen's function for random walk in four dimensions.
Preprint 94-03, Duke Univ.

\bibitem{Lawl04a}
\textsc{Lawler, G. F.} (2004).
Private communication.

\bibitem{MS93}
\textsc{Madras, N.} and \textsc{Slade, G.} (1993).
\textit{The Self-Avoiding Walk}.
Birkh{\"a}user, Boston.
\MR{1197356}

\bibitem{Mens86}
\textsc{Menshikov, M. V.} (1986).
Coincidence of critical points in percolation problems.
\textit{Soviet Math. Dokl.} \textbf{33} 856--859.
\MR{0852458}

\bibitem{Reis88b}
\textsc{Reisz, T.} (1988).
A convergence theorem for lattice {Feynman} integrals with
massless propagators.
\textit{Comm. Math. Phys.} \textbf{116} 573--606.
\MR{0943703}

\bibitem{Reis88a}
\textsc{Reisz, T.} (1988).
A power counting theorem for {Feynman} integrals on the lattice.
\textit{Comm. Math. Phys.} \textbf{116} 81--126.
\MR{0937362}

\bibitem{Sakai06}
\textsc{Sakai, A.} (2007).
Lace expansion for the Ising model.
\textit{Comm. Math. Phys.} \textbf{272} 283--344.

\bibitem{SKM93}
\textsc{Samko, S. G.}, \textsc{Kilbas, A. A.} and
\textsc{Marichev, O. I.} (1993).
\textit{Fractional Integrals and Derivatives}: \textit{Theory
and Applications}.
Gordon and Breach, New York.
\MR{1347689}

\bibitem{Slad87}
\textsc{Slade, G.} (1987).
The diffusion of self-avoiding random walk in high dimensions.
\textit{Comm. Math. Phys.} \textbf{110} 661--683.
\MR{0895223}

\bibitem{Slad04}
\textsc{Slade, G.} (2006).
\textit{The Lace Expansion and Its Applications.
Lecture Notes in Math.} \textbf{1879}.
Springer, Berlin.
\MR{2239599}

\bibitem{Soka82}
\textsc{Sokal, A. D.} (1982).
An alternate constructive approach to the $\varphi^4_3$
quantum field theory, and a possible destructive
approach to $\varphi^4_4$.
\textit{Ann. Inst. H. Poincar\'{e}}
\textit{Sect. A} (\textit{N.S.}) \textbf{37} 317--398.
\MR{0693644}

\bibitem{Uchi98}
\textsc{Uchiyama, K.} (1998).
Greens' function for random walks on ${Z}^{N}$.
\textit{Proc. London Math. Soc.} \textbf{77} 215--240.
\MR{1625467}

\end{thebibliography}
\end{document}